\documentclass{aa}

\usepackage{graphicx}
\usepackage{subfigure}
\usepackage{txfonts}
\usepackage{hyperref}

\begin{document}
	
	\title{SiO Outflows in the Most Luminous and Massive Protostellar Sources of the Southern Sky
	}
	
	\author{N. Guerra-Varas \inst{1}$^,$  \inst{2}$^,$  \inst{3}
		\and
		M. Merello \inst{1}
		\and
		L. Bronfman \inst{1}
		\and
		N. Duronea \inst{4}$^,$  \inst{5}
		\and
		D. Elia\inst{6}
		\and
		R. Finger \inst{1}
		\and
		E. Mendoza \inst{7}
	}

	\institute{Departamento de Astronomía, Universidad de Chile, Santiago, Chile\\
		\email{ \href{mailto:nguerra@ug.uchile.cl}{nguerra@ug.uchile.cl} }
		\and
		Dipartimento di Fisica, Università di Roma "Tor Vergata", Via della Ricerca Scientifica 1, I-00133, Roma, Italy
		\and
        Department of Astronomy, University of Belgrade - Faculty of Mathematics, Studentski trg 16, 11000 Belgrade, Serbia
        \and
		Instituto de Astrof\'isica de La Plata (UNLP-CONICET), La Plata, Argentina 
		\and
		Faculdad de Ciencias Astron\'omicas y Geof\'isicas, Universidad Nacional de La Plata, Paseo del Bosque s/n, 1900, La Plata, Argentina
		\and
		INAF - IAPS, via Fosso del Cavaliere, 100, I-00133 Roma, Italy
		\and
		Dept. Ciencias Integradas, Facultad de Ciencias Experimentales, Centro de Estudios Avanzados en F\'isica, Matem\'atica y Computaci\'on, Unidad Asociada GIFMAN, CSIC-UHU, Universidad de Huelva, Spain 
	}

	\date{July, 2023; Accepted to A\&A}
	
	\abstract
	{High-mass star formation remains far less understood than low-mass star formation. It entails the ejection of matter through molecular outflows, which disturb the protostellar clump. Studying these outflows and the shocked gas caused by them is key for a better understanding of this process.
	}
	{The present study aims to characterise the behaviour of molecular outflows in the most massive protostellar sources in the Southern Galaxy by looking for evolutionary trends and associating the presence of shocked gas with outflow activity.
	}
	{We present APEX SEPIA180 (Band 5) observations (beamwidth $\sim$36”) of SiO(4-3) molecular outflow candidates towards a well-selected sample of 32 luminous and dense clumps, which are candidates to harbouring Hot Molecular Cores. We study the emission of the SiO(4-3) line, which is an unambiguous tracer of shocked gas, and recent and active outflow activity, as well as the HCO$^+$(2-1) and H$^{13}$CO$^+$(2-1) lines.
	}
	{Results show that 78\% of our sample (25 sources) present SiO emission, revealing the presence of shocked gas. Nine of these sources are also found to have wings in the HCO$^+$(2-1) line, indicating outflow activity. The SiO emission of these 9 sources is generally more intense ($T_a > 1$ K) and wider ($\sim61$ km s$^{-1}$ FWZP) than the rest of the clumps with SiO detection ($\sim42$ km s$^{-1}$ FWZP), suggesting that the outflows in this group are faster and more energetic. 
    Three positive linear correlations are found: a weak one between the bolometric luminosity and outflow power, and two strong ones: between the outflow power and the rate of matter expulsion, and between the kinetic energy and outflow mass. These correlations suggest that more energetic outflows bear to mobilise more material. No correlation was found between the evolutionary stage indicator $L/M$ and SiO outflow properties, supporting that molecular outflows happen throughout the whole high-mass star formation process.
	}
	{We conclude that sources with both SiO emission and HCO$^+$ wings and sources with only SiO emission are in an advanced stage of evolution in the high-mass star formation process, and there is no clear evolutionary difference between them. The former present more massive and more powerful SiO outflows than the latter. Therefore, looking for more outflow signatures such as HCO$^+$ wings could help identify more massive and active massive star-forming regions in samples of similarly evolved sources, and could also help identify sources with older outflow activity.
	}
	
	
	\maketitle
	

    \nolinenumbers
 
	\section{Introduction} \label{intro}
	
	Massive stars are crucial to the evolution of the interstellar medium (ISM) and galaxies. However, they are difficult to observe and study because they are very scarce (about 1\% of stellar populations), have short timescales, large heliocentric distances, and are embedded in very complex environments, with high extinction and turbulence \cite[e.g.][]{Motte_2018}. Thus, high-mass star formation remains much less understood than low-mass star formation.
	
	The current picture for high-mass star formation takes place in massive dense cores (MDCs) embedded in massive clouds, called Massive Star-Forming (MSF) regions, and can be described by four main phases \citep{Tak_2004, Motte_2018, Elia_2021_HiGAL, Urquhart_2022, Jiao_2023}. Initially, in the quiescent or starless phase, the clump has no embedded objects and is not visible at 70 $\mu$m. Later, in the Young Stellar Object Phase (YSO), the clump has warmed up enough to be detected at 70 $\mu$m. In this stage, ($\sim 10^4$ yrs), there is an embedded cold and collapsing prestellar core. When a protostar appears (protostellar phase, $\sim 3 \times 10^5$ yrs), it feeds on inflow material and its mass increases until becoming a high-mass protostar. Within this phase, the Hot Molecular Core (HMC) stage can be identified. Then, as the protostar evolves and its temperature increases, it starts emitting ultraviolet (UV) radiation, which is quickly ionising the surrounding gas. This starts the Ultra-Compact (UC) HII region phase ($\sim 10^5$ - $10^6$ yrs).
 
	Massive star formation entails a greatly relevant feedback process: molecular outflows and jets, i.e., matter expulsion at high velocities due to angular momentum conservation during matter accretion \citep{Arce_2007, Bally_2016, Motte_2018}. This feedback process occurs throughout the whole of the high-mass formation process \citep{Li_ashes_2020, Yang_2022, Urquhart_2022}, and outflow features have been used as an indication of massive star formation activity \citep{Li_2019_dark_clouds, Liu_2020_chemistry}. In order to understand how massive stars form, the study of molecular outflows in MSF regions is imperative.
	
	The study of outflows heavily depends on the tracer used. There is no such thing as a perfect outflow tracer \citep{Bally_2016}. However, the SiO molecule has been found to be a very good tracer of outflow activity and shocked material and has been broadly used for that purpose \cite[e.g.][]{Beuther_2002, Lopez_2011, Bally_2016, Li_2019_survey, Liu_2021_SiO, Liu_2021, Liu_Rong_2022, De_Simone_2022}. Si falls onto the icy mantles of dust grains of the ISM. When hit by shocks of gas, the dust grains can sublimate to the gaseous state, releasing the Si. Thus, the Si in the gas phase can react with O to form SiO, making it observable in the millimetre (mm), sub-mm, and centimetre (cm) regimes via its rotational transitions \citep{Schilke_1997, Klaassen_2007, Gusdorf_2008}. Unlike other molecules, SiO has a key advantage. Its emission is not easily contaminated by excited ambient material, making it an unambiguous tracer of shocked material and thus outflow activity. Studying SiO emission can shed light on the kinematics of outflows, as well as relevant chemical processes that occur in shock environments such as the formation of complex organic molecules (COMs) \citep{Bally_2016, Li_ashes_2020, RojasGarcia_2022}. Both broad (Full Width Zero Power (FWZP) $\geq 20$ km s$^{-1}$) and narrow (FWZP $\leq 10$ km s$^{-1}$) spectral width SiO emission have been observed \citep{Garay_2010, Leurini_2014, Bally_2016, Csengeri_2016, Li_2019_survey, Zhu_2020}. It is thought that broad-width emission is due to high-velocity collimated shocks, while narrow-width emission is associated with less collimated low-velocity shocks, such as cloud-cloud collisions. It is possible to identify wings in the SiO spectral profile when it has a broad width, which are manifestations of SiO outflows. These have been extensively studied \citep{Beuther_2002, Liu_2021_SiO, Liu_2021}. Moreover, SiO outflow activity has been found to decrease and get less collimated over time \citep{Arce_2007, Sakai_2010, Lopez_2011}. However, there is currently no consensus on whether SiO abundance decreases over time \citep{Csengeri_2016, Li_2019_survey, Liu_Rong_2022}, and some works even conclude that SiO emission is much harder to interpret than just a shock tracer \citep{Widmann_2016}.
	
	Another very helpful outflow tracer is HCO$^+$ emission \citep{Myers_1996, Rawlings_2004, Klaassen_2007, Bally_2016, Li_2019, Liu_2020_chemistry, He_2021}. This species traces the surrounding material of protostar regions \citep{Rawlings_2004} and traces the disk material in low-mass star formation \citep{Dutrey_1997}. This species can trace both inflow and outflow motions \citep{Klaassen_2007}. When outflows are strong enough, they can be observed in broad high-velocity wings in the spectral profile \citep{Wu_2005, He_2021}. When HCO$^+$, and other species such as CO, are dragged outwards due to the outflow, their spectral profiles present wings. This is caused by the greater velocity gradient of the dragged material. However, the spectral profile of this species experiences significant absorption, which often complicates its analysis. Moreover, studying the emission of an HCO$^+$ isotopologue, H$^{13}$CO$^+$, can help distinguish between outflows and ambient dense gas, as it traces dense gas only (see Section \ref{subsec_clump_tracer}).

	SiO line emissions are associated with recent and active outflow of matter, as shock chemistry processes happen in $10^2$ to $10^4$ yrs, whilst the HCO$^+$ wing observations are associated with old `fossil' outflows \citep{Arce_2007, Klaassen_2007, Lopez_Sepulcre_2016, Li_2019}. The detection of only one of these phenomena is enough to indicate the presence of a molecular outflow. If both are detected, then one can confirm without ambiguity that there is ongoing matter expulsion \citep{Klaassen_2007}. Outflows are associated with an advanced stage of evolution. If a molecular core does not have an outflow, it may be because it has not reached this point yet. However, it is also possible that it has already experienced a significant loss of material, and/or accretion and outflow activity has ceased.

	In this work, we present SiO(2-1), HCO$^+$(2-1) and H$^{13}$CO$^+$(2-1) APEX Band 5 observations toward a well-selected sample of 32 very massive and luminous protostellar sources, which are amongst the brightest clumps in the Southern Milky Way. This study aims to answer the following research questions: How do SiO outflows behave in the most massive protostellar sources? Do any SiO outflow properties exhibit an evolutionary trend? Is the presence of shocked gas associated with outflow activity? What can different outflow signatures tell us about similarly evolved sources?
	
	This paper is organised as follows: in Sect. \ref{observations} we describe the studied sample, selection criteria and the observations. In Sect. \ref{analysis}, we analyse the data and describe how outflow and SiO detections were analysed. In Sect. \ref{results}, we calculate the outflows properties of the clumps and the SiO emission. In Sect. \ref{discussions}, we present relevant correlations, carry out a cross-check between SiO and outflow detections, and discuss possible evolutionary trends. Finally, in Sect. \ref{conclusions}, we provide a summary of our main results and present our conclusions.

	
	\section{Observations} \label{observations}
	
	\subsection{Source Selection} \label{subsec_source_sel}
	
	The studied sample consists of 32 protostellar clumps (HMC candidates) from the Hi-GAL catalogue of compact sources \citep{Elia_2021_HiGAL}, associated with CS(2-1) line emission from IRAS PSC sources with far infrared colours of Ultra Compact HII regions by \cite{Bronfman_1996}. Their characteristics are presented in Table \ref{tab:sources} (sources marked with a black diamond $\blacklozenge$ have saturated Hi-GAL observations; see Appendix \ref{ap_saturated}). They have been named with `HC' (Hot Core) and a number in increasing order for convenience.The sources were selected as follows:
	
	\begin{enumerate}
		\item Kinematic distances < 6 kpc (these were obtained from \cite{Reid_2009}, Hi-GAL distances were not used; see Appendix \ref{ap_distances}).
		
		\item Strong CS emission \citep{Bronfman_1996}: Main-beam temperature in CS $T_{MB}(\text{CS}) \geq 1.5$ K.
		
		\item Surface densities above the threshold for the formation of massive stars: $\Sigma > 0.2$ g cm$^{-2}$ \cite[e.g.][]{Butler_2012, Merello_2019}.
		
		\item Masses $> 100$ $M_{\odot} $ \citep{Elia_2021_HiGAL}.
		
		\item High values of the evolutionary stage indicator luminosity-to-mass ratio $L/M$. Values larger than $1$ are associated with internal protostellar heating and the appearance of new stars in massive clumps \citep{Lopez_2011, Molinari_2016}.
	\end{enumerate}
	
	Although low-mass star-forming clumps may also have a large value of $L/M$, the lower limit set on the mass of the clump together with a large $L/M$ ensures our sample will only consist of massive protostellar clumps.
	
\begin{figure}[]
		\centering
		\includegraphics[width=0.5\textwidth]{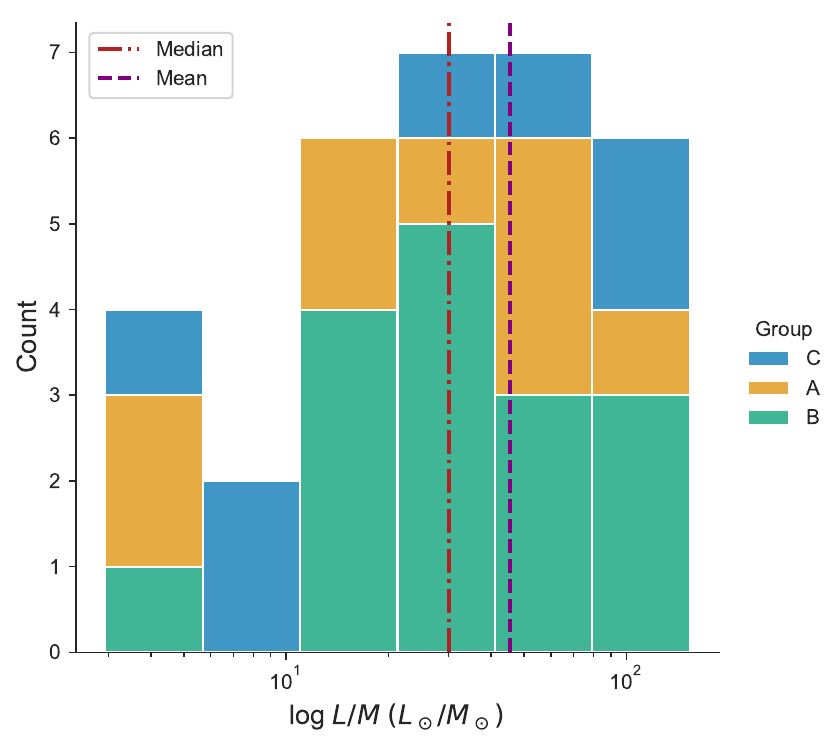}
		\caption{Histogram for the luminosity to mass ratio $L/M$ in logarithmic scale. The red and purple vertical lines denote the median ($30$ $L_{\odot}/M_{\odot}$) and the mean ($45$ $L_{\odot}/M_{\odot}$) respectively. Groups A, B and C are defined in Table \ref{tab:cross_check_groups}.}
		\label{fig: L/M hist}
	\end{figure}

	The sources in our sample are in an advanced stage of evolution, either in the protostellar or compact HII region phase. A histogram of the luminosity-to-mass ratio $L/M$, which acts as an evolutionary stage indicator, is presented in Fig. \ref{fig: L/M hist} (see Table \ref{tab:sources}). This parameter $L/M$ has a minimum value of $3$ $L_{\odot} / M_{\odot}$, a maximum of $154$ $L_{\odot} / M_{\odot}$, a mean of 45 $L_{\odot} / M_{\odot}$ and a median of 30 $L_{\odot} / M_{\odot}$.
	
\begin{figure}[]
		\centering
		\includegraphics[width=0.54\textwidth]{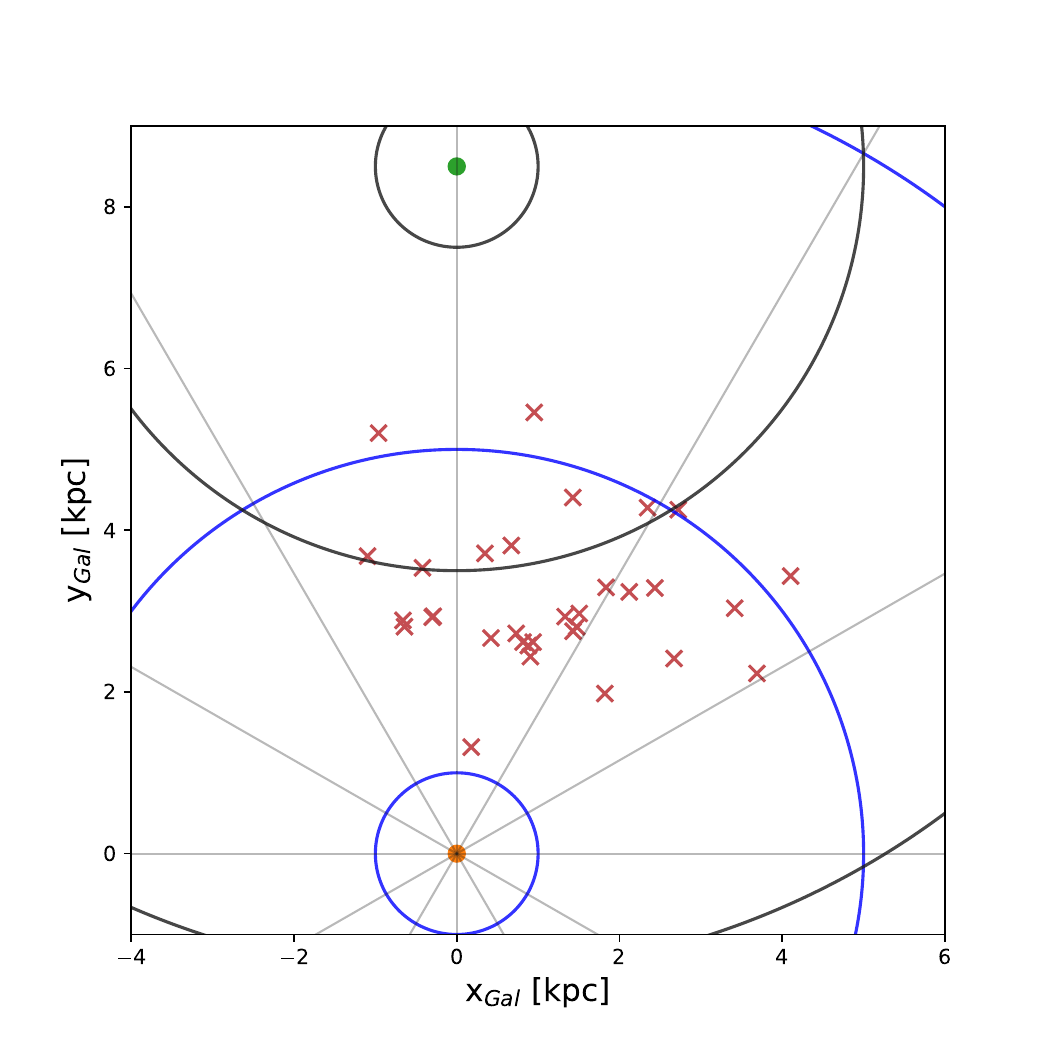}
		\caption{Galactic distribution of the sources. The orange point corresponds to the Sun and the green point to the Galactic centre. The red crosses are the sources studied in this paper. The blue circles are centred in the Sun and the black ones in the Galactic centre. They mark the 1 kpc, 5 kpc and 10 kpc distances.}
		\label{fig: mapa galaxia}
	\end{figure}

	The sources are very energetic, with bolometric luminosities up to $\sim 2.3 \times 10^5 L_{\odot}$. The selection criteria of the sources allowed us to work with luminous, strong and relatively near clumps only. They are among the most luminous sources in the southern sky, and some of them have been extensively studied (see Appendix \ref{ap_sources}). The continuum dust properties of the sources are shown in Table \ref{tab:sources_SED} (sources marked with a black diamond $\blacklozenge$ have saturated Hi-GAL observations; see Appendix \ref{ap_saturated}) and were obtained from a SED fit \citep{Elia_2021_HiGAL}. Compared with the general population of Hi-GAL protostellar sources \citep{Elia_2017_HiGAL, Elia_2021_HiGAL}, we see that our sample is much more massive and luminous. The median value for H$_2$ mass derived from dust $M_{dust}$ and bolometric luminosity $L_{Bol}$ for our sample is 828 $M_{\odot}$ and 19839 $L_{\odot}$ respectively, whilst for Hi-GAL protostellar sources, these values are equal to 464 $M_{\odot}$ and 1071 $L_{\odot}$ respectively. When comparing with the rest of the Hi-GAL sources within 6 kpc, the contrast is even bigger, since the mean bolometric luminosity is 206 $L_{\odot}$ and the mean dust mass is 88 $M_{\odot}$ in this region of the Galaxy. Finally, Fig. \ref{fig: mapa galaxia} shows the distribution in the Galactic plane of the sources presented in this work.
	
	\begin{table*}[]
		\centering
		\begin{tabular}{lcccccccc}
			\hline \hline
			Source & Hi-GAL Name & IRAS Name & $\alpha$ & $\delta$ & $D$ & $V_{LSR}$(CS) \\
                 &  &  & (J2000.0) & (J2000.0) & (kpc) & (km s$^{-1}$) \\ \hline
			HC01 &   HIGALBM5.6373+0.2352 & 17545-2357 & 17:57:35.0328 & -23:58:08.472 &       2.95 &    9.1 \\
			HC02 $\blacklozenge$ &   HIGALBM5.8856-0.3920 & 17574-2403 & 18:00:30.4080 & -24:03:59.663 &       2.94 &    9.3 \\
			HC03 &   HIGALBM6.7951-0.2571 & 17589-2312 & 18:01:57.7152 & -23:12:34.844 &       3.56 &   21.1 \\
			HC04 $\blacklozenge$ &  HIGALBM10.4631+0.0298 & 18056-1952 & 18:08:36.5496 & -19:52:14.534 &       5.29 &   68.9 \\
			HC05 &  HIGALBM12.8892+0.4895 & 18089-1732 & 18:11:51.4224 & -17:31:27.620 &       2.88 &   33.8 \\
			HC06 &  HIGALBM12.9083-0.2603 & 18117-1753 & 18:14:39.5184 & -17:52:00.710 &       2.96 &   36.6 \\
			HC07 &  HIGALBM16.5849-0.0505 & 18182-1433 & 18:21:09.0672 & -14:31:49.184 &       3.84 &   59.1 \\
			HC08 & HIGALBM301.1365-0.2259 & 12326-6245 & 12:35:35.1552 & -63:02:32.633 &       4.31 &  -39.4 \\
			HC09 & HIGALBM309.9212+0.4781 & 13471-6120 & 13:50:42.2184 & -61:35:12.466 &       5.35 &  -58.4 \\
			HC10 & HIGALBM311.6269+0.2898 & 14013-6105 & 14:04:55.0536 & -61:20:07.120 &       4.57 &  -56.2 \\
			HC11 & HIGALBM312.1074+0.3094 & 14050-6056 & 14:08:41.9400 & -61:10:42.391 &       3.60 &  -48.4 \\
			HC12 & HIGALBM317.4080+0.1102 & 14453-5912 & 14:49:07.8000 & -59:24:45.623 &       2.69 &  -39.9 \\
			HC13 & HIGALBM323.4594-0.0789 & 15254-5621 & 15:29:19.5696 & -56:31:21.918 &       4.09 &  -67.5 \\
			HC14 & HIGALBM326.7806-0.2411 & 15450-5431 & 15:48:55.2096 & -54:40:38.525 &       3.87 &  -64.7 \\
			HC15 & HIGALBM327.3918+0.1996 & 15464-5348 & 15:50:18.4992 & -53:57:05.612 &       5.05 &  -88.6 \\
			HC16 $\blacklozenge$ & HIGALBM330.8814-0.3656 & 16065-5158 & 16:10:20.4504 & -52:05:56.908 &       3.77 &  -62.2 \\
			HC17 & HIGALBM331.2786-0.1883 & 16076-5134 & 16:11:26.7480 & -51:41:56.044 &       4.88 &  -87.7 \\
			HC18 & HIGALBM332.2951-0.0938 & 16119-5048 & 16:15:45.4104 & -50:55:54.512 &       3.17 &  -48.4 \\
			HC19 & HIGALBM332.5443-0.1226 & 16132-5039 & 16:17:01.2000 & -50:46:46.643 &       3.10 &  -46.7 \\
			HC20 $\blacklozenge$ & HIGALBM333.1246-0.4244 & 16172-5028 & 16:20:58.1904 & -50:35:19.190 &       3.33 &  -51.2 \\
			HC21 & HIGALBM335.5848-0.2894 & 16272-4837 & 16:30:58.5120 & -48:43:52.183 &       3.22 &  -46.6 \\
			HC22 & HIGALBM339.6221-0.1209 & 16424-4531 & 16:46:06.1080 & -45:36:44.003 &       2.60 &  -33.2 \\
			HC23 & HIGALBM340.2740-0.2098 & 16452-4504 & 16:48:53.3160 & -45:10:20.420 &       2.78 &  -45.5 \\
			HC24 & HIGALBM341.1268-0.3456 & 16489-4431 & 16:52:33.2568 & -44:36:10.613 &       2.72 &  -40.6 \\
			HC25 & HIGALBM342.0590+0.4212 & 16489-4318 & 16:52:32.8656 & -43:23:45.553 &       4.63 &  -71.0 \\
			HC26 & HIGALBM342.7076+0.1257 & 16524-4300 & 16:56:03.1296 & -43:04:43.489 &       2.74 &  -41.9 \\
			HC27 & HIGALBM345.0035-0.2239 & 17016-4124 & 17:05:11.1168 & -41:29:05.665 &       2.82 &  -26.6 \\
			HC28 & HIGALBM350.0154+0.4332 & 17143-3700 & 17:17:45.4032 & -37:03:11.642 &       3.87 &  -31.5 \\
			HC29 & HIGALBM350.1024+0.0805 & 17160-3707 & 17:19:27.3216 & -37:11:05.788 &       5.54 &  -69.5 \\
			HC30 & HIGALBM351.0412-0.3354 & 17204-3636 & 17:23:50.2800 & -36:38:57.138 &       2.70 &  -18.0 \\
			HC31 & HIGALBM352.3155-0.4422 & 17244-3536 & 17:27:48.6096 & -35:39:11.207 &       1.33 &  -10.8 \\
			HC32 & HIGALBM354.6619+0.4800 & 17271-3309 & 17:30:22.6008 & -33:11:17.668 &       3.73 &  -20.9 \\
			\hline
		\end{tabular}
		\caption{List of the studied sample sources.}
		\label{tab:sources}
	\end{table*}

	\begin{table}[]
		\centering
		\begin{tabular}{lcccc}
			\hline \hline
			Source & $L_{bol}$ & $M_{dust}$ & $T_{dust}$ & ${L/M}$ \\
                 & ($10^3 L_{\odot}$) & ($M_{\odot}$) & (K) & ($L_{\odot}/M_{\odot}$) \\\hline
			HC01 &  11.14 &   394.0 &    25.8 &       28.0 \\
			HC02 $\blacklozenge$ &  16.75 &  2017.0 & 25.0 & 72.0 \\
			HC03 &  20.57 &  1141.0 &    23.1 &       18.0 \\
			HC04 $\blacklozenge$ &   0.77 &  8675.0 & 25.0 & 25.0 \\
			HC05 &  22.04 &   540.0 &    33.4 &       41.0 \\
			HC06 &  42.28 &  1328.0 &    22.6 &       32.0 \\
			HC07 &  17.47 &   535.0 &    31.5 &       33.0 \\
			HC08 & 227.44 &  2803.0 &    33.5 &       81.0 \\
			HC09 & 222.95 &  1477.0 &    34.4 &      151.0 \\
			HC10 & 121.44 &  1458.0 &    30.4 &       83.0 \\
			HC11 &  68.39 &   701.0 &    25.6 &       98.0 \\
			HC12 &   2.70 &   785.0 &    18.3 &        3.0 \\
			HC13 & 142.72 &   929.0 &    31.5 &      154.0 \\
			HC14 &   8.67 &   799.0 &    19.4 &       11.0 \\
			HC15 &  13.55 &  1074.0 &    22.9 &       13.0 \\
			HC16 $\blacklozenge$ &  56.05 &  1962.0 & 25.0 & 3.0 \\
			HC17 & 100.00 &  1400.0 &    32.0 &       71.0 \\
			HC18 &  14.98 &   949.0 &    22.4 &       16.0 \\
			HC19 &   2.22 &   756.0 &    18.8 &        3.0 \\
			HC20 $\blacklozenge$ & 114.35 &  4980.0 & 25.0 & 3.0 \\
			HC21 &  19.10 &  1608.0 &    23.3 &       12.0 \\
			HC22 &  13.86 &   321.0 &    24.1 &       43.0 \\
			HC23 &   4.19 &   341.0 &    33.4 &       12.0 \\
			HC24 &   6.12 &   390.0 &    20.0 &       16.0 \\
			HC25 &  23.05 &  1001.0 &    23.2 &       23.0 \\
			HC26 &  24.66 &   585.0 &    28.8 &       42.0 \\
			HC27 &  67.66 &  1030.0 &    34.0 &       66.0 \\
			HC28 &  33.62 &   230.0 &    40.0 &      146.0 \\
			HC29 & 162.23 &  2083.0 &    27.5 &       78.0 \\
			HC30 &  11.50 &   456.0 &    27.0 &       25.0 \\
			HC31 &   4.56 &   105.0 &    25.9 &       43.0 \\
			HC32 &   6.34 &   859.0 &    20.6 &        7.0 \\
			\hline
		\end{tabular}
		\caption{Clump properties.}
		\label{tab:sources_SED}
	\end{table}
	
	\subsection{APEX Observations} \label{subsec_instr}
	
	Observations were made using the SEPIA180 instrument at the Atacama Pathfinder Experiment (APEX)\footnote{\href{https://www.apex-telescope.org/ns/instruments/sepia/sepia180/}{https://www.apex-telescope.org/ns/instruments/sepia/sepia180/}} in the Atacama Desert in Chile \citep{Belitsky_2018_APEX}, for 10 nights between November 2018 and October 2019. The observations were made in single-pointing mode. For each source, a region free from emission was taken as reference from the dust maps at 250 $\mu$m by Herschel, typically at $300-500 ''$ from the source. The system temperature was 174.30 K on average.
	
	The band was tuned at 172 GHz in the LSB to observe the H$^{13}$CO$^+$(2-1) (173.507 GHz) and SiO(4-3) (173.688 GHz) lines, and centred at the HCO$^+$(2-1) line (178.375 GHz). The spectral resolution used was of $\Delta V = 0.2$ km s$^{-1}$, which gives a typical RMS noise temperature of $25$ mK, and the main beam efficiency $\eta_{MB}$ was equal to $0.83$. APEX observations have an antenna temperature uncertainty of about 20 \% (APEX staff, private communication). The frequency of the transitions used was obtained from the SPLATALOGUE\footnote{\href{https://splatalogue.online//advanced1.php}{https://splatalogue.online//advanced1.php}} database. This frequency range corresponds to the Band 5 of the Atacama Large Millimetre/Sub-millimetre Array.
	
	The beam angular size can be calculated with the frequency $f$ of a spectral line with $\theta '' = 7.8'' \times 800 / f$. For these observations, this results in a beam angular size of approximately $36''$.
	
	The CLASS software from the GILDAS\footnote{\href{https://www.iram.fr/IRAMFR/GILDAS/}{https://www.iram.fr/IRAMFR/GILDAS/}} software package was used to reduce the spectra. It is used to process single–dish spectra and is oriented towards the processing of a large number of spectra. For every spectral line, the baseline was subtracted at first order and then centred to the $V_{LSR}$ of the corresponding source.
	
	
	\section{Analysis} \label{analysis}
	
\begin{figure}[h]
		\centering
		\includegraphics[width=0.3\textwidth]{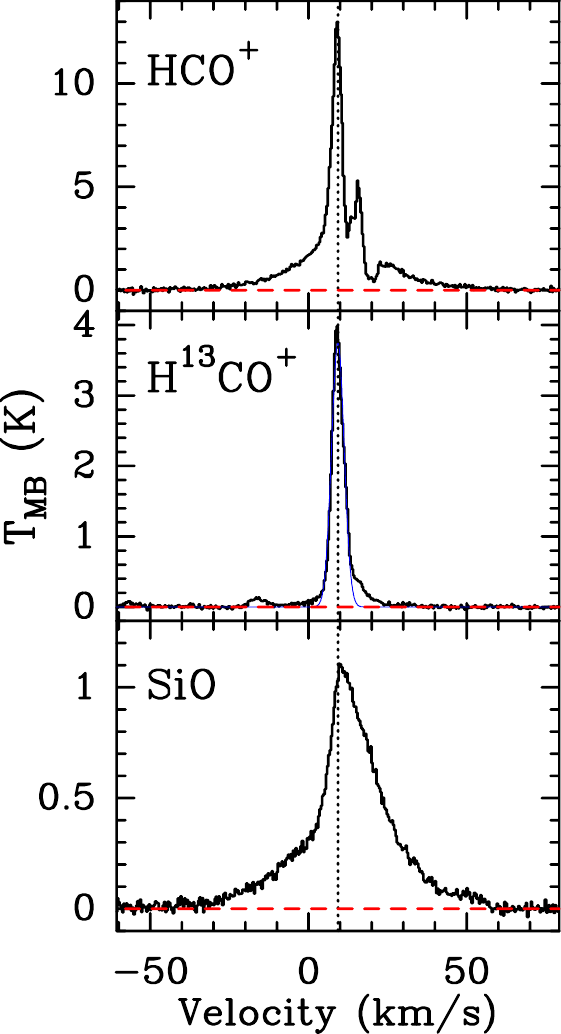}
		\caption{Spectral Profiles for the source HC02, Group A. The red dotted line is at FWZP, and the blue line in the H$^{13}$CO$^+$ panel shows the Gaussian fit.
		}
		\label{fig: HC02 lines}
	\end{figure}

	The SiO(4-3), H$^{13}$CO$^+$(2-1) and HCO$^+$(2-1) spectral profiles of the HC02 source are presented in Fig. \ref{fig: HC02 lines} as an example, where the red dotted line is at FWZP, and the blue line in the H$^{13}$CO$^+$ panel shows the Gaussian fit (see Appendix \ref{spectral_profiles} for the rest, in Figs. \ref{fig: HC01 lines} to \ref{fig: HC32 lines}).
	
	Results for the HCO$^+$(2-1), H$^{13}$CO$^+$(2-1) and SiO(4-3) spectral lines are displayed in Table \ref{tab:hcop_data}, \ref{tab:h13cop_data} and \ref{tab:sio_data} respectively. We have calculated the spatial beam size $R$ with the angular beam size $\theta$ and distance to the source $D$:
	
	\begin{equation}
		R \text{ (pc) } =  \theta \left ( \text{$''$} \right ) \cdot D \text{ (pc)} \cdot \frac{\pi}{180} \frac{1}{3600}
	\end{equation}
	\label{radius}
	
	Note that since our observations do not resolve the singular clump, the spatial beam size is diluted and introduces uncertainties. We compared the beam size of the sources at the H$^{13}$CO$^+$(2-1) line with their beam-deconvolved sizes measured at 250 $\mu$m continuum emission obtained from the Hi-GAL catalogue of compact sources \citep{Elia_2021_HiGAL}. The H$^{13}$CO$^+$ line traces cold surrounding material, and, on the other hand, cold dust is expected to emit at around 250 $\mu$m, i.e. they trace mostly the same dense gas \cite[e.g.][]{Elia_2017_HiGAL}. Thus, we computed the ratio between the clump sizes in 250 $\mu$m and in the H$^{13}$CO$^+$(2-1) line. This indicates how beam dilution affects our results. We found that the beam size in our APEX observations is on average $2.35$ times larger than the size of the clump at 250 $\mu$m. On the other hand, the SiO(4-3) line has an angular resolution of 36'', which corresponds to 0.087 pc at a distance of 1 kpc. The distances to our sources range from 1.33 kpc to 5.54 kpc. These uncertainties will propagate linearly to the column densities to be calculated in Sect. \ref{subsubsec_LTE_mass}.
	
	\subsection{Clump Tracer Emission} \label{subsec_clump_tracer}
	
	The HCO$^+$ and H$^{13}$CO$^+$ species are commonly used as dense gas tracers. They are treated as linear rigid rotors \citep{Mladenovic_HCO_2017} and are expected to present a Gaussian profile. In spite of being isotopologues, these lines present different behaviours. At a kinetic temperature of 20 K, the HCO$^+$(2-1) and H$^{13}$CO$^+$(2-1) lines have optically thin critical densities of $4.2 \times 10^5$ cm$^{-3}$ and $3.8 \times 10^5$ cm$^{-3}$ respectively, and effective excitation densities of $1.7 \times 10^3$ cm$^{-3}$ and $7.1 \times 10^4$ cm$^{-3}$ respectively \citep{Shirley_2015}. The HCO$^+$ has a much lower effective density than its isotopologue, which indicates that HCO$^+$ traces regions with lower density, such as outflows and jets, which are usually less dense than the surrounding material. This is why HCO$^+$ traces both outflows and the surrounding material, whilst H$^{13}$CO$^+$ only traces the latter. The differences between these two lines are visually clear in their spectral profiles (see Fig. \ref{fig: HC02 lines} and Appendix \ref{spectral_profiles}). HCO$^+$ emission experiences significant absorption and self-absorption \citep{Klaassen_2007}, thus it is not possible to quantify properties of the clump using this spectral profile only. In addition, H$^{13}$CO$^+$ emission is optically thin and does not present absorption. Therefore, the Gaussian parameters fitted to the H$^{13}$CO$^+$(2-1) spectral profiles were used to calculate the properties of the cores for all the sources, and the velocity of the H$^{13}$CO$^+$(2-1) Gaussian peak was used to centre the standard of rest velocity V$_{LSR}$ of the sources. The parameters of the HCO$^+$(2-1) spectral lines are displayed in Table \ref{tab:hcop_data}, and the results from the Gaussian fit to the H$^{13}$CO$^+$(2-1) spectral profile are displayed in Table \ref{tab:h13cop_data}.
	
	We note that if the outflow is strong enough, it is possible that the H$^{13}$CO$^+$ line presents wings as well, like is the case for the source e.g. HC20 (see Fig. \ref{fig: HC20 lines}). Since the sources in our sample are extreme, it is reasonable to see this feature. The wings in the H$^{13}$CO$^+$ line are much smaller than the ones in the HCO$^+$ line, and it is still possible to properly fit a Gaussian curve.
 
	
	\begin{table*}[]
		\centering
		\begin{tabular}{cccccc}
			\hline \hline
			Source & RMS & Area & FWZP & Peak $T_{MB}$ & Wings \\ 
                 & (K) & (K km s$^{-1}$) & (km s$^{-1}$) & (K) &  \\ \hline
			HC01 &           0.04 &       11.9 &       23.97 &       2.73 &             \\
			HC02 &           0.10 &      124.7 &       92.12 &      13.00 &   \checkmark \\
			HC03 &           0.09 &       11.7 &       19.86 &       2.35 &             \\
			HC04 &           0.10 &       50.8 &       29.45 &       4.74 &             \\
			HC05 &           0.11 &        7.3 &       19.86 &       2.15 &             \\
			HC06 &           0.11 &       25.5 &       46.58 &       2.37 &   \checkmark \\
			HC07 &           0.12 &       16.1 &       16.10 &       4.52 &             \\
			HC08 &           0.03 &       61.4 &       53.43 &       6.83 &   \checkmark \\
			HC09 &           0.03 &       13.9 &       26.37 &       3.29 &             \\
			HC10 &           0.03 &        6.4 &       27.06 &       2.17 &             \\
			HC11 &           0.03 &        5.1 &       18.84 &       1.24 &             \\
			HC12 &           0.04 &        9.0 &       19.86 &       2.49 &             \\
			HC13 &           0.05 &       28.7 &       27.06 &       5.74 &             \\
			HC14 &           0.03 &       17.3 &       28.13 &       4.12 &             \\
			HC15 &           0.02 &       24.8 &       31.17 &       3.46 &             \\
			HC16 &           0.02 &       44.2 &       55.82 &       4.58 &   \checkmark \\
			HC17 &           0.03 &       23.7 &       53.77 &       2.20 &   \checkmark \\
			HC18 &           0.04 &       11.4 &       40.75 &       2.02 &   \checkmark \\
			HC19 &           0.05 &        8.3 &       10.96 &       2.55 &             \\
			HC20 &           0.05 &      104.3 &       45.89 &      11.14 &   \checkmark \\
			HC21 &           0.05 &       45.2 &       35.27 &       9.37 &   \checkmark \\
			HC22 &           0.09 &       17.8 &       23.29 &       3.32 &             \\
			HC23 &           0.03 &       17.0 &       27.06 &       3.25 &             \\
			HC24 &           0.04 &        9.8 &       19.18 &       2.67 &             \\
			HC25 &           0.05 &        4.6 &       19.52 &       1.21 &             \\
			HC26 &           0.03 &       21.5 &       20.21 &       5.43 &             \\
			HC27 &           0.04 &       20.1 &       52.40 &       2.44 &   \checkmark \\
			HC28 &           0.02 &       15.5 &       24.66 &       2.28 &             \\
			HC29 &           0.02 &       33.3 &       29.45 &       4.51 &             \\
			HC30 &           0.03 &       19.3 &       28.08 &       4.11 &             \\
			HC31 &           0.03 &       14.4 &       14.73 &       2.88 &             \\
			HC32 &           0.03 &       12.4 &       20.20 &       3.45 &             \\
			
			\hline
		\end{tabular}
		\caption{HCO$^+$ spectral line data.}
		\label{tab:hcop_data}
	\end{table*}
	
	\begin{table*}[]
		\centering
		\begin{tabular}{ccccccc}
			\hline \hline               
			Source &  $R$ &  RMS &  Area &  Width & Position &  Peak $T_{MB}$ \\ 
                 &  (pc) &  (K) &  (K km s$^{-1}$) &  (km s$^{-1}$) & (km s$^{-1}$) &  (K) \\ \hline
			HC01 &          0.26 &             0.02 &          2.4 &          2.29 &         8.4 &         1.00 \\
			HC02 &          0.26 &             0.03 &         18.7 &          4.68 &         9.3 &         3.75 \\
			HC03 &          0.31 &             0.02 &          3.9 &          2.83 &        21.0 &         1.29 \\
			HC04 &          0.46 &             0.03 &          7.5 &          7.09 &        66.8 &         0.99 \\
			HC05 &          0.25 &             0.02 &          4.0 &          3.80 &        33.1 &         0.99 \\
			HC06 &          0.26 &             0.02 &          7.0 &          3.91 &        37.3 &         1.67 \\
			HC07 &          0.33 &             0.02 &          2.9 &          3.23 &        59.5 &         0.84 \\
			HC08 &          0.38 &             0.01 &          6.8 &          4.58 &       -39.5 &         1.40 \\
			HC09 &          0.47 &             0.02 &          3.8 &          3.09 &       -57.9 &         1.16 \\
			HC10 &          0.40 &             0.02 &          4.4 &          3.19 &       -54.9 &         1.30 \\
			HC11 &          0.31 &             0.01 &          1.3 &          2.96 &       -48.1 &         0.43 \\
			HC12 &          0.23 &             0.02 &          1.8 &          1.99 &       -40.2 &         0.86 \\
			HC13 &          0.36 &             0.03 &          4.2 &          2.94 &       -68.0 &         1.34 \\
			HC14 &          0.34 &             0.03 &          1.6 &          2.52 &       -64.9 &         0.59 \\
			HC15 &          0.44 &             0.03 &          3.1 &          3.80 &       -88.7 &         0.77 \\
			HC16 &          0.33 &             0.03 &          5.8 &          4.40 &       -62.7 &         1.24 \\
			HC17 &          0.425 &             0.03 &          4.0 &          4.60 &       -87.8 &         3.84 \\
			HC18 &          0.28 &             0.03 &          2.8 &          2.99 &       -48.5 &         0.88 \\
			HC19 &          0.27 &             0.03 &          1.1 &          1.67 &       -47.0 &         0.60 \\
			HC20 &          0.29 &             0.03 &         22.4 &          5.48 &       -53.2 &         3.84 \\
			HC21 &          0.28 &             0.02 &          7.8 &          3.54 &       -46.6 &         2.07 \\
			HC22 &          0.23 &             0.02 &          2.7 &          2.44 &       -34.3 &         1.04 \\
			HC23 &          0.24 &             0.01 &          2.8 &          2.85 &       -46.2 &         0.92 \\
			HC24 &          0.24 &             0.02 &          1.9 &          2.23 &       -40.8 &         0.78 \\
			HC25 &          0.40 &             0.01 &          2.2 &          4.19 &       -71.0 &         0.48 \\
			HC26 &          0.24 &             0.02 &          4.1 &          3.04 &       -40.8 &         1.27 \\
			HC27 &          0.25 &             0.02 &          7.4 &          6.30 &       -27.0 &         1.10 \\
			HC28 &          0.34 &             0.01 &          2.6 &          3.74 &       -31.5 &         0.65 \\
			HC29 &          0.48 &             0.01 &          5.1 &          3.72 &       -69.4 &         1.28 \\
			HC30 &          0.24 &             0.01 &          3.8 &          3.09 &       -18.0 &         1.16 \\
			HC31 &          0.12 &             0.01 &          4.1 &          3.05 &       -10.6 &         1.27 \\
			HC32 &          0.325 &             0.01 &          1.1 &          2.30 &       -20.5 &         0.44 \\
			\hline
		\end{tabular}
		\caption{H$^{13}$CO$^+$ Spectral Line Data.}
		\label{tab:h13cop_data}
	\end{table*}
	
	\begin{table*}[]
		\centering
		\begin{tabular}{ccccccccccc}
			\hline \hline
			Source & $R$ & RMS & \multicolumn{3}{c}{Area} & \multicolumn{3}{c}{Width} & Position & Peak \\
                 & (pc) & (mK) & \multicolumn{3}{c}{(K km s$^{-1}$)} & \multicolumn{3}{c}{(km s$^{-1}$)} & (km s$^{-1}$) & (K) \\\hline
			& & & Full & Red Wing & Blue Wing & FWZP & Red Wing & Blue Wing & & \\ \cline{4-6} \cline{7-9}
			
			HC01 &       0.26 &          0.02 &            0.4 &           - &            - &           33.90 &           - &            - &          10.1 &      0.05 \\
			HC02 &       0.26 &          0.028 &           26.8 &           7.5 &           14.6 &           92.12 &          40.11 &           47.33 &          10.0 &      1.11 \\
			HC03 &       0.31 &          0.02 &            1.0 &           0.4 &            0.5 &           27.05 &          13.59 &           10.64 &          21.1 &      0.17 \\
			HC04 &       0.46 &          0.03 &            8.4 &           1.1 &            3.8 &           45.55 &          11.51 &           26.95 &          67.2 &      0.59 \\
			HC05 &       0.25 &          0.02 &            2.3 &           0.5 &            1.2 &           43.15 &          16.95 &           22.41 &          33.1 &      0.22 \\
			HC06 &       0.26 &          0.02 &            5.0 &           2.1 &            1.7 &           49.32 &          24.06 &           21.35 &          36.9 &      0.39 \\
			HC07 &       0.33 &          0.02 &            2.5 &           0.6 &            1.2 &           47.26 &          16.56 &           27.47 &          59.4 &      0.26 \\
			HC08 &       0.38 &          0.01 &           13.5 &           5.5 &            6.1 &           74.66 &          35.58 &           34.50 &         -39.1 &      1.57 \\
			HC09 &       0.47 &          0.01 &            1.0 &           0.4 &            0.3 &           36.64 &          19.13 &           14.42 &         -58.1 &      0.12 \\
			HC10 &       0.40 &          0.01 &            0.8 &           0.1 &            0.3 &           19.18 &           4.88 &           11.11 &         -53.5 &      0.15 \\
			HC11 &       0.31 &          0.014 &            0.1 &           - &            - &           14.73 &           - &            - &         -50.1 &      0.06 \\
			HC12 &       0.23 &          0.02 &            1.3 &           0.4 &            0.6 &           30.82 &          11.74 &           17.09 &         -39.2 &      0.25 \\
			HC13 &       0.36 &          0.03 &            0.0 &           - &            - &            0.00 &           - &            - &         -67.5 &      0.08 \\
			HC14 &       0.34 &          0.03 &            0.0 &           - &            - &            0.00 &           - &            - &         -64.7 &      0.08 \\
			HC15 &       0.44 &          0.03 &            2.4 &           0.6 &            0.8 &           29.80 &          11.34 &           14.66 &         -87.9 &      0.34 \\
			HC16 &       0.33 &          0.03 &            6.5 &           2.4 &            2.2 &           42.81 &          20.25 &           18.17 &         -62.9 &      0.48 \\
			HC17 &       0.425 &          0.03 &           12.3 &           3.3 &            5.9 &           56.85 &          26.33 &           25.93 &         -86.7 &      0.78 \\
			HC18 &       0.28 &          0.03 &           15.6 &           1.3 &            0.5 &           36.99 &          19.59 &           14.40 &         -49.4 &      0.21 \\
			HC19 &       0.27 &          0.03 &            0.0 &           - &            - &            0.00 &           - &            - &           0.0 &      0.07 \\
			HC20 &       0.29 &          0.02 &           11.8 &           1.3 &            6.4 &           50.34 &          18.90 &           25.97 &         -53.3 &      0.94 \\
			HC21 &       0.28 &          0.02 &            2.3 &           2.9 &            4.5 &           60.96 &          23.95 &           33.46 &         -46.3 &      0.97 \\
			HC22 &       0.23 &          0.02 &           19.1 &           0.8 &            2.0 &           87.67 &          27.45 &           57.78 &         -34.6 &      0.15 \\
			HC23 &       0.24 &          0.01 &            1.6 &           0.6 &            0.7 &           44.86 &          19.48 &           22.53 &         -45.8 &      0.15 \\
			HC24 &       0.24 &          0.01 &            1.5 &           0.6 &            0.6 &           46.92 &          21.63 &           23.06 &         -40.3 &      0.15 \\
			HC25 &       0.40 &          0.02 &            0.9 &           0.3 &            0.3 &           31.51 &          13.29 &           14.02 &         -70.3 &      0.09 \\
			HC26 &       0.24 &          0.02 &            1.5 &           0.4 &            0.6 &           30.14 &           9.84 &           17.25 &         -40.5 &      0.22 \\
			HC27 &       0.25 &          0.02 &           17.6 &           6.1 &            6.9 &           74.66 &          34.50 &           33.86 &         -25.3 &      1.02 \\
			HC28 &       0.34 &          0.01 &            0.8 &           0.3 &            0.2 &           31.85 &          15.56 &           12.55 &         -30.8 &      0.10 \\
			HC29 &       0.48 &          0.01 &            2.1 &           0.8 &            0.6 &           42.47 &          22.89 &           15.86 &         -68.5 &      0.16 \\
			HC30 &       0.24 &          0.01 &            0.0 &           0.9 &            0.2 &           45.21 &          29.28 &           12.83 &          18.7 &      0.15 \\
			HC31 &       0.12 &          0.013 &            0.3 &           - &            - &           17.81 &           - &            - &         -12.5 &      0.05 \\
			HC32 &       0.325 &          0.01 &            0.3 &           - &            - &           33.56 &           - &            - &         -19.2 &      0.04 \\
			
			\hline
		\end{tabular}
		\caption{SiO Spectral Line Data.}
		\label{tab:sio_data}
	\end{table*}
	
	Since wings in the HCO$^+$ spectral profile are a very good indicator of outflows, its (2-1) spectral line was used to determine the presence of molecular outflows through the detection of wings. In the literature, spectral lines with wings are usually modelled by a Lorentz distribution. However, this line presents significant absorption in our data, which makes any curve fit far too inaccurate. Thus, HCO$^+$ wings were checked for as follows: the sum of the widths of the red and blue wings was obtained by subtracting the baseline width of the Gaussian fit of the H$^{13}$CO$^+$ line from the HCO$^+$ FWZP; if the sum of the widths is larger than 25 km s$^{-1}$, then the given source is deemed to have HCO$^+$ wings. As modelled by \cite{Gusdorf_2008} (see also \cite{Leurini_2014}), shocks trigger erosion of grains and SiO can be efficiently produced only if shock speeds are above 25 km s$^{-1}$. Results show (see Table \ref{tab:hcop_data}) that 28\% of our sample (9 sources) present wings in the HCO$^+$ spectral profile (e.g. see Fig. \ref{fig: HC08 lines}). However, we note that, due to the lack of spatial information, there is significant uncertainty in these results.
	
	\subsection{SiO Detection} \label{subsec_sio_det}
	
	We checked whether the sources present SiO emission, which directly traces outflows and shocked gas. Emission at $5 \sigma$ or more was considered. SiO emission was found in 78\% of our sample (25 sources). Results are displayed in Table \ref{tab:sio_data}.
	
	The material ejected by bipolar outflows conforms two lobes, as modelled in \cite{Rawlings_2004}. Depending on the inclination angle, this produces mainly two features: the red and blue wings. While centred at the velocity of each source, the width from the Gaussian fit to the H$^{13}$CO$^+$(2-1) line was subtracted from the SiO(4-3) spectral profile width to obtain the width of the red and blue wings. With these measurements (see Table \ref{tab:sio_data}) the velocity integrated intensity of the SiO wings were obtained and further used to calculate the physical properties of the outflows.
	
	There are a few sources that have particularly intense SiO emission, such as the source HC02 (associated to IRAS 17574-2403). The peak of the emission of these sources is $\geq 20$ times the RMS and they have $T_{peak} > 0.7$ K. For a more detailed description and summary of previous works done on these sources, see Appendix \ref{ap_sources}.
	
	
	According to these results, we have divided the sample into three groups, as seen in Table \ref{tab:cross_check_groups}: 9 sources present both SiO detection and HCO$^+$ wings, 16 sources present SiO emission and no HCO$^+$ wings, and the remaining 7 present neither.
	
	\begin{table}[h]
		\centering
		\begin{tabular}{cccc}
			\hline
			Group & SiO Detection             & HCO$^+$ Wings   & No. of Sources          \\ \hline
			A     & \checkmark & \checkmark   & 9  \\
			B     & \checkmark & -            & 16 \\
			C     & -          & -            & 7  \\ \hline
		\end{tabular}
		\caption{Grouping of the sources.}
		\label{tab:cross_check_groups}
	\end{table}
	
	
	\section{Results} \label{results}
	
	We calculated the following properties of the clumps and their outflows.
	
	\subsection{Properties of the Clump} \label{sub_sec_clump}
	
	\subsubsection{Virial Masses} \label{subsubsec_vir_mass}
	
	The Gaussian width of the H$^{13}$CO$^+$(2-1) spectral line was used to calculate the virial mass of the clump. Assuming that it is gravitationally bound, so that the virial theorem applies, the virial mass can be calculated as follows \cite[e.g.][]{Merello_2013}:
	
	\begin{align}
		\left (  \frac{M_{\text{virial}}}{M_{\odot}}  \right ) = 210 \left (  \frac{\Delta V}{\text{km s}^{-1}}  \right )^2 \left (  \frac{R}{\text{pc} } \right )
	\end{align}
	\label{M_vir}
	
	\noindent where $\Delta V$ is the line's Gaussian width, and $R$ is the spatial beam size of the source.
	
	Results show (see Table \ref{tab:mass_vir_LTE_LM}) that these sources have very high virial masses, reaching up to $4.9 \times 10^3 M_{\odot}$. However, the beam of the observation instrument APEX is larger than the clump, thus the observations include more material.
	
	\subsubsection{LTE Mass} \label{subsubsec_LTE_mass}
	
	The Local Thermodynamic Equilibrium (LTE) mass of the clump was calculated using the LTE formalism, assuming the emission is optically thin. The H$^{13}$CO$^+$ is a linear and rigid rotor molecule. Thus, the column density $N_J$ was calculated as follows \cite[e.g.][]{Garden_1991, Sanhueza_2012}:
	\begin{align}
		\begin{aligned}
			N_J &= \frac{3 h}{8 \pi^3 \mu^2} \frac{U(T_{ex}) }{(J + 1)} \frac{\exp{ \left ( \frac{E_J}{k T_{ex}} \right )}}{ \left [ 1 - \exp{ \left ( \frac{-h \nu}{k T_{ex}} \right )} \right ] }\\
			&\times  \frac{1}{\left (  J(T_{ex}) - J(T_{bg})  \right )} \int T_{MB} d\nu
		\end{aligned}
	\end{align}
	\label{N_J_col_dens_Sanhueza}
	
	Here, $k$ is the Boltzmann constant, $h$ is the Planck constant, $\mu = 3.89 \times 10^{-18}$ (D) is the electric dipole momenta and $E_J = h B J (J + 1)$ is the energy in the level $J$ (in this case, $J = 1$), where $B = 43377.165$ (MHz) is the rotational constant. The intensity $J(T)$ is defined for a given frequency $\nu$ as:
	
	\begin{equation}
		J(T) = \frac{h \nu}{k} \frac{1}{\exp{\left ( \frac{h \nu}{k T} - 1  \right ) }}
	\end{equation}
	\label{intensity_J(T)}
	
	Finally, $U(T)$ is the partition function:
	\begin{align}
		U(T) = \sum_{J=0}^\infty g_J \exp{\left [ \frac{- h B J (J + 1)}{k T_{ex}} \right ]} \approx \frac{k}{h B} \left ( T_{ex} + \frac{h B}{3 k} \right ) ,
	\end{align}
	\label{part_func_U(T)}
	
	\noindent where $g_J = 2J + 1$ is the rotational degeneracy. A cosmic background temperature $T_{bg}$ of $2.75$ K was considered. The excitation temperature $T_{ex}$ was set to a standard value of $30$ K, which is close to the median of the dust temperatures \cite[e.g.][]{Elia_2017_HiGAL} (26.4 $\pm$ 6.0 K). The total column density in units of cm$^{-2}$ can be obtained by multiplying $N_J$ by the abundance of H$^{13}$CO$^+$: $ \left [ \text{H}_2 / \text{H}^{13}\text{CO}^+ \right ] = 3.0 \times 10^{10}$ \citep{Blake_1987}. Finally, the LTE mass is obtained by multiplying the total column density by the area of the source:
	
	\begin{align}
		M_{\text{LTE}} = N_{\text{tot}} \pi R^2
	\end{align}
	
	\noindent where $R$ is the radius of the beam. These results are also shown in Table \ref{tab:mass_vir_LTE_LM}.
	
	We compared our LTE mass estimates to the clump mass derived from dust emission obtained by Hi-GAL \citep{Elia_2017_HiGAL, Elia_2021_HiGAL}. The ratio between the dust and LTE masses has a mean of 2.11, a median of 1.64 and a standard deviation of 1.34. The origin of this uncertainty can come from the assumed temperature as well as the H13CO+ abundance. Furthermore, the dust masses from Hi-GAL used here were calculated by fitting a SED which was scaled to the angular size at 250 $\mu$m \citep{Elia_2013, Motte_2010} (we note that for the four saturated sources of the sample, the mass used here was obtained using the flux at 500 $\mu$m instead; see Appendix \ref{ap_saturated} for the treatment of saturated sources). The H$^{13}$CO$^+$ line and emission at 250 $\mu$m trace mostly the same dense gas and their angular sizes are within a factor of 2.35 on average (see Section \ref{analysis}), which is consistent with the difference seen between the masses. This indicates our estimates are sound.
	
	The uncertainty of the molecular observations is $\sim$ 20 \%, which affects the Gaussian fit parameters, and propagates to the rest of the clump's properties. Because of the propagating error, the column density can be considered to have $10 - 20$ \% of uncertainty, whilst the mass uncertainty is  $< 30$ \%. The distances are also a source of uncertainty, which affects the beam radius and masses (see Appendix \ref{ap_distances}). 
	
	\begin{table}[]
		\centering
		\begin{tabular}{cccccc}
			\hline \hline
			Source  & $M_{vir}$ & $N_{tot}$ & $M_{LTE}$ & $M_{vir} / M_{LTE}$ & Group \\
                 & ($M_{\odot}$) & ($10^{22}$ cm$^{-2}$) & ($M_{\odot}$) &  & \\ \hline
			HC01 & 284 &   5.36 & 242 & 1.17 & C \\
			HC02 & 1181 &  40.99 & 1839 & 0.64 & A \\
			HC03 & 521 &   8.50 & 559 & 0.93 & B \\
			HC04 & 4867 &  16.39 & 2381 & 2.04 & B \\
			HC05 & 760 &   8.73 & 376 & 2.02 & B \\
			HC06 & 828 &  15.24 & 693 & 1.19 & A \\
			HC07 & 735 &   6.34 & 485 & 1.51 & B \\
			HC08 & 1656 &  14.91 & 1438 & 1.15 & A \\
			HC09 & 936 &   8.37 & 1245 & 0.75 & B \\
			HC10 & 849 &   9.66 & 1047 & 0.81 & B \\
			HC11 & 579 &   2.95 & 199 & 2.91 & C \\
			HC12 & 195 &   4.00 & 150 & 1.30 & B \\
			HC13 & 645 &   9.17 & 796 & 0.81 & C \\
			HC14 & 450 &   3.49 & 271 & 1.66 & C \\
			HC15 & 1332 &   6.81 & 901 & 1.48 & B \\
			HC16 & 1334 &  12.69 & 937 & 1.42 & A \\
			HC17 & 1887 &   8.78 & 1086 & 1.74 & A \\
			HC18 & 519 &   6.16 & 322 & 1.61 & A \\
			HC19 & 158 &   2.35 & 117 & 1.35 & C \\
			HC20 & 1829 &  49.04 & 2823 & 0.65 & A \\
			HC21 & 741 &  178 & 920 & 0.81 & A \\
			HC22 & 284 &   5.92 & 208 & 1.37 & B \\
			HC23 & 414 &   6.14 & 247 & 1.68 & B \\
			HC24 & 247 &   4.07 & 156 & 1.58 & B \\
			HC25 & 1488 &   4.73 & 526 & 2.83 & B \\
			HC26 & 465 &   9.02 & 352 & 1.32 & B \\
			HC27 & 2050 &  16.19 & 669 & 3.07 & A \\
			HC28 & 993 &   5.67 & 441 & 2.25 & B \\
			HC29 & 1400 &  11.12 & 1772 & 0.79 & B \\
			HC30 & 473 &   8.39 & 318 & 1.49 & B \\
			HC31 & 227 &   9.06 &  83 & 2.73 & C \\
			HC32 & 360 &   2.38 & 172 & 2.09 & C \\
			\hline
		\end{tabular}
		\caption{Properties obtained from H$^{13}$CO$+$ emission.}
		\label{tab:mass_vir_LTE_LM}
	\end{table}
	
	\subsubsection{Virial and LTE Mass Comparison} \label{subsubsec_LTE_vs_vir_mass}
	
	The ratio between the virial and LTE mass can provide information about the gravitational stability of the source. High values of this ratio indicate that the internal kinetic energy of the source is too high and that it is not massive enough to be gravitationally stable. Since the material traced by H$^{13}$CO$^+$ emission should be both gravitationally bound and in local thermodynamic equilibrium, the virial and LTE mass should match within a factor of $\sim 2$ when magnetic effects are not considered. Results are presented in Table \ref{tab:mass_vir_LTE_LM}. The ratio between the virial and LTE mass has a mean value of 1.54, a median of 1.45, a minimum of 0.64 and a maximum of 3.07. Only 8 sources have a ratio of 2.00 or higher. Thus, the virial and LTE masses are in good agreement. 
	
	\subsection{Properties of the SiO Outflows} \label{sub_outflow}
	
	\subsubsection{Outflow Mass} \label{subsubsec_outflow_mass}

	We calculated the LTE mass for the SiO outflows in the red and blue wings. With an assumption of optically thin thermal SiO(4-3) emission in LTE \cite[e.g.][]{Liu_2021_SiO}, the column densities $N_J$, $N_{tot}$ and the LTE mass were calculated as described in Sect. \ref{subsubsec_LTE_mass} using a rotational constant $B$ of $21711.96$ (MHz), an electric dipole moment $\mu$ of $3.10 \times 10^{-18}$ (D), an excitation temperature $T_{ex}$ of $18$ K and an SiO abundance $\left [ \text{H}_2 / \text{SiO} \right ]$ of $10^9$ \citep{Liu_2021_SiO, Sanhueza_2012}, standard values for this type of sources. If a higher $T_{ex}$ of 25 K or 30 K was used, the values of the outflow masses would increase by 16.32\% and 29.42\% respectively. Since SiO abundances can drastically vary in different MSF regions (in up to six orders of magnitude, \cite{Ziurys_1989, Martin_Pintado_1992, Li_2019_survey}), using an inadequate choice for this parameter could artificially modify the results. Here, we use a standard value for $\left [ \text{H}_2 / \text{SiO} \right ]$, accepted for protostellar objects. Results are displayed in Table \ref{tab:outflow_properties}.
	
	\subsubsection{Dynamical Properties} \label{subsubsec_outflow_properties}
	
	Further characteristics of the outflow were computed as follows \citep{Beuther_2002}: momentum $P_{out}$, kinetic energy $E_{kin}$, characteristic timescale $t$, mechanical force or flow of momentum $F$, rate of matter expulsion $\dot{M}$ and mechanical luminosity or power $L_m$.
	
	\begin{align} \label{eq:momentum}
		P_{out} = M_r V_r + M_b V_b
	\end{align}
	
	\begin{align} \label{eq:Ekin}
		E_{kin} = \frac{1}{2} M_r V_r^2 + \frac{1}{2} M_b V_b^2
	\end{align}
	
	\begin{align} \label{eq:timescale}
		t = \frac{2 R}{\left ( V_r + V_b \right )}
	\end{align}
	
	\begin{align} \label{eq:force}
		F_m = \frac{P_{\text{out}}}{t}
	\end{align}
	
	\begin{align} \label{eq:Mdot}
		\dot{M}_{out} = \frac{M_{\text{out}}}{t}
	\end{align}
	
	\begin{align} \label{eq:mech_lum}
		L_m = \frac{E_{\text{kin}}}{t}
	\end{align}
	
	Here, $R$ is the radius of the source, $M_{\text{out}} = M_r + M_b$ is the total outflow mass, and $V_b$ and $V_r$ are the maximum velocities of the outflows, which corresponds to $| V_{LSR} - V_i|$ \citep{Beuther_2002, Liu_2021}, where $V_i$ is the velocity at each the extreme of the SiO emission baseline. Results are displayed in Table \ref{tab:outflow_properties}.
	
	\begin{sidewaystable*}[]
		\centering
		\begin{tabular}{ccccccccccccc}
			\hline \hline
			Source & $N_R$ & $N_B$ & $M_R$ & $M_B$ & $M_{out}$ & $P_{out}$ & $E_{kin}$ & $t$ & $\dot{M}$ & $F$ & $L_m$ & Group \\ 
                 & \multicolumn{2}{c}{($10^{20}$ cm$^{-2}$)} & \multicolumn{3}{c}{($M_{\odot}$)} & ($M_{\odot}$ km s$^{-1}$) & ($10^{46}$ erg) & ($10^4$ yr) & ($10^{-4} M_{\odot}$ yr$^{-1}$) & ($10^{-3} M_{\odot}$ km s$^{-1}$ yr$^{-1}$) & ($L_{\odot}$) &  \\    \hline
			HC02 & 77.02 & 150.54 & 34.5 & 67.4 & 101.9 & 4813 & 183 & 0.5 & 187 & 884.91 & 2784 & A \\
			HC03 & 4.34 & 4.64 & 2.9 & 3.0 & 5.9 & 79 & 1 & 2.2 & 3 & 3.54 & 5 & B \\
			HC04 & 11.78 & 38.75 & 17.1 & 56.2 & 73.3 & 1971 & 17 & 2.0 & 37 & 99.56 & 69 & B \\
			HC05 & 4.87 & 11.93 & 2.1 & 5.1 & 7.2 & 164 & 3 & 1.1 & 6 & 14.42 & 19 & B \\
			HC06 & 21.79 & 17.54 & 9.9 & 8.0 & 17.9 & 443 & 12 & 1.0 & 17 & 43.30 & 97 & A \\
			HC07 & 6.47 & 12.68 & 4.9 & 9.7 & 14.6 & 372 & 5 & 1.4 & 11 & 26.83 & 29 & B \\
			HC08 & 57.04 & 63.19 & 54.9 & 60.8 & 115.7 & 4316 & 165 & 1.0 & 118 & 438.71 & 1390 & A \\
			HC09 & 3.83 & 3.25 & 5.7 & 4.8 & 10.5 & 194 & 4 & 2.5 & 4 & 7.81 & 15 & B \\
			HC10 & 1.36 & 2.81 & 1.5 & 3.0 & 4.5 & 48 & 0.2 & 4.1 & 1 & 1.19 & 0.4 & B \\
			HC12 & 3.96 & 6.05 & 1.5 & 2.3 & 3.8 & 60 & 1 & 1.5 & 3 & 4.03 & 3 & B \\
			HC15 & 5.73 & 7.82 & 7.6 & 10.3 & 17.9 & 271 & 3 & 2.9 & 6 & 9.39 & 9 & B \\
			HC16 & 24.69 & 22.75 & 18.2 & 16.7 & 34.9 & 749 & 17 & 1.5 & 23 & 49.91 & 97 & A \\
			HC17 & 33.47 & 60.78 & 41.3 & 75.0 & 116.3 & 3299 & 95 & 1.5 & 80 & 225.54 & 537 & A \\
			HC18 & 13.31 & 5.63 & 6.9 & 2.9 & 9.9 & 193 & 4 & 1.5 & 7 & 13.19 & 25 & A \\
			HC20 & 13.77 & 65.74 & 7.9 & 37.8 & 45.7 & 1255 & 21 & 1.1 & 41 & 111.37 & 156 & A \\
			HC21 & 29.53 & 46.83 & 15.9 & 25.2 & 41.0 & 1294 & 27 & 0.9 & 46 & 143.79 & 249 & A \\
			HC22 & 8.19 & 20.50 & 2.9 & 7.2 & 10.0 & 506 & 8 & 0.5 & 20 & 100.10 & 135 & B \\
			HC23 & 5.70 & 7.14 & 2.3 & 2.9 & 5.1 & 116 & 2 & 1.1 & 5 & 11.00 & 18 & B \\
			HC24 & 6.02 & 6.47 & 2.3 & 2.5 & 4.8 & 112 & 2 & 1.0 & 5 & 11.38 & 21 & B \\
			HC25 & 2.93 & 3.15 & 3.3 & 3.5 & 6.8 & 107 & 2 & 2.5 & 3 & 4.25 & 5 & B \\
			HC26 & 3.94 & 5.99 & 1.5 & 2.3 & 3.9 & 61 & 0.5 & 1.5 & 2 & 3.95 & 3 & B \\
			HC27 & 63.10 & 71.16 & 26.0 & 29.3 & 55.3 & 2064 & 78 & 0.6 & 86 & 320.61 & 1004 & A \\
			HC28 & 2.73 & 1.71 & 2.1 & 1.3 & 3.4 & 56 & 1 & 2.1 & 2 & 2.70 & 4 & B \\
			HC29 & 8.74 & 6.45 & 13.9 & 10.2 & 24.1 & 526 & 15 & 2.2 & 11 & 23.64 & 55 & B \\
			HC30 & 9.51 & 2.36 & 3.6 & 0.9 & 4.5 & 124 & 4 & 1.0 & 4 & 12.14 & 35 & B \\
			\hline
		\end{tabular}
		\caption{Properties of the SiO outflows.}
		\label{tab:outflow_properties}
	\end{sidewaystable*}
	
	The uncertainty of the observations (20\%; see Section \ref{subsec_instr}) propagates to these calculated properties. The outflow column density and mass have about 10 - 20\% and $<30$\% of uncertainty, respectively. The momentum and kinetic energy have the same uncertainty as the mass, whilst the force, rate of matter expulsion and mechanical luminosity have an additional source of uncertainty due to the dynamical timescales, which comes from the distances and the instrumental error ($\sim$20\%).
	
	A statistical description of these results is shown in Table \ref{tab:outflow_stats}. The typical rate of matter expulsion for mid and early massive protostars ranges from $10^{-5}$ to a few times $10^{-3} M_{\odot}$ yr$^{-1}$, and the typical momentum from $10^{-4}$ to $10^{-2}$ $M_{\odot}$ km s$^{-1}$ \citep{Arce_2007}. Our results for the rate of matter expulsion are well within this typical range. The momentum is generally higher, with a mean of $922$ $M_{\odot}$ km s$^{-1}$, indicating that the outflows in our sample are generally very fast and/or massive. Furthermore, the SiO outflows studied here present properties that are within the range of results presented in other works but are more massive and powerful. This is the case for the sample of protostellar candidates studied by \cite{Beuther_2002}, which were traced with CO(2-1), as well as the sample of massive young stellar objects (MYSOs) and compact HII regions traced with $^{12}$CO(2-1) and $^{13}$CO(2-1) by \cite{Maud_2015}. Similarly, our sample has more massive and powerful SiO outflows than other studies that have traced them with the SiO(5-4) spectral line, such as the sample of infrared dark clouds (IRDCs) studied by \cite{Liu_2021_SiO}, for which our results are larger by up to three orders of magnitude. Thus, not only our results are in good accordance with other works in the literature, but our sources stand out because of their powerful outflows.
	
	\begin{table*}[]
		\centering
		\begin{tabular}{cccccccc}
			\hline \hline
			& $M_{out}$ & $P_{out}$ & $E_{kin}$ & $t$ & $\dot{M}$ & $F$    & $L_m$  \\ 
                & ($M_{\odot}$) & ($M_{\odot}$ km s$^{-1}$) & ($10^{46}$ erg) & ($10^4$ yr) & ($10^{-4} M_{\odot}$ yr$^{-1}$) & ($10^{-3} M_{\odot}$ km s$^{-1}$ yr$^{-1}$)    & ($L_{\odot}$)  \\ \hline
			Mean 	& 30  & 928  & 27  & 1.6 & 29  & 102.69 & 271  \\
			STD. 	& 36  & 1358 & 50  & 0.8 & 45  & 196.34 & 622  \\
			Min. 	& 3   & 48 	 & 0.2 & 0.5 & 1   & 1.19 	& 0.4    \\
			25\% 	& 5   & 112  & 2   & 1.0 & 4   & 7.81 	& 9    \\
			Median 	& 11  & 271  & 4   & 1.5 & 7   & 14.42 	& 29   \\
			75\% 	& 41  & 1255 & 17  & 2.1 & 37  & 100.10 & 135  \\
			Max. 	& 116 & 4813 & 183 & 4.1 & 187 & 884.91 & 2784 \\
			
			\hline
		\end{tabular}
		\caption{Statistics of the properties of the outflow.}
		\label{tab:outflow_stats}
	\end{table*}
	
	
	\section{Discussions} \label{discussions}
	
	\subsection{Correlations}  \label{subsec_correlations}
	
	We looked for correlations between the properties of the outflows and the properties of their hosting clump. We found three relevant correlations: between the bolometric luminosity and outflow power, between the outflow power and the rate of matter expulsion, and between the kinetic energy of the outflow and outflow mass. Their rank correlation coefficients were calculated and the Python package \texttt{sklearn.linear\_model} \citep{Pedregosa_2012_sklearn} was used to perform a linear regression in order to find the slope of these correlations. Our results show that, for these three correlations, the rank correlation coefficients are positive, as well as the slopes.
 
\begin{figure}[h]
		\centering
		\includegraphics[width=0.5\textwidth]{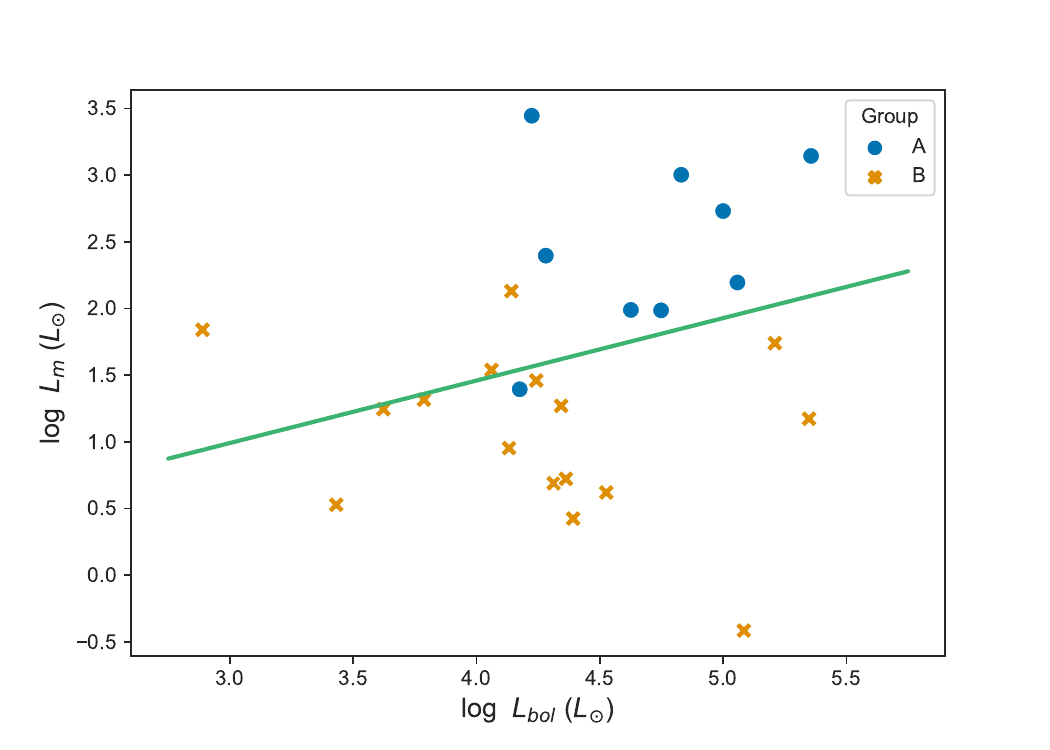}
		\caption{Bolometric luminosity ($L_{\odot}$) vs. outflow power $L_m$ ($L_{\odot}$) in logarithmic scale.}
		\label{fig: Lbol_vs_LSiO_OLS}
	\end{figure}

	A weak linear trend between the bolometric luminosity and the outflow power shown in Fig. \ref{fig: Lbol_vs_LSiO_OLS} was found. This correlation tells us that, given that a source has a molecular outflow, the more luminous the source, the more energetic its outflow will be. Therefore, it is possible that it will power more massive and energetic outflows. However, this correlation has a large amount of dispersion, it is not possible to make definitive conclusions. Other studies have also found this correlation: \cite{Wu_2004}, \cite{Liu_2021} and \cite{Maud_2015} found it in the $^{12}$CO and $^{13}$CO lines, and \cite{Liu_2021} also found it on the HCO$^+$ and CS lines. It has also been found in other SiO transitions by \cite{Csengeri_2016, Li_2019_survey, Liu_2021_SiO}. All of these species are outflow tracers and the linear regressions done have always given a positive slope. Thus, in spite of being a weak correlation in our sample, it is still positive and is in agreement with the trends found by these other works. Thus, the idea that more luminous sources are able to produce more powerful shocks \citep{Codella_1999} cannot be discarded. Since this has been found in various SiO transitions and outflow tracers, and now in the SiO(4-3) one as well, it is a possibly universal trend to molecular outflows in MSF regions.
	
\begin{figure}[h]
		\centering
		\includegraphics[width=0.5\textwidth]{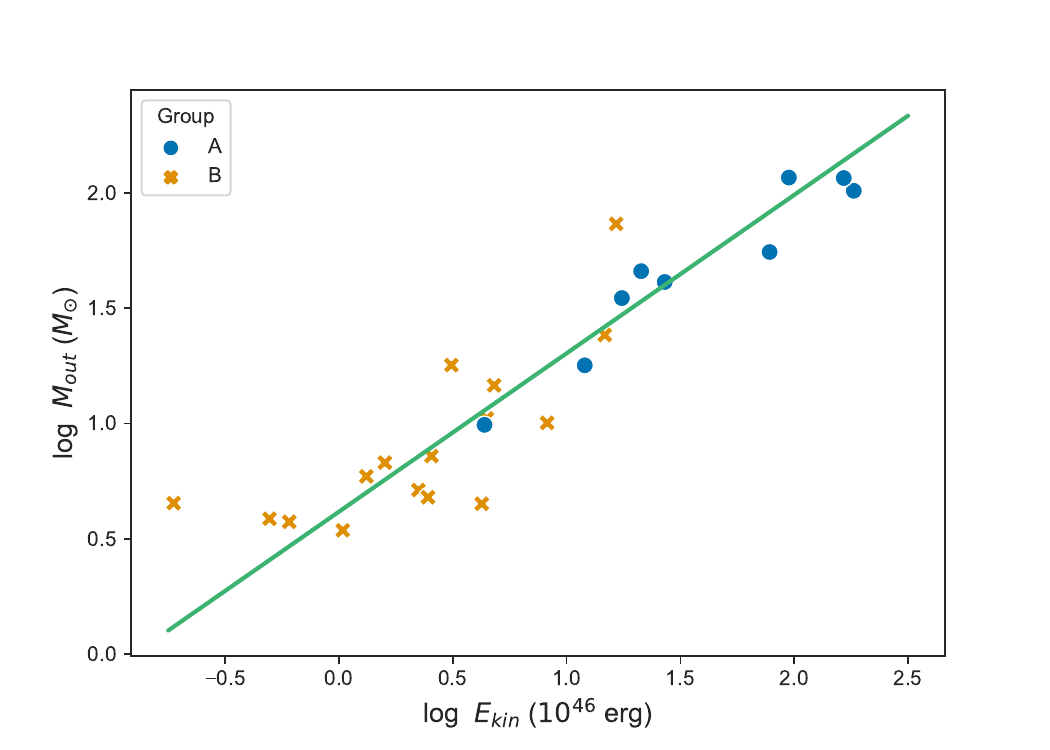}
		\caption{Kinetic energy ($10^{46}$ erg) vs. outflow mass ($M_{\odot}$) in logarithmic scale. The Spearman $\rho$ rank correlation coefficient is equal to 0.93 and the slope is equal to 0.69.
		}
		\label{fig: Ekin_vs_Mout_OLS}
	\end{figure}

	As shown in Fig. \ref{fig: Ekin_vs_Mout_OLS}, there is a close linear relationship between the kinetic energy and the outflow mass. This correlation has a Spearman $\rho$ coefficient of 0.93 and a slope of 0.69. It is strong and suggests that the more massive an outflow is, the faster it will be.
	
	\begin{figure}[h]
		\centering
		\includegraphics[width=0.5\textwidth]{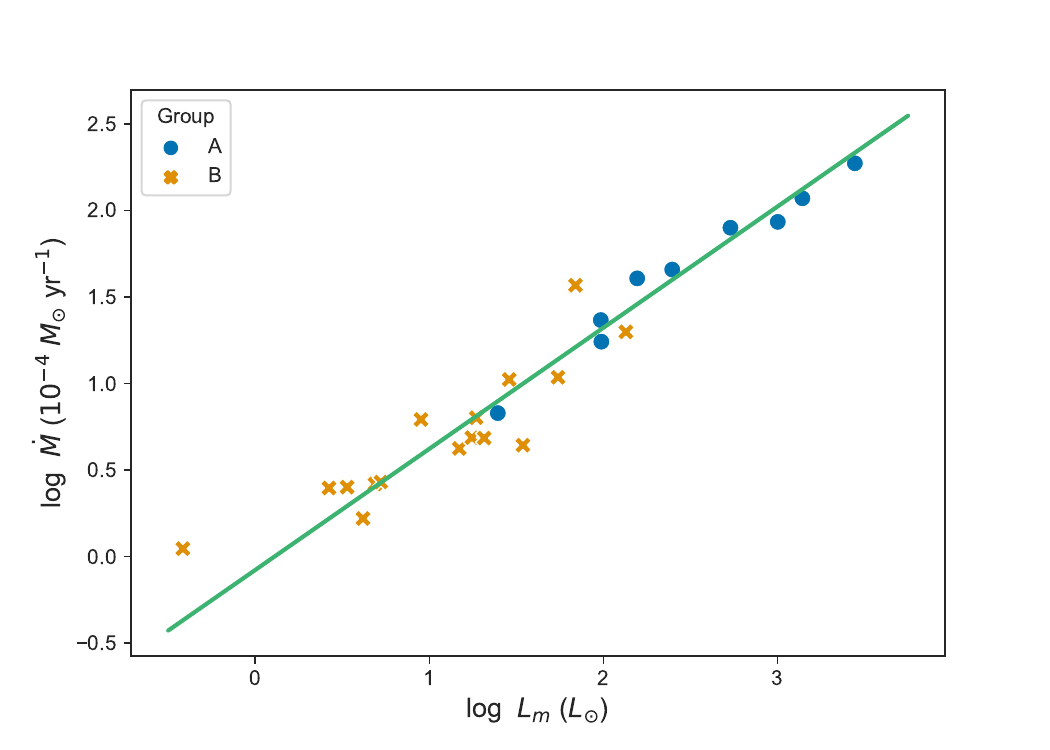}
		\caption{Outflow power $L_m$ ($L_{\odot}$) vs. rate of matter expulsion $\dot{M}$ ($10^{-4} M_{\odot}$ yr$^{-1}$) in logarithmic scale. The Spearman $\rho$ rank correlation coefficient is equal to 0.97 and the slope is equal to 0.70.
		}
		\label{fig: LSiO_vs_Mdot_OLS}
	\end{figure}
	
	Note that this correlation is physically equivalent to one between the outflow power and the rate of mass expulsion, as these parameters are proportional to the inverse of the characteristic timescale (see Equations \ref{eq:Mdot} and \ref{eq:mech_lum}):
	
	\begin{equation}
		t = \frac{M_{out}}{\dot{M}_{out}} = \frac{E_{kin}}{L_m}
	\end{equation}
	
	Thus, this correlation carries the uncertainties of the dynamical timescale previously discussed. However, it is still useful to analyse this trend as it provides intuitive physical significance and a direct comparison with other works is possible. In our sample, this is a strong relation: its Spearman $\rho$ rank correlation coefficient is equal to 0.97 and its slope is equal to 0.70 (see Fig. \ref{fig: LSiO_vs_Mdot_OLS}). This relationship tells us that the more powerful the outflow is, the faster it ejects matter. Therefore, in addition to expelling more material, energetic outflows expel this material faster. This trend has also been found by other studies, such as \cite{Beuther_2002} and \cite{Liu_2021} in the CO line and its isotopologues. \cite{Liu_2021} found this trend in other outflow tracers too, such as HCO$^+$ and CS, indicating that this trend is common to outflows independent of the tracer used. Finding this trend in SiO here further supports this idea.
	
	
	The correlations found here suggest that, since sources with a larger bolometric luminosity are more energetic, their outflows also are. These energetic outflows bear to mobilise more material. Thus, their rate of mass loss is high, which translates into large outflow masses. These correlations agree with the dependency suggested by \cite{Liu_2021} \cite[see also][]{Wu_2004}: bolometric luminosity $\rightarrow$  outflow energy $\rightarrow$  rate of mass loss $\rightarrow$ outflow mass.
	
	\subsection{Evolutionary Stage Indicator L/M} \label{subsec_LM}
	
	The luminosity to mass ratio $L/M$ can be used as an evolutionary stage indicator \cite[e.g.][]{Molinari_2016, Elia_2021_HiGAL, Urquhart_2022}. In the first stages of star formation, the core is cold and actively accreting matter, so $L/M \leq 1$. Eventually, the temperature and luminosity increase, while the mass remains virtually unchanged. This results in a value of $L/M \geq 1$. It has been found that the more luminous sources in the Galaxy usually have $L/M \geq 10$ \citep{Lopez_2011, Molinari_2016}. The sources in the sample studied here are in an advanced stage of the star-forming process, as $L/M$ has values ranging from $3$ $L_{\odot} / M_{\odot}$ up to $177$ $L_{\odot} / M_{\odot}$, with a mean of 58 $L_{\odot} / M_{\odot}$ and a median of 41 $L_{\odot} / M_{\odot}$ (see Table \ref{tab:sources_SED}). Furthermore, we looked for correlations between $L/M$ and the SiO outflow properties and found that they are not correlated. Other works have also failed to find such correlations \citep{Liu_Rong_2022, Maud_2015, Liu_2021}. This strongly suggests that molecular outflows are present throughout the whole of the high-mass star formation process \citep{Csengeri_2016, Urquhart_2022}.
 
	\subsection{Cross-Check between SiO and HCO$^+$ Wings Detections} \label{subsec_cross_check}
	
	All of the sources that present wings in the HCO$^+$ line, also present SiO emission above $5 \sigma$. We compared these sources (Group A) with the rest of the clumps with SiO detections (Group B) (see Table \ref{tab:cross_check_groups}). Table \ref{tab:comparison_A_vs_B} shows a statistical comparison of the spectral parameters, outflow properties and the evolutionary stage indicator $L/M$ between these groups.
	
	SiO spectral profiles with FWZP $> 20$ km s$^{-1}$ are considered broad line widths due to high-velocity shocks \citep{Duarte_Cabral_2014, Zhu_2020}. We note that all the sources in both Groups A and B exhibit broad line widths, with a minimum FWZP of 19.18 km s$^{-1}$, a mean value of 47.15 km s$^{-1}$, a median value of 44.86 km s$^{-1}$ and a maximum of 92.12 km s$^{-1}$. This means that all of the sources where SiO was detected have experienced recent outflow activity, which has manifested in high-velocity shocks, typically associated with collimated jets, and it is strong enough to produce SiO outflows \citep{Csengeri_2016, Li_2019_survey, Liu_Rong_2022}. This confirms the sources in both Groups A and B are protostellar \citep{Zhu_2020}.
 
	Furthermore, the SiO spectral profile of the sources in Group A is wider and more intense, indicating that the outflows in this group are more massive, faster and more energetic than those in Group B \citep{Schilke_1997, Duarte_Cabral_2014, Liu_Rong_2022} (see Table \ref{tab:comparison_A_vs_B}). Their mean outflow mass is larger than those in Group B by a factor of 4.8, their mean momentum by a factor of 7, their mean kinetic energy by a factor of 14.7, their dynamic timescale by a factor of 0.58, their rate of matter expulsion by a factor of 8.7, their outflow force by a factor of 12, and their outflow power by a factor of 25.1. Sources in Group A are possibly more collimated too, as broad-width SiO detection is associated with them. Further studies of the spatial morphology of these outflows could confirm this trend.
	
	Even though SiO outflow activity has been found to decrease and get less collimated over time \citep{Arce_2007, Sakai_2010, Lopez_2011}, results show that sources in Group A and B are virtually in the same stage of evolution because their evolutionary stage indicator $L/M$ have similar values. Table \ref{tab:comparison_A_vs_B} shows that the evolutionary stage indicator $L/M$ has a mean value of 40 ($L_{\odot} / M_{\odot}$) and 47 ($L_{\odot} / M_{\odot}$), and a median of 32 ($L_{\odot} / M_{\odot}$) and 29 ($L_{\odot} / M_{\odot}$) for Groups A and B, respectively. Thus, sources in Group A and B are most likely equally evolved, even within the protostellar phase \citep{Motte_2018}. The fact that sources in Group A have more intense and massive SiO outflows, but both groups are in the same evolutionary stage, suggests that the presence of wings in the HCO$^+$ spectral profile could help identify sources with stronger SiO outflows. Therefore, checking for wings in the HCO$^+$ line could help identify more massive and active MSF regions in samples of similarly evolved sources. Alternatively, since HCO$^+$ emission is associated with old outflow activity \citep{Li_2019}, sources in Group B might have just recently started exhibiting strong outflow activity, allowing for SiO outflows to be detected, and are too new to produce wings in the HCO$^+$ spectral profile. Meanwhile, sources in Group A might have started exhibiting outflow activity long enough time ago for wings in the HCO$^+$ spectral profile, a signal of old outflows, to form. In this scenario, SiO outflows detected from the sources in Group A have been active for a longer time, which might be a consequence of their large mass, power and luminosity. Thus, checking for wings in the HCO$^+$ line might also help identify sources whose outflows have been active for longer. Further research with larger samples and better resolution could provide insight into which of these two scenarios is more likely and significant.
 
	\begin{sidewaystable}[]
		\centering
		\begin{tabular}{ccccccccccccccccccccccc}
			\hline \hline
			& \multicolumn{2}{c}{FWZP} & \multicolumn{2}{c}{Peak} & \multicolumn{2}{c}{Full Area} & \multicolumn{2}{c}{$M_{out}$} & \multicolumn{2}{c}{$P_{out}$} & \multicolumn{2}{c}{$E_{kin}$} & \multicolumn{2}{c}{$t$} & \multicolumn{2}{c}{$\dot{M}$} & \multicolumn{2}{c}{$F$} & \multicolumn{2}{c}{$L_m$} &  \multicolumn{2}{c}{$L/M$} \\
			& \multicolumn{2}{c}{(km s$^{-1}$)} & \multicolumn{2}{c}{(K)} & \multicolumn{2}{c}{(K km s$^{-1}$)} & \multicolumn{2}{c}{($M_{\odot}$)} & \multicolumn{2}{c}{($M_{\odot}$ km s$^{-1}$)} & \multicolumn{2}{c}{($10^{46}$ erg)} & \multicolumn{2}{c}{($10^4$ yr)} & \multicolumn{2}{c}{($10^{-4} M_{\odot}$ yr$^{-1}$)} & \multicolumn{2}{c}{($10^{-3} M_{\odot}$ km s$^{-1}$ yr$^{-1}$)} & \multicolumn{2}{c}{($L_{\odot}$)} & \multicolumn{2}{c}{($L_{\odot} / M_{\odot}$)} \\ \hline
			Group  & A & B  & A& B   & A& B   & A   & B  & A& B   & A   & B  & A   & B   & A   & B  & A  & B  & A& B& A  & B \\ \hline
			Mean   & 59.86 & 40.01 & 0.8 & 0.2 & 12.4 & 3.0  & 60  & 13 & 2047 & 298  & 67  & 4   & 1.1 & 1.9 & 67  & 8  & 247.93 & 21.00  & 704  & 27  & 40 & 47  \\
			STD    & 17.69 & 15.29 & 0.4 & 0.1 & 7.4  & 4.7  & 41  & 17 & 1705 & 472  & 68  & 5   & 0.4 & 0.9 & 58  & 9  & 276.95 & 31.61  & 910  & 35  & 33 & 45  \\
			Min.   & 36.99 & 19.18 & 0.2 & 0.1 & 2.3  & 0.04 & 10  & 3  & 193  & 48   & 4   & 0.2 & 0.5 & 0.5 & 7   & 1  & 13.19  & 1.19   & 25   & 0.4 & 3  & 3   \\
			25\%   & 49.32 & 30.65 & 0.5 & 0.2 & 6.5  & 1.0  & 35  & 5  & 749  & 75   & 17  & 1   & 0.9 & 1.1 & 23  & 3  & 49.91  & 4.01   & 97   & 5   & 12 & 17  \\
			Median & 56.85 & 39.56 & 0.9 & 0.2 & 12.3 & 1.5  & 46  & 6  & 1294 & 120  & 27  & 3   & 1.0 & 1.8 & 46  & 5  & 143.79 & 10.20  & 249  & 17  & 32 & 29  \\
			75\%   & 74.66 & 45.30 & 1.0 & 0.2 & 15.6 & 2.3  & 102 & 12 & 3299 & 296  & 95  & 4   & 1.5 & 2.3 & 86  & 7  & 320.61 & 16.73  & 1004 & 31  & 71 & 52  \\
			Max.   & 92.12 & 87.67 & 1.6 & 0.6 & 26.8 & 19.1 & 116 & 73 & 4813 & 1971 & 183 & 17  & 1.5 & 4.1 & 187 & 37 & 884.91 & 100.10 & 2784 & 135 & 81 & 151 \\ \hline
			\hline
		\end{tabular}
		\caption{Statistics of the spectral parameters and outflow properties of sources from Groups A and B.}
		\label{tab:comparison_A_vs_B}
	\end{sidewaystable}
	
	
	\section{Summary and Conclusions} \label{conclusions}
	
	The aim of this work was to study shocked matter and SiO outflows in very massive, luminous and powerful protostellar sources. We characterised them, searched for any evolutionary trends they might exhibit, associated the presence of shocked gas with outflow activity, and cross-checked SiO emission with other outflow signatures. We used single-dish APEX/SEPIA Band 5 observations of 32 of the brightest massive protostellar sources in the Southern Galaxy. We studied their outflow activity using the SiO and HCO$^+$ tracers, as well as their general clump properties with H$^{13}$CO$^+$ emission. The following is a summary of our main results and conclusions:

	\begin{enumerate}
		
		\item The SiO emission detection rate above $5 \sigma$ is 78\% (25 sources). All SiO detections have a broad line width due to high-velocity shocks, which confirms these sources are well-evolved into the protostellar phase. Furthermore, 28\% of our sample (9 sources) shows wings in the HCO$^+$ spectral profile.
		
		\item We calculated the dynamical properties of the SiO outflows. Results show they are very massive, fast and energetic. Outflow mass has a minimum value of $3$ $M_{\odot}$, a median of $10$ $M_{\odot}$, a mean of $30$ $M_{\odot}$, and a maximum value of $116$ $M_{\odot}$. Outflow momentum has a minimum value of $50$ $M_{\odot}$ km s$^{-1}$, a median of $271$ $M_{\odot}$ km s$^{-1}$, a mean of $922$ $M_{\odot}$ km s$^{-1}$, and a maximum value of $4812$ $M_{\odot}$ km s$^{-1}$. The kinetic energy of the outflow has a minimum value of $0.1 \times 10^{46}$ erg, a median of $4 \times 10^{46}$ erg, a mean of $27 \times 10^{46}$ erg, and a maximum value of $183 \times 10^{46}$ erg. Outflow power has a minimum value of $1.24 \times10^{-3} M_{\odot}$ km s$^{-1}$ yr$^{-1}$, a median of $14.24 \times 10^{-3} M_{\odot}$ km s$^{-1}$ yr$^{-1}$, a mean of $102.25 \times 10^{-3} M_{\odot}$ km s$^{-1}$ yr$^{-1}$, and a maximum value of $884.85 \times 10^{-3} M_{\odot}$ km s$^{-1}$ yr$^{-1}$.
		
		\item We found three positive linear correlations involving SiO outflows properties: a weak one between the bolometric luminosity and outflow power (Fig. \ref{fig: Lbol_vs_LSiO_OLS}), a strong one between the outflow power and the rate of matter expulsion (Fig. \ref{fig: LSiO_vs_Mdot_OLS}), and between the kinetic energy and outflow mass (Fig. \ref{fig: Ekin_vs_Mout_OLS}). They are both positive correlations. The second and third have very low dispersion, but the former has high dispersion. These correlations suggest that the more energetic the outflow, the more material it bears to mobilise.
		
		\item We did not find any correlations between the evolutionary stage indicator $L/M$ and SiO outflow properties. This agrees with the idea that molecular outflows are a ubiquitous phenomenon throughout the process of high-mass star formation.

		\item We performed a cross-check between SiO detection and the presence of wings in the HCO$^+$ line. Sources in Group A (sources that present both features; see Table \ref{tab:cross_check_groups}) have more massive, faster, more energetic and more collimated SiO outflows than sources in Group B (sources that only exhibit SiO emission). Sources in both groups are in an advanced stage of evolution in the high-mass star formation process, and there is no clear evolutionary difference between them. Thus, since SiO emission is such a good tracer of outflow activity, checking for wings in the HCO$^+$ line could help identify more massive and active MSF regions in samples of similarly evolved sources. Alternatively, checking for wings in the HCO$^+$ line might help identify sources whose outflows have been active for longer, since HCO$^+$ traces old outflow activity, whilst SiO traces recent and active outflows.
  
	\end{enumerate}

	Our findings show potential for further studies of sources in an advanced stage of evolution ($L/M > 10$) and their molecular outflows.
	
	
	\begin{acknowledgements}

            We thank the anonymous referee for their helpful and insightful comments.
		
		This publication is based on data acquired with the Atacama Pathfinder Experiment (APEX) under programme ID [TP.C-NNNN(R)]. APEX is a collaboration between the Max-Planck-Institut fur Radioastronomie, the European Southern Observatory, and the Onsala Space Observatory.
		
		M.M. acknowledges support from ANID, Programa de Astronomía - Fondo ALMA-CONICYT, project 3119AS0001. L.B. and R.F. gratefully acknowledge support by the ANID BASAL projects ACE210002 and FB210003, FONDEF ID21I10359 and FONDECYT 1221662. E.M. acknowledges support under the grant "Maria Zambrano" from the UHU funded by the Spanish Ministry of Universities and the "European Union NextGenerationEU".
		
	\end{acknowledgements}
	
	\bibliographystyle{aa}
	\bibliography{references.bib}

\begin{thebibliography}{96}
\expandafter\ifx\csname natexlab\endcsname\relax\def\natexlab#1{#1}\fi

\bibitem[{{Acord} {et~al.}(1997){Acord}, {Walmsley}, \&
  {Churchwell}}]{HC02_Acord_1997}
{Acord}, J.~M., {Walmsley}, C.~M., \& {Churchwell}, E. 1997, \apj, 475, 693

\bibitem[{Araya {et~al.}(2005)Araya, Hofner, Kurtz, Bronfman, \&
  DeDeo}]{HC08_Araya_2005}
Araya, E., Hofner, P., Kurtz, S., Bronfman, L., \& DeDeo, S. 2005, \apjs, 157,
  279

\bibitem[{{Arce} {et~al.}(2007){Arce}, {Shepherd}, {Gueth}, {Lee}, {Bachiller},
  {Rosen}, \& {Beuther}}]{Arce_2007}
{Arce}, H.~G., {Shepherd}, D., {Gueth}, F., {et~al.} 2007, in Protostars and
  Planets V, ed. B.~{Reipurth}, D.~{Jewitt}, \& K.~{Keil}, 245

\bibitem[{{Bally}(2016)}]{Bally_2016}
{Bally}, J. 2016, \araa, 54, 491

\bibitem[{Baug {et~al.}(2020)Baug, Wang, Liu, Tang, Zhang, Li, Eswaraiah, Liu,
  Tej, Goldsmith, Bronfman, Qin, T{\'{o}}th, Li, \& Kim}]{HC17_Baug_2020}
Baug, T., Wang, K., Liu, T., {et~al.} 2020, \apj, 890, 44

\bibitem[{{Belitsky} {et~al.}(2018){Belitsky}, {Lapkin}, {Fredrixon},
  {Meledin}, {Sundin}, {Billade}, {Ferm}, {Pavolotsky}, {Rashid}, {Strandberg},
  {Desmaris}, {Ermakov}, {Krause}, {Olberg}, {Aghdam}, {Shafiee}, {Bergman},
  {De Beck}, {Olofsson}, {Conway}, {De Breuck}, {Immer}, {Yagoubov},
  {Montenegro-Montes}, {Torstensson}, {P{\'e}rez-Beaupuits}, {Klein}, {Boland},
  {Baryshev}, {Hesper}, {Barkhof}, {Adema}, {Bekema}, \&
  {Koops}}]{Belitsky_2018_APEX}
{Belitsky}, V., {Lapkin}, I., {Fredrixon}, M., {et~al.} 2018, \aap, 612, A23

\bibitem[{{Beuther} {et~al.}(2002){Beuther}, {Schilke}, {Menten}, {Walmsley},
  {Sridharan}, \& {Wyrowski}}]{Beuther_2002}
{Beuther}, H., {Schilke}, P., {Menten}, K.~M., {et~al.} 2002, in Astronomical
  Society of the Pacific Conference Series, Vol. 267, Hot Star Workshop III:
  The Earliest Phases of Massive Star Birth, ed. P.~{Crowther}, 341

\bibitem[{{Blake} {et~al.}(1987){Blake}, {Sutton}, {Masson}, \&
  {Phillips}}]{Blake_1987}
{Blake}, G.~A., {Sutton}, E.~C., {Masson}, C.~R., \& {Phillips}, T.~G. 1987,
  \apj, 315, 621

\bibitem[{{Bronfman} {et~al.}(1996){Bronfman}, {Nyman}, \&
  {May}}]{Bronfman_1996}
{Bronfman}, L., {Nyman}, L.~A., \& {May}, J. 1996, \aaps, 115, 81

\bibitem[{Butler \& Tan(2012)}]{Butler_2012}
Butler, M.~J. \& Tan, J.~C. 2012, \apj, 754, 5

\bibitem[{{Codella} {et~al.}(1999){Codella}, {Bachiller}, \&
  {Reipurth}}]{Codella_1999}
{Codella}, C., {Bachiller}, R., \& {Reipurth}, B. 1999, \aap, 343, 585

\bibitem[{{Csengeri} {et~al.}(2016){Csengeri}, {Leurini}, {Wyrowski},
  {Urquhart}, {Menten}, {Walmsley}, {Bontemps}, {Wienen}, {Beuther}, {Motte},
  {Nguyen-Luong}, {Schilke}, {Schuller}, {Zavagno}, \& {Sanna}}]{Csengeri_2016}
{Csengeri}, T., {Leurini}, S., {Wyrowski}, F., {et~al.} 2016, \aap, 586, A149

\bibitem[{{De Buizer, J. M.} {et~al.}(2009){De Buizer, J. M.}, {Redman, R. O.},
  {Longmore, S. N.}, {Caswell, J.}, \& {Feldman, P. A.}}]{HC17_DeBuizer_2009}
{De Buizer, J. M.}, {Redman, R. O.}, {Longmore, S. N.}, {Caswell, J.}, \&
  {Feldman, P. A.} 2009, A\&A, 493, 127

\bibitem[{{De Simone} {et~al.}(2022){De Simone}, {Codella}, {Ceccarelli},
  {L{\'o}pez-Sepulcre}, {Neri}, {Rivera-Ortiz}, {Busquet}, {Caselli},
  {Bianchi}, {Fontani}, {Lefloch}, {Oya}, \& {Pineda}}]{De_Simone_2022}
{De Simone}, M., {Codella}, C., {Ceccarelli}, C., {et~al.} 2022, \mnras, 512,
  5214

\bibitem[{{Dedes, C.} {et~al.}(2011){Dedes, C.}, {Leurini, S.}, {Wyrowski, F.},
  {Schilke, P.}, {Menten, K. M.}, {Thorwirth, S.}, \& {Ott,
  J.}}]{HC08_Dedes_2010}
{Dedes, C.}, {Leurini, S.}, {Wyrowski, F.}, {et~al.} 2011, A\&A, 526, A59

\bibitem[{{Duarte-Cabral, A.} {et~al.}(2014){Duarte-Cabral, A.}, {Bontemps,
  S.}, {Motte, F.}, {Gusdorf, A.}, {Csengeri, T.}, {Schneider, N.}, \& {Louvet,
  F.}}]{Duarte_Cabral_2014}
{Duarte-Cabral, A.}, {Bontemps, S.}, {Motte, F.}, {et~al.} 2014, A\&A, 570, A1

\bibitem[{{Duronea, N. U.} {et~al.}(2021){Duronea, N. U.}, {Cichowolski, S.},
  {Bronfman, L.}, {Mendoza, E.}, {Finger, R.}, {Suad, L. A.}, {Corti, M.}, \&
  {Reynoso, E. M.}}]{HC08_Duronea_2020}
{Duronea, N. U.}, {Cichowolski, S.}, {Bronfman, L.}, {et~al.} 2021, A\&A, 646,
  A103

\bibitem[{{Dutrey} {et~al.}(1997){Dutrey}, {Guilloteau}, \&
  {Guelin}}]{Dutrey_1997}
{Dutrey}, A., {Guilloteau}, S., \& {Guelin}, M. 1997, \aap, 317, L55

\bibitem[{{Elia} {et~al.}(2021){Elia}, {Merello}, {Molinari}, {Schisano},
  {Zavagno}, {Russeil}, {M{\`e}ge}, {Martin}, {Olmi}, {Pestalozzi}, {Plume},
  {Ragan}, {Benedettini}, {Eden}, {Moore}, {Noriega-Crespo}, {Paladini},
  {Palmeirim}, {Pezzuto}, {Pilbratt}, {Rygl}, {Schilke}, {Strafella}, {Tan},
  {Traficante}, {Baldeschi}, {Bally}, {Giorgio}, {Fiorellino}, {Liu}, {Piazzo},
  \& {Polychroni}}]{Elia_2021_HiGAL}
{Elia}, D., {Merello}, M., {Molinari}, S., {et~al.} 2021, \mnras, 504, 2742

\bibitem[{Elia {et~al.}(2013)Elia, Molinari, Fukui, Schisano, Olmi, Veneziani,
  Hayakawa, Pestalozzi, Schneider, Benedettini, di~Giorgio, Ikhenaode, Mizuno,
  Onishi, Pezzuto, Piazzo, Polychroni, Rygl, Yamamoto, \& Maruccia}]{Elia_2013}
Elia, D., Molinari, S., Fukui, Y., {et~al.} 2013, \apj, 772, 45

\bibitem[{{Elia} {et~al.}(2017){Elia}, {Molinari}, {Schisano}, {Pestalozzi},
  {Pezzuto}, {Merello}, {Noriega-Crespo}, {Moore}, {Russeil}, {Mottram},
  {Paladini}, {Strafella}, {Benedettini}, {Bernard}, {Di Giorgio}, {Eden},
  {Fukui}, {Plume}, {Bally}, {Martin}, {Ragan}, {Jaffa}, {Motte}, {Olmi},
  {Schneider}, {Testi}, {Wyrowski}, {Zavagno}, {Calzoletti}, {Faustini},
  {Natoli}, {Palmeirim}, {Piacentini}, {Piazzo}, {Pilbratt}, {Polychroni},
  {Baldeschi}, {Beltr{\'a}n}, {Billot}, {Cambr{\'e}sy}, {Cesaroni},
  {Garc{\'\i}a-Lario}, {Hoare}, {Huang}, {Joncas}, {Liu}, {Maiolo}, {Marsh},
  {Maruccia}, {M{\`e}ge}, {Peretto}, {Rygl}, {Schilke}, {Thompson},
  {Traficante}, {Umana}, {Veneziani}, {Ward-Thompson}, {Whitworth}, {Arab},
  {Bandieramonte}, {Becciani}, {Brescia}, {Buemi}, {Bufano}, {Butora},
  {Cavuoti}, {Costa}, {Fiorellino}, {Hajnal}, {Hayakawa}, {Kacsuk}, {Leto}, {Li
  Causi}, {Marchili}, {Martinavarro-Armengol}, {Mercurio}, {Molinaro},
  {Riccio}, {Sano}, {Sciacca}, {Tachihara}, {Torii}, {Trigilio}, {Vitello}, \&
  {Yamamoto}}]{Elia_2017_HiGAL}
{Elia}, D., {Molinari}, S., {Schisano}, E., {et~al.} 2017, \mnras, 471, 100

\bibitem[{Fern{\'{a}}ndez-L{\'{o}}pez
  {et~al.}(2021)Fern{\'{a}}ndez-L{\'{o}}pez, Sanhueza, Zapata, Stephens, Hull,
  Zhang, Girart, Koch, Cort{\'{e}}s, Silva, Tatematsu, Nakamura, Guzm{\'{a}}n,
  Luong, Ccolque, Tang, \& Chen}]{HC02_Fernandez_Lopez_2021}
Fern{\'{a}}ndez-L{\'{o}}pez, M., Sanhueza, P., Zapata, L.~A., {et~al.} 2021,
  \apj, 913, 29

\bibitem[{Figuer{\^{e}}do {et~al.}(2005)Figuer{\^{e}}do, Blum, Damineli, \&
  Conti}]{HC20_Figueredo_2005}
Figuer{\^{e}}do, E., Blum, R.~D., Damineli, A., \& Conti, P.~S. 2005, \aj, 129,
  1523

\bibitem[{Garay {et~al.}(2002)Garay, Brooks, Mardones, Norris, \&
  Burton}]{HC21_Garay_2002}
Garay, G., Brooks, K.~J., Mardones, D., Norris, R.~P., \& Burton, M.~G. 2002,
  \apj, 579, 678

\bibitem[{{Garay} {et~al.}(2010){Garay}, {Mardones}, {Bronfman}, {May},
  {Chavarr{\'\i}a}, \& {Nyman}}]{Garay_2010}
{Garay}, G., {Mardones}, D., {Bronfman}, L., {et~al.} 2010, \apj, 710, 567

\bibitem[{{Garden} {et~al.}(1991){Garden}, {Hayashi}, {Gatley}, {Hasegawa}, \&
  {Kaifu}}]{Garden_1991}
{Garden}, R.~P., {Hayashi}, M., {Gatley}, I., {Hasegawa}, T., \& {Kaifu}, N.
  1991, \apj, 374, 540

\bibitem[{{G{\'o}mez} {et~al.}(2014){G{\'o}mez}, {Wyrowski}, {Schuller},
  {Menten}, \& {Ballesteros-Paredes}}]{HC17_Gomez_2014}
{G{\'o}mez}, L., {Wyrowski}, F., {Schuller}, F., {Menten}, K.~M., \&
  {Ballesteros-Paredes}, J. 2014, \aap, 561, A148

\bibitem[{{Gusdorf, A.} {et~al.}(2008){Gusdorf, A.}, {Cabrit, S.}, {Flower, D.
  R.}, \& {Pineau des For\^ets, G.}}]{Gusdorf_2008}
{Gusdorf, A.}, {Cabrit, S.}, {Flower, D. R.}, \& {Pineau des For\^ets, G.}
  2008, A\&A, 482, 809

\bibitem[{He {et~al.}(2021)He, Henkel, Zhou, Esimbek, Stutz, Liu, Ji, Li, Wu,
  Tang, Komesh, \& Sailanbek}]{He_2021}
He, Y.-X., Henkel, C., Zhou, J.-J., {et~al.} 2021, \apjs, 253, 2

\bibitem[{{Henning} {et~al.}(2000){Henning}, {Lapinov}, {Schreyer}, {Stecklum},
  \& {Zinchenko}}]{HC08_Henning_2000}
{Henning}, T., {Lapinov}, A., {Schreyer}, K., {Stecklum}, B., \& {Zinchenko},
  I. 2000, \aap, 364, 613

\bibitem[{{Jiao} {et~al.}(2023){Jiao}, {Wang}, {Pillai}, {Baug}, {Zhang}, \&
  {Xu}}]{Jiao_2023}
{Jiao}, W., {Wang}, K., {Pillai}, T. G.~S., {et~al.} 2023, \apj, 945, 81

\bibitem[{{Klaassen} \& {Wilson}(2007)}]{Klaassen_2007}
{Klaassen}, P.~D. \& {Wilson}, C.~D. 2007, \apj, 663, 1092

\bibitem[{{Krishnan} {et~al.}(2013){Krishnan}, {Ellingsen}, {Voronkov}, \&
  {Breen}}]{HC27_Krishnan_2013}
{Krishnan}, V., {Ellingsen}, S.~P., {Voronkov}, M.~A., \& {Breen}, S.~L. 2013,
  \mnras, 433, 3346

\bibitem[{Lee {et~al.}(2001)Lee, Walsh, Burton, \& Ashley}]{HC17_Lee_2001}
Lee, J.-K., Walsh, A., Burton, M., \& Ashley, M. 2001, \mnras, 324, 1102

\bibitem[{{Lee} {et~al.}(2002){Lee}, {Walsh}, \& {Burton}}]{HC17_Lee_2002}
{Lee}, J.~K., {Walsh}, A.~J., \& {Burton}, M.~G. 2002, in Cosmic Masers: From
  Proto-Stars to Black Holes, ed. V.~{Migenes} \& M.~J. {Reid}, Vol. 206,
  175--178

\bibitem[{{Leurini} {et~al.}(2014){Leurini}, {Codella}, {L{\'o}pez-Sepulcre},
  {Gusdorf}, {Csengeri}, \& {Anderl}}]{Leurini_2014}
{Leurini}, S., {Codella}, C., {L{\'o}pez-Sepulcre}, A., {et~al.} 2014, \aap,
  570, A49

\bibitem[{{Leurini, S.} {et~al.}(2015){Leurini, S.}, {Wyrowski, F.},
  {Wiesemeyer, H.}, {Gusdorf, A.}, {G\"usten, R.}, {Menten, K. M.}, {Gerin,
  M.}, {Levrier, F.}, {H\"ubers, H. W.}, {Jacobs, K.}, {Ricken, O.}, \&
  {Richter, H.}}]{HC02_Leurini_2015}
{Leurini, S.}, {Wyrowski, F.}, {Wiesemeyer, H.}, {et~al.} 2015, A\&A, 584, A70

\bibitem[{Li {et~al.}(2019{\natexlab{a}})Li, Zhou, Esimbek, He, Baan, Li, Wu,
  Tang, Ji, Komesh, \& et~al.}]{Li_2019}
Li, Q., Zhou, J., Esimbek, J., {et~al.} 2019{\natexlab{a}}, \mnras, 488,
  4638–4647

\bibitem[{Li {et~al.}(2020)Li, Sanhueza, Zhang, Nakamura, Lu, Wang, Liu,
  Tatematsu, Jackson, Silva, Guzmán, Sakai, Izumi, Tafoya, Li, Contreras,
  Morii, \& Kim}]{Li_ashes_2020}
Li, S., Sanhueza, P., Zhang, Q., {et~al.} 2020, \apj, 903, 119

\bibitem[{Li {et~al.}(2019{\natexlab{b}})Li, Wang, Fang, Zhang, Li, Zhang, Li,
  Zhu, \& Zeng}]{Li_2019_survey}
Li, S., Wang, J., Fang, M., {et~al.} 2019{\natexlab{b}}, \apj, 878, 29

\bibitem[{{Li} {et~al.}(2019){Li}, {Zhang}, {Pillai}, {Stephens}, {Wang}, \&
  {Li}}]{Li_2019_dark_clouds}
{Li}, S., {Zhang}, Q., {Pillai}, T., {et~al.} 2019, \apj, 886, 130

\bibitem[{Liu {et~al.}(2020)Liu, Sanhueza, Liu, Zavagno, Tang, Wu, \&
  Zhang}]{Liu_2020_chemistry}
Liu, H.-L., Sanhueza, P., Liu, T., {et~al.} 2020, \apj, 901, 31

\bibitem[{{Liu} {et~al.}(2021){Liu}, {Tan}, {Marvil}, {Kong}, {Rosero},
  {Caselli}, \& {Cosentino}}]{Liu_2021_SiO}
{Liu}, M., {Tan}, J.~C., {Marvil}, J., {et~al.} 2021, \apj, 921, 96

\bibitem[{{Liu} {et~al.}(2022){Liu}, {Liu}, {Chen}, {Liu}, {Wang}, {Li}, {Lee},
  {Liu}, {Juvela}, {Garay}, {Dewangan}, {Soam}, {Bronfman}, {He}, {Eswaraiah},
  {Zhang}, {Zhang}, {Xu}, {T{\'o}th}, {Shen}, {Li}, {Wu}, {Qin}, {Ren},
  {Zhang}, {Tej}, {Goldsmith}, {Baug}, {Luo}, {Zhou}, \&
  {Zhang}}]{Liu_Rong_2022}
{Liu}, R., {Liu}, T., {Chen}, G., {et~al.} 2022, \mnras, 511, 3618

\bibitem[{{Liu, D.J.} {et~al.}(2021){Liu, D.J.}, {Xu}, {Li}, {Zheng}, {Lu},
  {Hao}, {Lin}, {Bian}, \& {Liu}}]{Liu_2021}
{Liu, D.J.}, D.-J., {Xu}, Y., {Li}, Y.-J., {et~al.} 2021, \apjs, 253, 15

\bibitem[{Lo {et~al.}(2007)Lo, Cunningham, Bains, Burton, \&
  Garay}]{HC20_Lo_2007}
Lo, N., Cunningham, M., Bains, I., Burton, M.~G., \& Garay, G. 2007, \mnras:
  Letters, 381, L30

\bibitem[{Lo {et~al.}(2014)Lo, Cunningham, Jones, Bronfman, Cortes, Simon,
  Lowe, Fissel, \& Novak}]{HC20_Lo_2014}
Lo, N., Cunningham, M.~R., Jones, P.~A., {et~al.} 2014, \apj, 797, L17

\bibitem[{{L{\'o}pez-Sepulcre} {et~al.}(2016){L{\'o}pez-Sepulcre}, {Watanabe},
  {Sakai}, {Furuya}, {Saruwatari}, \& {Yamamoto}}]{Lopez_Sepulcre_2016}
{L{\'o}pez-Sepulcre}, A., {Watanabe}, Y., {Sakai}, N., {et~al.} 2016, \apj,
  822, 85

\bibitem[{{L\'opez-Sepulcre, A.} {et~al.}(2011){L\'opez-Sepulcre, A.},
  {Walmsley, C. M.}, {Cesaroni, R.}, {Codella, C.}, {Schuller, F.}, {Bronfman,
  L.}, {Carey, S. J.}, {Menten, K. M.}, {Molinari, S.}, \& {Noriega-Crespo,
  A.}}]{Lopez_2011}
{L\'opez-Sepulcre, A.}, {Walmsley, C. M.}, {Cesaroni, R.}, {et~al.} 2011, A\&A,
  526, L2

\bibitem[{{Martin-Pintado} {et~al.}(1992){Martin-Pintado}, {Bachiller}, \&
  {Fuente}}]{Martin_Pintado_1992}
{Martin-Pintado}, J., {Bachiller}, R., \& {Fuente}, A. 1992, \aap, 254, 315

\bibitem[{Maud {et~al.}(2015)Maud, Moore, Lumsden, Mottram, Urquhart, \&
  Hoare}]{Maud_2015}
Maud, L.~T., Moore, T. J.~T., Lumsden, S.~L., {et~al.} 2015, \mnras, 453,
  645–665

\bibitem[{Merello {et~al.}(2013)Merello, Bronfman, Garay, Nyman, II, \&
  Walmsley}]{Merello_2013}
Merello, M., Bronfman, L., Garay, G., {et~al.} 2013, \apj, 774, 38

\bibitem[{Merello {et~al.}(2018)Merello, Molinari, Rygl, Evans, Elia, Schisano,
  Traficante, Shirley, Svoboda, \& Goldsmith}]{Merello_2019}
Merello, M., Molinari, S., Rygl, K. L.~J., {et~al.} 2018, \mnras, 483, 5355

\bibitem[{Mladenovi\'c(2017)}]{Mladenovic_HCO_2017}
Mladenovi\'c, M. 2017, AIP, 147

\bibitem[{{Molinari} {et~al.}(2016){Molinari}, {Merello}, {Elia}, {Cesaroni},
  {Testi}, \& {Robitaille}}]{Molinari_2016}
{Molinari}, S., {Merello}, M., {Elia}, D., {et~al.} 2016, \apjl, 826, L8

\bibitem[{Morales {et~al.}(2009)Morales, Mardones, Garay, Brooks, \&
  Pineda}]{HC27_Morales_2009}
Morales, E. F.~E., Mardones, D., Garay, G., Brooks, K.~J., \& Pineda, J.~E.
  2009, \apj, 698, 488–501

\bibitem[{Motte {et~al.}(2018)Motte, Bontemps, \& Louvet}]{Motte_2018}
Motte, F., Bontemps, S., \& Louvet, F. 2018, \araa, 56, 41–82

\bibitem[{{Motte} {et~al.}(2007){Motte}, {Bontemps}, {Schilke}, {Schneider},
  {Menten}, \& {Brogui{\`e}re}}]{Motte_2007}
{Motte}, F., {Bontemps}, S., {Schilke}, P., {et~al.} 2007, \aap, 476, 1243

\bibitem[{{Motte, F.} {et~al.}(2010){Motte, F.}, {Zavagno, A.}, {Bontemps, S.},
  {Schneider, N.}, {Hennemann, M.}, {Di Francesco, J.}, {Andr\'e, Ph.},
  {Saraceno, P.}, {Griffin, M.}, {Marston, A.}, {Ward-Thompson, D.}, {White,
  G.}, {Minier, V.}, {Men'shchikov, A.}, {Hill, T.}, {Abergel, A.}, {Anderson,
  L. D.}, {Aussel, H.}, {Balog, Z.}, {Baluteau, J.-P.}, {Bernard, J.-Ph.},
  {Cox, P.}, {Csengeri, T.}, {Deharveng, L.}, {Didelon, P.}, {di Giorgio,
  A.-M.}, {Hargrave, P.}, {Huang, M.}, {Kirk, J.}, {Leeks, S.}, {Li, J. Z.},
  {Martin, P.}, {Molinari, S.}, {Nguyen-Luong, Q.}, {Olofsson, G.}, {Persi,
  P.}, {Peretto, N.}, {Pezzuto, S.}, {Roussel, H.}, {Russeil, D.}, {Sadavoy,
  S.}, {Sauvage, M.}, {Sibthorpe, B.}, {Spinoglio, L.}, {Testi, L.}, {Teyssier,
  D.}, {Vavrek, R.}, {Wilson, C. D.}, \& {Woodcraft, A.}}]{Motte_2010}
{Motte, F.}, {Zavagno, A.}, {Bontemps, S.}, {et~al.} 2010, A\&A, 518, L77

\bibitem[{Murphy {et~al.}(2010)Murphy, Cohen, Ekers, Green, Wark, \&
  Moss}]{HC08_Murphy_2010}
Murphy, T., Cohen, M., Ekers, R.~D., {et~al.} 2010, \mnras, 405, 1560

\bibitem[{Myers {et~al.}(1996)Myers, Mardones, Tafalla, Williams, \&
  Wilner}]{Myers_1996}
Myers, P.~C., Mardones, D., Tafalla, M., Williams, J.~P., \& Wilner, D.~J.
  1996, \apj, 465, L133

\bibitem[{Nicholas {et~al.}(2011{\natexlab{a}})Nicholas, Rowell, Burton, Walsh,
  Fukui, Kawamura, Longmore, \& Keto}]{HC02_Nicholas_2011_12mm}
Nicholas, B., Rowell, G., Burton, M.~G., {et~al.} 2011{\natexlab{a}}, \mnras,
  411, 1367

\bibitem[{Nicholas {et~al.}(2011{\natexlab{b}})Nicholas, Rowell, Burton, Walsh,
  Fukui, Kawamura, \& Maxted}]{HC02_Nicholas_2011_7mm}
Nicholas, B.~P., Rowell, G., Burton, M.~G., {et~al.} 2011{\natexlab{b}},
  \mnras, 419, 251

\bibitem[{Olguin {et~al.}(2021)Olguin, Sanhueza, Guzm{\'{a}}n, Lu, Saigo,
  Zhang, Silva, Chen, Li, Ohashi, Nakamura, Sakai, \& Wu}]{HC21_Olguin_2021}
Olguin, F.~A., Sanhueza, P., Guzm{\'{a}}n, A.~E., {et~al.} 2021, \apj, 909, 199

\bibitem[{{Ossenkopf} \& {Henning}(1994)}]{Ossenkopf_Henning_1994}
{Ossenkopf}, V. \& {Henning}, T. 1994, \aap, 291, 943

\bibitem[{Pedregosa {et~al.}(2012)Pedregosa, Varoquaux, Gramfort, Michel,
  Thirion, Grisel, Blondel, Müller, Nothman, Louppe, Prettenhofer, Weiss,
  Dubourg, Vanderplas, Passos, Cournapeau, Brucher, Perrot, \&
  Duchesnay}]{Pedregosa_2012_sklearn}
Pedregosa, F., Varoquaux, G., Gramfort, A., {et~al.} 2012, Journal of Machine
  Learning Research

\bibitem[{Rawlings {et~al.}(2004)Rawlings, Redman, Keto, \&
  Williams}]{Rawlings_2004}
Rawlings, J. M.~C., Redman, M.~P., Keto, E., \& Williams, D.~A. 2004, \mnras,
  351, 1054–1062

\bibitem[{Reid {et~al.}(2019)Reid, Menten, Brunthaler, Zheng, Dame, Xu, Li,
  Sakai, Wu, Immer, \& et~al.}]{Reid_2019}
Reid, M.~J., Menten, K.~M., Brunthaler, A., {et~al.} 2019, \apj, 885, 131

\bibitem[{{Reid} {et~al.}(2014){Reid}, {Menten}, {Brunthaler}, {Zheng}, {Dame},
  {Xu}, {Wu}, {Zhang}, {Sanna}, {Sato}, {Hachisuka}, {Choi}, {Immer},
  {Moscadelli}, {Rygl}, \& {Bartkiewicz}}]{Reid_2014}
{Reid}, M.~J., {Menten}, K.~M., {Brunthaler}, A., {et~al.} 2014, \apj, 783, 130

\bibitem[{Reid {et~al.}(2009)Reid, Menten, Zheng, Brunthaler, Moscadelli, Xu,
  Zhang, Sato, Honma, Hirota, Hachisuka, Choi, Moellenbrock, \&
  Bartkiewicz}]{Reid_2009}
Reid, M.~J., Menten, K.~M., Zheng, X.~W., {et~al.} 2009, \apj, 700, 137

\bibitem[{{Rojas-Garc{\'\i}a} {et~al.}(2022){Rojas-Garc{\'\i}a},
  {G{\'o}mez-Ruiz}, {Palau}, {Orozco-Aguilera}, {Dagostino}, \&
  {Kurtz}}]{RojasGarcia_2022}
{Rojas-Garc{\'\i}a}, O.~S., {G{\'o}mez-Ruiz}, A.~I., {Palau}, A., {et~al.}
  2022, \apjs, 262, 13

\bibitem[{{Sakai} {et~al.}(2010){Sakai}, {Sakai}, {Hirota}, \&
  {Yamamoto}}]{Sakai_2010}
{Sakai}, T., {Sakai}, N., {Hirota}, T., \& {Yamamoto}, S. 2010, \apj, 714, 1658

\bibitem[{Sanhueza {et~al.}(2012)Sanhueza, Jackson, Foster, Garay, Silva, \&
  Finn}]{Sanhueza_2012}
Sanhueza, P., Jackson, J.~M., Foster, J.~B., {et~al.} 2012, \apj, 756, 60

\bibitem[{{Schilke} {et~al.}(1997){Schilke}, {Walmsley}, {Pineau des Forets},
  \& {Flower}}]{Schilke_1997}
{Schilke}, P., {Walmsley}, C.~M., {Pineau des Forets}, G., \& {Flower}, D.~R.
  1997, \aap, 321, 293

\bibitem[{Shirley(2015)}]{Shirley_2015}
Shirley, Y.~L. 2015, \pasp, 127, 299–310

\bibitem[{Sollins {et~al.}(2004)Sollins, Hunter, Battat, Beuther, Ho, Lim, Liu,
  Ohashi, Sridharan, Su, Zhao, \& Zhang}]{HC02_Sollins_2004}
Sollins, P.~K., Hunter, T.~R., Battat, J., {et~al.} 2004, \apj, 616, L35

\bibitem[{Su {et~al.}(2009)Su, Liu, Wang, Chen, \& Chen}]{HC02_Su_2009}
Su, Y.-N., Liu, S.-Y., Wang, K.-S., Chen, Y.-H., \& Chen, H.-R. 2009, \apj,
  704, L5

\bibitem[{Tang {et~al.}(2009)Tang, Ho, Girart, Rao, Koch, \&
  Lai}]{HC02_Tang_2009}
Tang, Y.-W., Ho, P. T.~P., Girart, J.~M., {et~al.} 2009, \apj, 695, 1399

\bibitem[{{Testi} {et~al.}(1998){Testi}, {Felli}, {Persi}, \&
  {Roth}}]{HC27_Testi_2009}
{Testi}, L., {Felli}, M., {Persi}, P., \& {Roth}, M. 1998, \aaps, 129, 495

\bibitem[{{Urquhart} {et~al.}(2022){Urquhart}, {Wells}, {Pillai}, {Leurini},
  {Giannetti}, {Moore}, {Thompson}, {Figura}, {Colombo}, {Yang}, {K{\"o}nig},
  {Wyrowski}, {Menten}, {Rigby}, {Eden}, \& {Ragan}}]{Urquhart_2022}
{Urquhart}, J.~S., {Wells}, M.~R.~A., {Pillai}, T., {et~al.} 2022, \mnras, 510,
  3389

\bibitem[{{van der Tak}(2004)}]{Tak_2004}
{van der Tak}, F.~F.~S. 2004, in Star Formation at High Angular Resolution, ed.
  M.~G. {Burton}, R.~{Jayawardhana}, \& T.~L. {Bourke}, Vol. 221, 59

\bibitem[{{van der Tak, F. F. S.} {et~al.}(2019){van der Tak, F. F. S.},
  {Shipman, R. F.}, {Jacq, T.}, {Herpin, F.}, {Braine, J.}, \& {Wyrowski,
  F.}}]{HC21_van_der_Tak_2019}
{van der Tak, F. F. S.}, {Shipman, R. F.}, {Jacq, T.}, {et~al.} 2019, A\&A,
  625, A103

\bibitem[{{Voronkov} {et~al.}(2014){Voronkov}, {Caswell}, {Ellingsen}, {Green},
  \& {Breen}}]{HC27_Voronkov_2014}
{Voronkov}, M.~A., {Caswell}, J.~L., {Ellingsen}, S.~P., {Green}, J.~A., \&
  {Breen}, S.~L. 2014, \mnras, 439, 2584

\bibitem[{{Widmann} {et~al.}(2016){Widmann}, {Beuther}, {Schilke}, \&
  {Stanke}}]{Widmann_2016}
{Widmann}, F., {Beuther}, H., {Schilke}, P., \& {Stanke}, T. 2016, \aap, 589,
  A29

\bibitem[{Wiles {et~al.}(2016)Wiles, Lo, Redman, Cunningham, Jones, Burton, \&
  Bronfman}]{HC20_Wiles_2016}
Wiles, B., Lo, N., Redman, M.~P., {et~al.} 2016, \mnras, 458, 3429

\bibitem[{{Wu} {et~al.}(2005){Wu}, {Zhu}, {Wei}, {Xu}, {Zhang}, \&
  {Fiege}}]{Wu_2005}
{Wu}, Y., {Zhu}, M., {Wei}, Y., {et~al.} 2005, \apjl, 628, L57

\bibitem[{{Wu, Y.} {et~al.}(2004){Wu, Y.}, {Wei, Y.}, {Zhao, M.}, {Shi, Y.},
  {Yu, W.}, {Qin, S.}, \& {Huang, M.}}]{Wu_2004}
{Wu, Y.}, {Wei, Y.}, {Zhao, M.}, {et~al.} 2004, A\&A, 426, 503

\bibitem[{{Yang} {et~al.}(2022){Yang}, {Urquhart}, {Wyrowski}, {Thompson},
  {K{\"o}nig}, {Colombo}, {Menten}, {Duarte-Cabral}, {Schuller}, {Csengeri},
  {Eden}, {Barnes}, {Traficante}, {Bronfman}, {Sanchez-Monge}, {Ginsburg},
  {Cesaroni}, {Lee}, {Beuther}, {Medina}, {Mazumdar}, \& {Henning}}]{Yang_2022}
{Yang}, A.~Y., {Urquhart}, J.~S., {Wyrowski}, F., {et~al.} 2022, \aap, 658,
  A160

\bibitem[{{Yu} \& {Wang}(2015)}]{HC27_Yu_2015}
{Yu}, N. \& {Wang}, J.-J. 2015, \mnras, 451, 2507

\bibitem[{Yu {et~al.}(2018)Yu, Xu, \& Wang}]{HC21_Yu_2018}
Yu, N.-P., Xu, J.-L., \& Wang, J.-J. 2018, Research in Astronomy and
  Astrophysics, 18, 015

\bibitem[{Zahorecz {et~al.}(2017)Zahorecz, Jimenez-Serra, Testi, Immer,
  Fontani, Caselli, Wang, \& Toth}]{HC02_Zahorecz_2017_SEPIA}
Zahorecz, S., Jimenez-Serra, I., Testi, L., {et~al.} 2017, A\&A, 602, L3

\bibitem[{Zapata {et~al.}(2020)Zapata, Ho, Fern{\'{a}}ndez-L{\'{o}}pez,
  Ccolque, Rodr{\'{\i}}guez, Reyes-Vald{\'{e}}s, Bally, Palau, Saito, Sanhueza,
  Rivera-Ortiz, \& Rodr{\'{\i}}guez-Gonz{\'{a}}lez}]{HC02_Zapata_2020_ALMA}
Zapata, L.~A., Ho, P. T.~P., Fern{\'{a}}ndez-L{\'{o}}pez, M., {et~al.} 2020,
  \apj, 902, L47

\bibitem[{Zapata {et~al.}(2019)Zapata, Ho, Guzmán~Ccolque, Fernández-Lopéz,
  Rodríguez, Bally, Sanhueza, Palau, \& Saito}]{HC02_Zapata_2019}
Zapata, L.~A., Ho, P. T.~P., Guzmán~Ccolque, E., {et~al.} 2019, \mnras:
  Letters, 486, L15

\bibitem[{{Zhu} {et~al.}(2020){Zhu}, {Wang}, {Liu}, {Kim}, {Zhu}, \&
  {Li}}]{Zhu_2020}
{Zhu}, F.-Y., {Wang}, J.-Z., {Liu}, T., {et~al.} 2020, \mnras, 499, 6018

\bibitem[{{Zinchenko} {et~al.}(2000){Zinchenko}, {Henkel}, \&
  {Mao}}]{HC17_HC27_Zinchenko_2000}
{Zinchenko}, I., {Henkel}, C., \& {Mao}, R.~Q. 2000, \aap, 361, 1079

\bibitem[{{Ziurys} {et~al.}(1989){Ziurys}, {Friberg}, \&
  {Irvine}}]{Ziurys_1989}
{Ziurys}, L.~M., {Friberg}, P., \& {Irvine}, W.~M. 1989, \apj, 343, 201

\end{thebibliography}
	
	\begin{appendix}
		
		\section{Treatment of Saturated Sources} \label{ap_saturated}
		
		There are four sources in our sample with saturated observations (with a $\blacklozenge$ in Table \ref{tab:sources} and \ref{tab:sources_SED}): HC02, HC04, HC16 and HC20. Their dust mass $M_{d}$ was estimated at $500$ $\mu$m \cite[e.g.][]{Merello_2013, Motte_2007}:
		
		\begin{align} \label{eq:mass_saturated}
			M_d = \frac{F_{\nu} D^2}{\kappa_{\nu} B_{\nu}(T_d)}     ,
		\end{align}
		
		\noindent where $D$ is the distance to the source, $F_{\nu}$ is the flux density, $\kappa_{\nu}$ is the opacity and $B_{\nu}(T_d)$ is Planck's function, at frequency $\nu = 599.585$ GHz. The dust temperature $T_{d}$ was set to $25$ K. The opacity $\kappa_{500}$ for dense clumps with typical conditions has a value of $5.04$ cm$^2$ g$^{-1}$ at $500$ $\mu$m \citep{Ossenkopf_Henning_1994}.
		
		The bolometric luminosity $L_{bol}$ was calculated using the flux density at $70$ $\mu$m, which has been found to be a good estimator for the total luminosity of protostellar objects when using the following empirical relation \citep{Elia_2017_HiGAL}:
		
		\begin{align} \label{eq:luminosity_saturated}
			L_{bol} (L_{\odot}) = 2.56 \cdot {F_{70}}^{1.00} (Jy)
		\end{align}
		
		The flux densities at $500$ $\mu$m and $70$ $\mu$m were obtained from the Hi-GAL catalogue \cite{Elia_2021_HiGAL}.
		
		
		\section{Distances} \label{ap_distances}
		
		The distances used for the selection of the sources $D_{sel}$ are the kinematic distances from \cite{Reid_2009} using the local standard of rest velocity V$_{LSR}$ taken from \cite{Bronfman_1996}. Once the selection was made, the distances to the sources were corrected as seen in Table \ref{tab:distances}. The distance measured through trigonometric parallax from \cite{Reid_2019} was used for the sources that have it. For the sources that do not have a distance calculated with parallax, the kinematic distances from \cite{Reid_2014} were used instead. This left us with distances with the least possible uncertainty. These corrected distances were the ones used in the calculation of the properties of the sources.
		
		\begin{table}[]
			\centering
			\begin{tabular}{ccccc}
				\hline \hline
				Source & $D_{sel}$ & $D_{new}$ & Type \\
                 & (kpc) & (kpc) &  \\ \hline
				HC01 &           2.33 &       2.95 &  Parallax \\
				HC02 &           2.25 &       2.94 &  Parallax \\
				HC03 &           3.64 &       3.56 &  Parallax \\
				HC04 &           5.49 &       5.29 &  Parallax \\
				HC05 &           3.42 &       2.88 &  Parallax \\
				HC06 &           3.61 &       2.96 &  Parallax \\
				HC07 &           4.27 &       3.84 &  Parallax \\
				HC08 &           4.31 &       4.31 & Kinematic \\
				HC09 &           5.35 &       5.35 & Kinematic \\
				HC10 &           4.30 &       4.57 & Kinematic \\
				HC11 &           3.37 &       3.60 & Kinematic \\
				HC12 &           2.53 &       2.69 & Kinematic \\
				HC13 &           4.04 &       4.09 & Kinematic \\
				HC14 &           3.81 &       3.87 & Kinematic \\
				HC15 &           5.08 &       5.05 & Kinematic \\
				HC16 &           3.68 &       3.77 & Kinematic \\
				HC17 &           4.90 &       4.88 & Kinematic \\
				HC18 &           3.05 &       3.17 & Kinematic \\
				HC19 &           3.00 &       3.10 & Kinematic \\
				HC20 &           3.22 &       3.33 & Kinematic \\
				HC21 &           3.13 &       3.22 & Kinematic \\
				HC22 &           2.60 &       2.60 &  Parallax \\
				HC23 &           3.37 &       2.78 &  Parallax \\
				HC24 &           3.17 &       2.72 &  Parallax \\
				HC25 &           3.26 &       4.63 & Kinematic \\
				HC26 &           3.37 &       2.74 &  Parallax \\
				HC27 &           2.29 &       2.82 & Kinematic \\
				HC28 &           3.78 &       3.87 & Kinematic \\
				HC29 &           5.61 &       5.54 & Kinematic \\
				HC30 &           2.78 &       2.70 &  Parallax \\
				HC31 &           2.16 &       1.33 &  Parallax \\
				HC32 &           4.15 &       3.73 &  Parallax \\
				
				\hline
			\end{tabular}
			\caption{Distances to the sources in the studied sample.}
			\label{tab:distances}
		\end{table}
		
		
		\section{Notes on Particular Sources} \label{ap_sources}
		
		Some of the sources in our sample were found to be of particular interest because the intensity of their SiO outflows stands out from the rest (peak $> 0.7$ K). They are all of potential interest for future work. A brief description of the studies that have been carried out about these sources is shown here.
		
		\subsection{HC02: HIGALBM5.8856-0.3920, IRAS 17574-2403}
		
		This source (see Fig. \ref{fig: HC02 lines}) presents a particularly intense outflow. Despite not being the most massive, it has the maximum rate of matter expulsion calculated here. In addition, its spectral profiles shown in Fig. \ref{fig: HC02 lines} suggest that there is a second cloud at $15.7$ km s$^{-1}$, and absorption at $\sim 20$ km s$^{-1}$. It has been extensively studied since its discovery \citep{HC02_Acord_1997}. \cite{HC02_Zahorecz_2017_SEPIA} observed this source with APEX in the Band 5 and found that it does not present D$_2$CO emission, suggesting that it is past the early stages of evolution. It was mapped at 12 mm and 7 mm by \cite{HC02_Nicholas_2011_12mm} and \cite{HC02_Nicholas_2011_7mm} respectively, and its [OI] spectra were resolved by \cite{HC02_Leurini_2015}. Its magnetic fields were studied by \cite{HC02_Tang_2009} and \cite{HC02_Fernandez_Lopez_2021}. \cite{HC02_Su_2009} reported a dense and hot-molecular cocoon in the environment of the source. Furthermore, its outflow has been extensively studied. The outflow of this source was mapped in the SiO(5-4) line by \cite{HC02_Sollins_2004}, traced in the SiO(8-7) line by \cite{Klaassen_2007}, and \cite{HC02_Zapata_2019} found indications that it is an explosive multi-polar outflow. This was confirmed in \cite{HC02_Zapata_2020_ALMA}, where a graph and animation modelling the source are presented, clearly showing the multiple lobes.
		
		\subsection{HC08: HIGALBM301.1365-0.2259, IRAS 12326-6245}
		
\begin{figure}[h]
			\centering
			\includegraphics[width=0.3\textwidth]{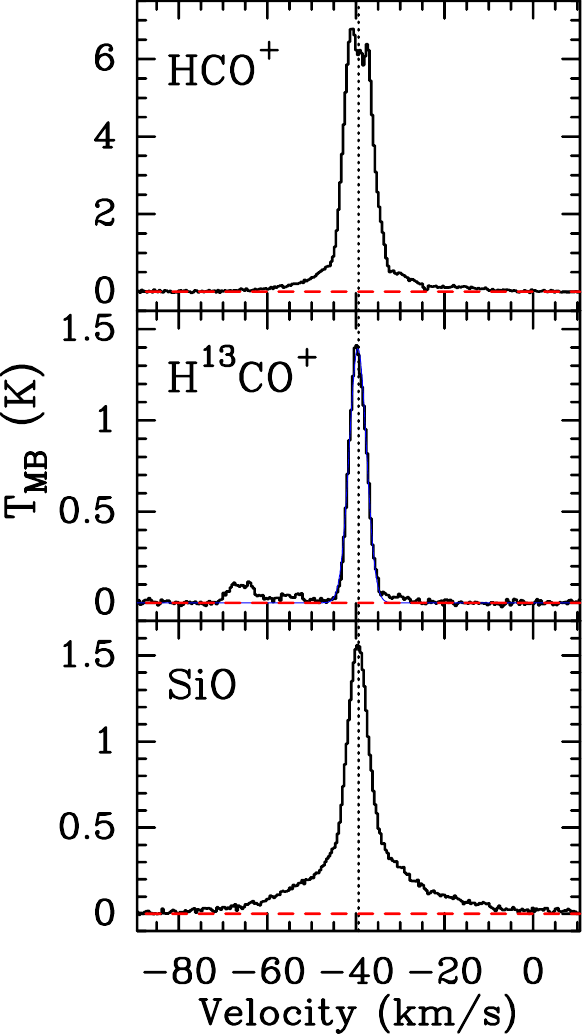}
			\caption{Spectral Profiles for the source HC08, Group A. The red dotted line is at FWZP, and the blue line in the H$^{13}$CO$^+$ panel shows the Gaussian fit. 
			}
			\label{fig: HC08 lines}
		\end{figure}

		This source (see Fig. \ref{fig: HC08 lines}) presents the most intense ($T_{peak} = 1.57$ K) SiO emission of our sample and the most massive outflow ($M_{out} = 116 M_{\odot}$). There are multiple infrared and radio sources deeply embedded in this clump \citep{HC08_Henning_2000, HC08_Murphy_2010}, and it presents one of the strongest molecular outflows in the southern sky \citep{HC08_Araya_2005}. This suggests it is already in an advanced stage of evolution (\cite{HC08_Dedes_2010}). Finally, \cite{HC08_Duronea_2020} present a study of the IR bubble S169 associated with this source, and conclude that star formation is actively taking place.
		
		\subsection{HC17: HIGALBM331.2786-0.1883, IRAS 16076-5134}
		
\begin{figure}[h]
			\centering
			\includegraphics[width=0.3\textwidth]{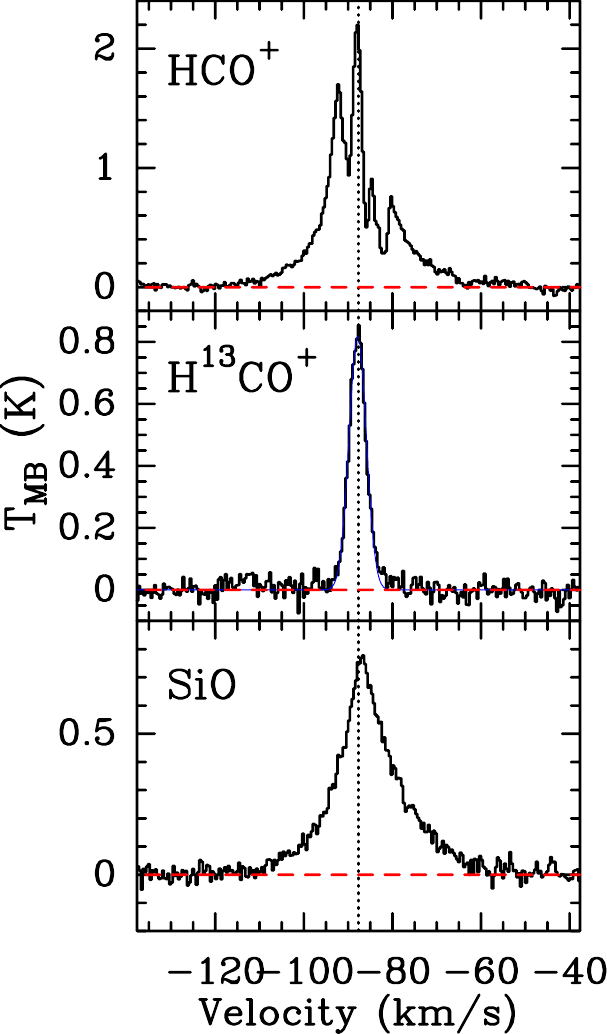}
			\caption{Spectral Profiles for the source HC17, Group A. The red dotted line is at FWZP, and the blue line in the H$^{13}$CO$^+$ panel shows the Gaussian fit. 
			}
			\label{fig: HC17 lines}
		\end{figure}

		This source presents one of the strongest outflows in our sample, which is clearly seen in both its HCO$^+$ and SiO lines (see Fig. \ref{fig: HC17 lines}). Its HCO$^+$ line presents significant absorption and the SiO profile is one of the widest in our sample (FWZP = 52.26 km s$^{-1}$). It has been an object of interest in numerous studies. \cite{HC17_HC27_Zinchenko_2000} present C$^{18}$O and HNCO observations and find that HNCO emission is most likely related to shocks of matter, similar to SiO emission. They detected HNCO emission in this source. \cite{HC17_Lee_2001, HC17_Lee_2002} reported shock-excited H$_2$ emission, in addition to an outflow powered by a deeply embedded object in the former. Furthermore, \cite{HC17_DeBuizer_2009} present single-dish SiO(6–5) observations and evidence of outflow, and \cite{HC17_Baug_2020} present images of the CO outflow observed with ALMA.
		
		\subsection{HC20: HIGALBM333.1246-0.4244, IRAS 16172-5028 }
		
\begin{figure}[h]
			\centering
			\includegraphics[width=0.3\textwidth]{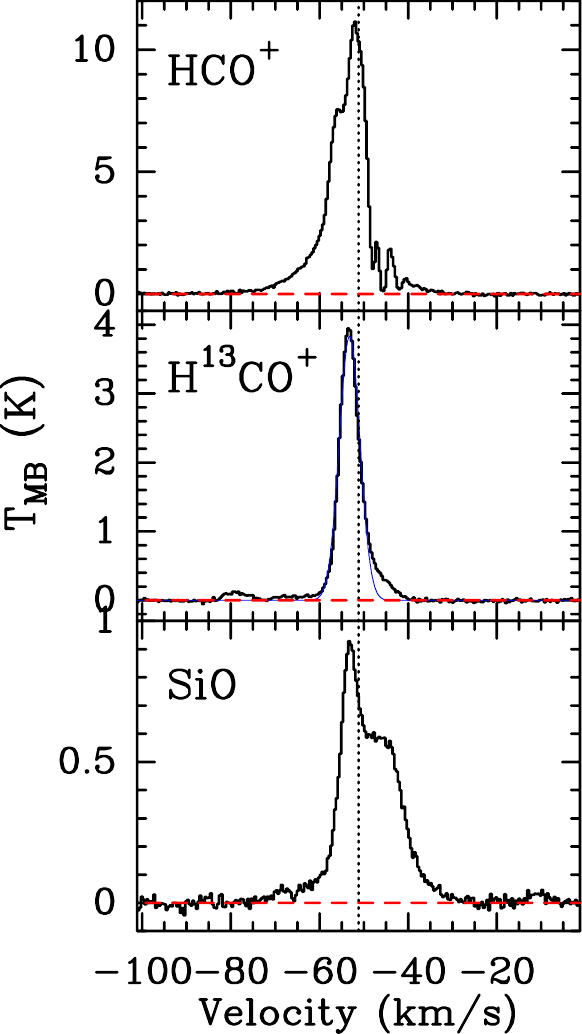}
			\caption{Spectral Profiles for the source HC20, Group A. The red dotted line is at FWZP, and the blue line in the H$^{13}$CO$^+$ panel shows the Gaussian fit. 
			}
			\label{fig: HC20 lines}
		\end{figure}

		The SiO intensity and the clear asymmetry of the spectral lines of this source make it particularly interesting. As is clear in Fig. \ref{fig: HC20 lines}, its HCO$^+$ line experiences more absorption on the blue side than on the red side. The H$^{13}$CO$^+$ even shows a wing on the red side, indicating that the outflow is particularly intense. The SiO line has a particular profile: its intensity decreases at the same velocity as the HCO$^+$ absorption and the H$^{13}$CO$^+$ wing. This suggests that there might be a second colder cloud at around 50 km s$^{-1}$. \cite{RojasGarcia_2022} map this source at 216.968 GHz and also report the possibility of a second source nearby. They also present a chemical profile of this source, where several COMs and SiO are detected. This source was also studied by \cite{HC20_Lo_2007}, where strong SiO and water maser emissions were detected. \cite{HC20_Lo_2014} present a [CI] mapping of the source. Moreover, \cite{HC20_Figueredo_2005} show this source contains an embedded OB star cluster in very early evolutionary stages, and the presence of infall was confirmed by \cite{HC20_Wiles_2016}.
		
		\subsection{HC21: HIGALBM335.5848-0.2894, IRAS 16272-4837}
		
\begin{figure}[h]
			\centering
			\includegraphics[width=0.3\textwidth]{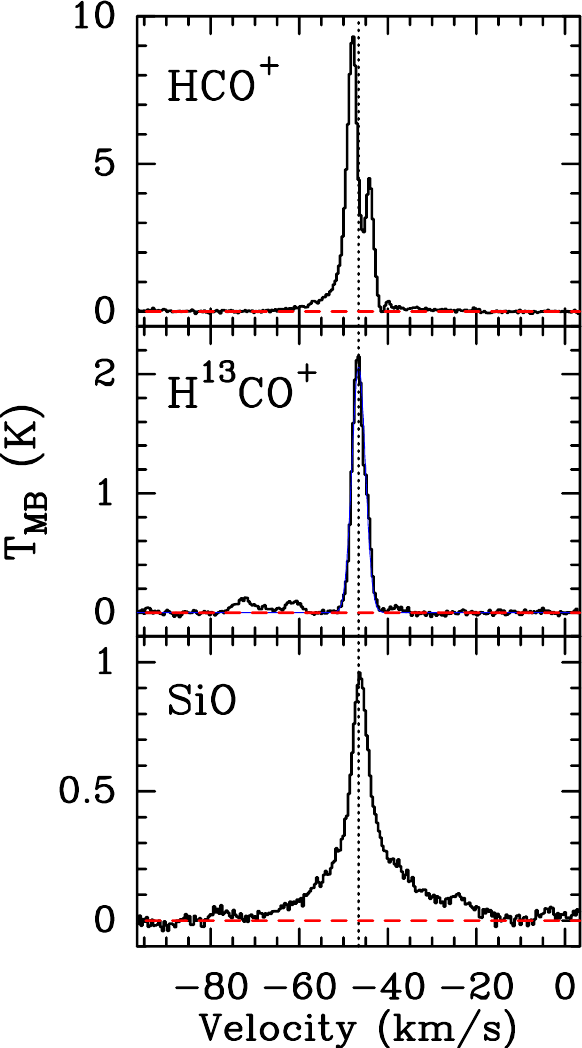}
			\caption{Spectral Profiles for the source HC21, Group A. The red dotted line is at FWZP, and the blue line in the H$^{13}$CO$^+$ panel shows the Gaussian fit. 
			}
			\label{fig: HC21 lines}
		\end{figure}

		This source presents one of the most intense ($97.0$ km s$^{-1}$) and wide ($57.41$ km s$^{-1}$) SiO profiles in the sample, as clearly seen in Fig. \ref{fig: HC21 lines}. The studies done on this source include the discovery of water emission toward this source \citep{HC21_van_der_Tak_2019}, which has also been used as an evolutionary indicator because it can only be detected when the inner hot gas has reached a temperature of at least 300 K \citep{Tak_2004}. Emission of HNCO, another shock tracer, was found by \cite{HC21_Yu_2018}. Moreover, \cite{HC21_Garay_2002} suggest that there is a young massive protostar embedded in this clump and it is still undergoing an intense accretion phase. This corresponds well with our findings and further suggests this is a HMC. Finally, ALMA observations of this source are presented by \cite{HC21_Olguin_2021}, where evidence of infall and rotation motions was found.
		
		\subsection{HC27: HIGALBM345.0035-0.2239, IRAS 17016-4124 }
		
\begin{figure}[h]
			\centering
			\includegraphics[width=0.3\textwidth]{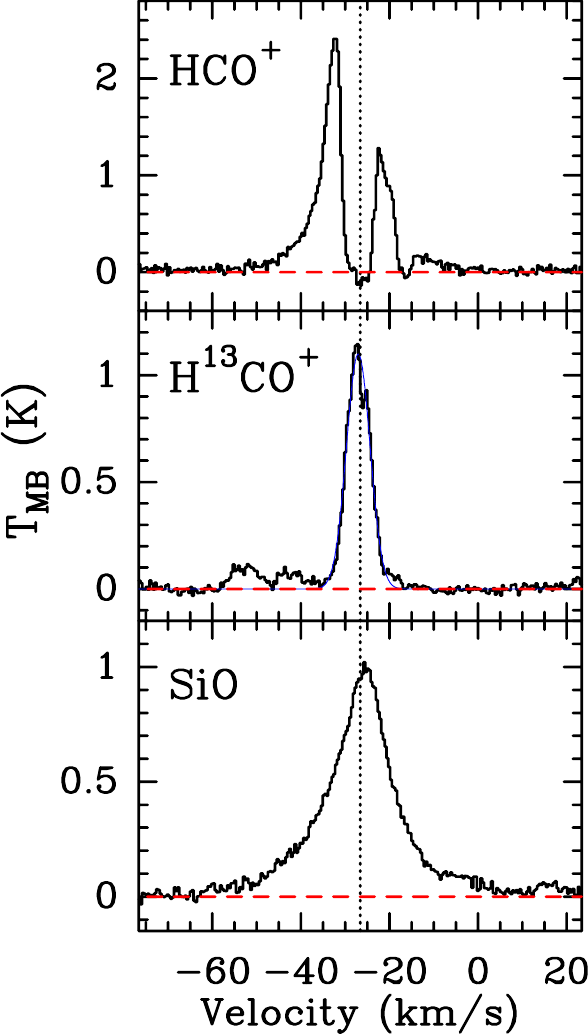}
			\caption{Spectral Profiles for the source HC27, Group B. The red dotted line is at FWZP, and the blue line in the H$^{13}$CO$^+$ panel shows the Gaussian fit. 
			}
			\label{fig: HC27 lines}
		\end{figure}

		This source presents a very wide (68.36 km s$^{-1}$) and intense (1.02 K) SiO emission (see Fig. \ref{fig: HC27 lines}). In addition, its HCO$^+$ profile presents significant self-absorption and absorption. These characteristics make it a particularly interesting source. Several studies have conducted observations towards this source. \cite{HC17_HC27_Zinchenko_2000} present C$^{18}$O and HNCO emission observations, where it was found that high-velocity gas enhances the abundance of HNCO. Furthermore, mid-infrared observations are presented by \cite{HC27_Morales_2009}, and \cite{HC27_Testi_2009} present H$_2$O maser emission detection. \cite{HC27_Krishnan_2013} and \cite{HC27_Voronkov_2014} have found methanol maser emission as well. This source has been mapped with Large APEX BOlometer CAmera (LABOCA) at APEX by \cite{HC17_Gomez_2014}. Finally, \cite{HC27_Yu_2015} present MALT90 (Millimetre Astronomy Legacy Team 90 GHz) observations of the N$_2$H$^+$, H$^{13}$CO$^+$, HCO$^+$, HNC, C$_2$H, HC$_3$N and SiO spectral lines, and conclude that the sources in their sample are associated with dense clumps and recent outflow activity.

		\section{Spectral Profiles at the SiO(4-3), H$^{13}$CO$^+$(2-1) and HCO$^+$(2-1) Lines} \label{spectral_profiles}
		
		\begin{figure}[h]
			\centering
			\includegraphics[width=0.3\textwidth]{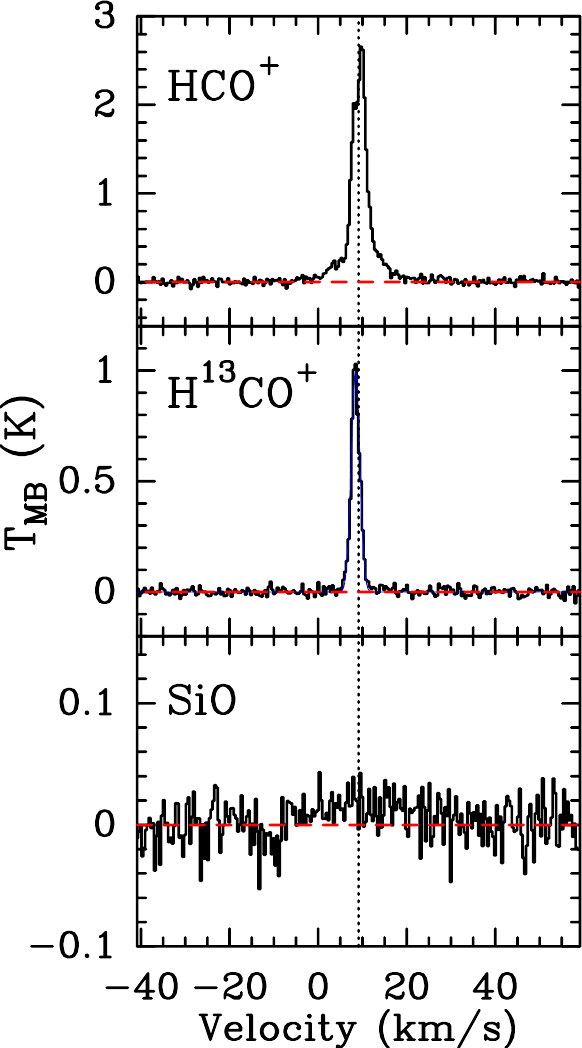}
			\caption{Spectral profiles for the source HC01, Group C. The red dotted line is at FWZP, and the blue line in the H$^{13}$CO$^+$ panel shows the Gaussian fit.
			}
			\label{fig: HC01 lines}
		\end{figure}
		
		\begin{figure}[h]
			\centering
			\includegraphics[width=0.3\textwidth]{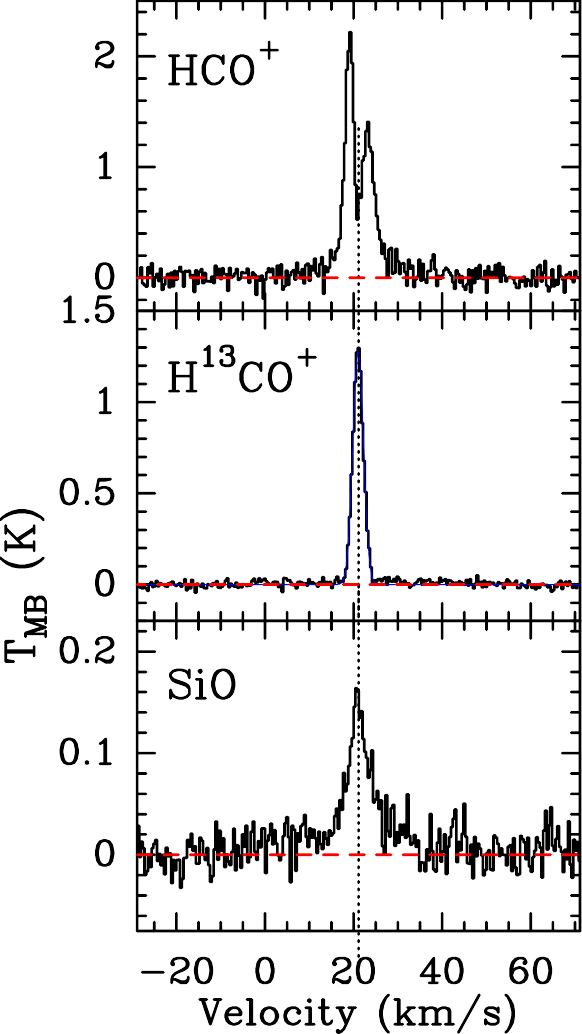}
			\caption{Spectral Profiles for the source HC03, Group B. The red dotted line is at FWZP, and the blue line in the H$^{13}$CO$^+$ panel shows the Gaussian fit. 
			}
			\label{fig: HC03 lines}
		\end{figure}
		
		\begin{figure}[h]
			\centering
			\includegraphics[width=0.3\textwidth]{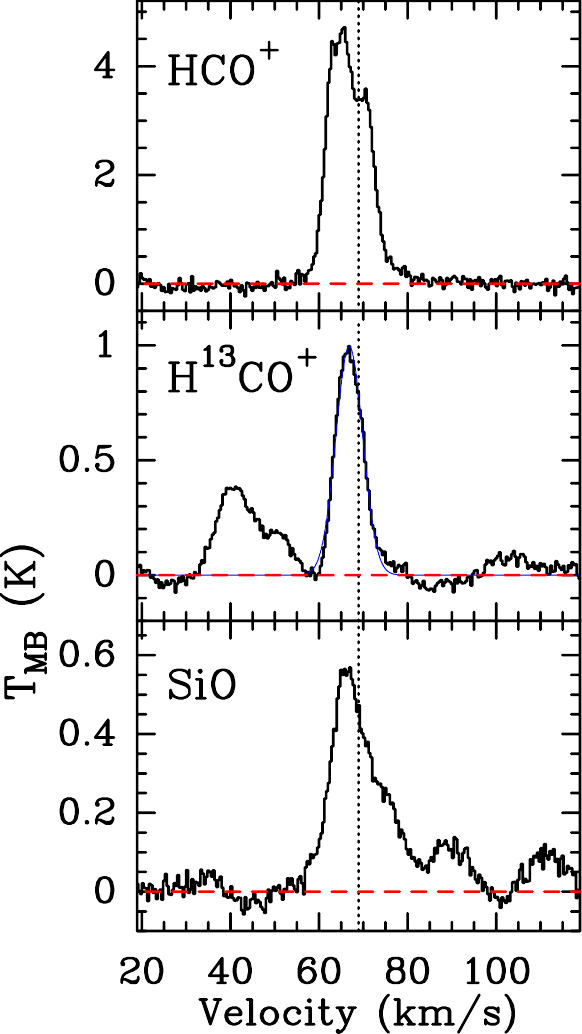}
			\caption{Spectral Profiles for the source HC04, Group B. The red dotted line is at FWZP, and the blue line in the H$^{13}$CO$^+$ panel shows the Gaussian fit. 
			}
			\label{fig: HC04 lines}
		\end{figure}
		
		\begin{figure}[h]
			\centering
			\includegraphics[width=0.3\textwidth]{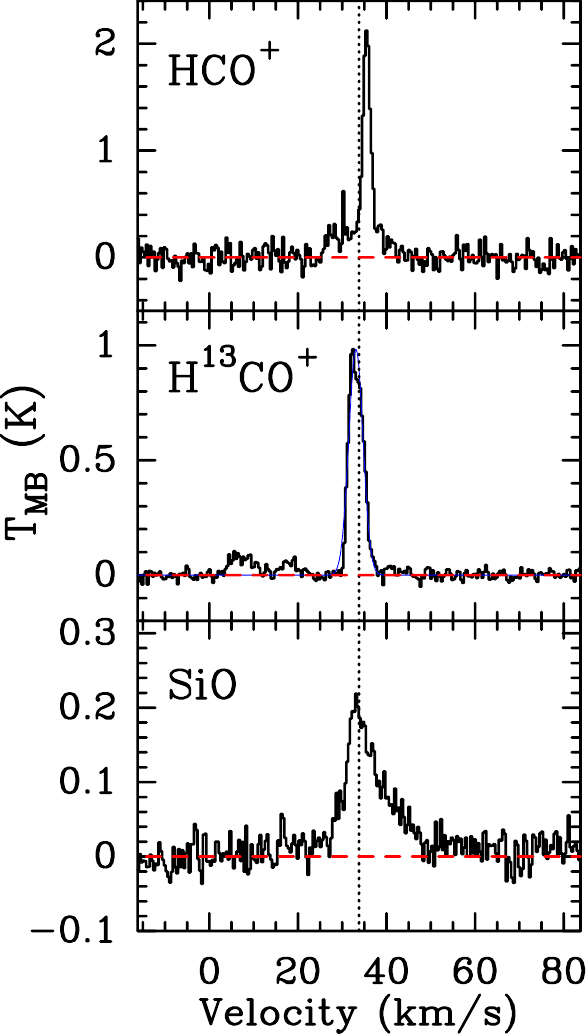}
			\caption{Spectral Profiles for the source HC05, Group B. The red dotted line is at FWZP, and the blue line in the H$^{13}$CO$^+$ panel shows the Gaussian fit. 
			}
			\label{fig: HC05 lines}
		\end{figure}
		
		\begin{figure}[h]
			\centering
			\includegraphics[width=0.3\textwidth]{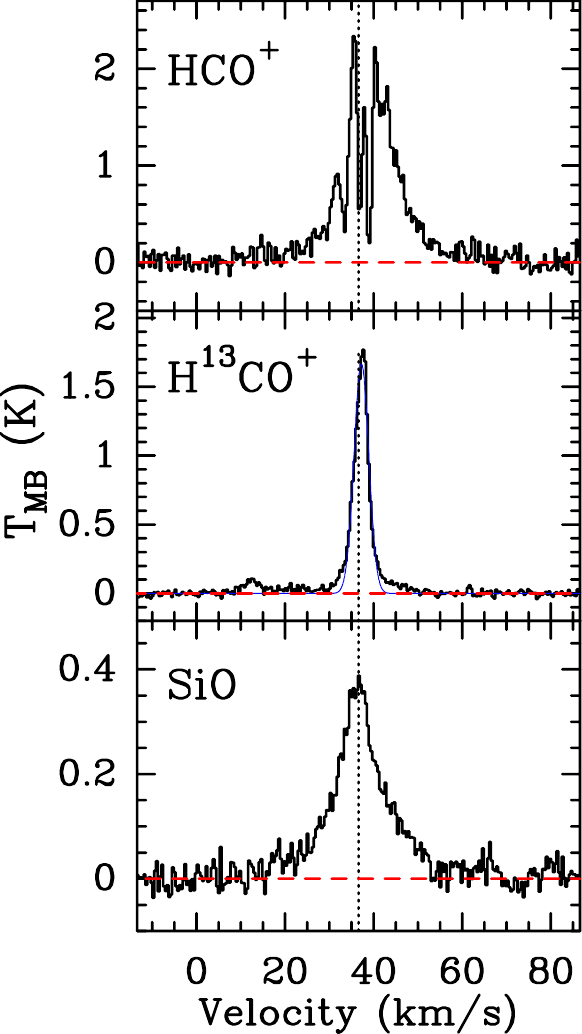}
			\caption{Spectral Profiles for the source HC06, Group B. The red dotted line is at FWZP, and the blue line in the H$^{13}$CO$^+$ panel shows the Gaussian fit. 
			}
			\label{fig: HC06 lines}
		\end{figure}
		
		\begin{figure}[h]
			\centering
			\includegraphics[width=0.3\textwidth]{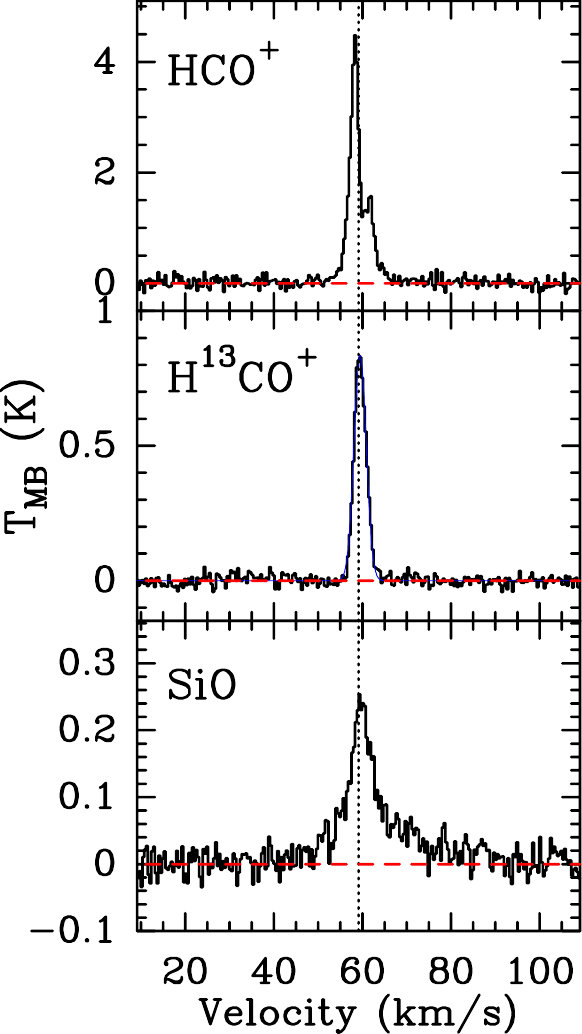}
			\caption{Spectral Profiles for the source HC07, Group B. The red dotted line is at FWZP, and the blue line in the H$^{13}$CO$^+$ panel shows the Gaussian fit. 
			}
			\label{fig: HC07 lines}
		\end{figure}
		
		\begin{figure}[h]
			\centering
			\includegraphics[width=0.3\textwidth]{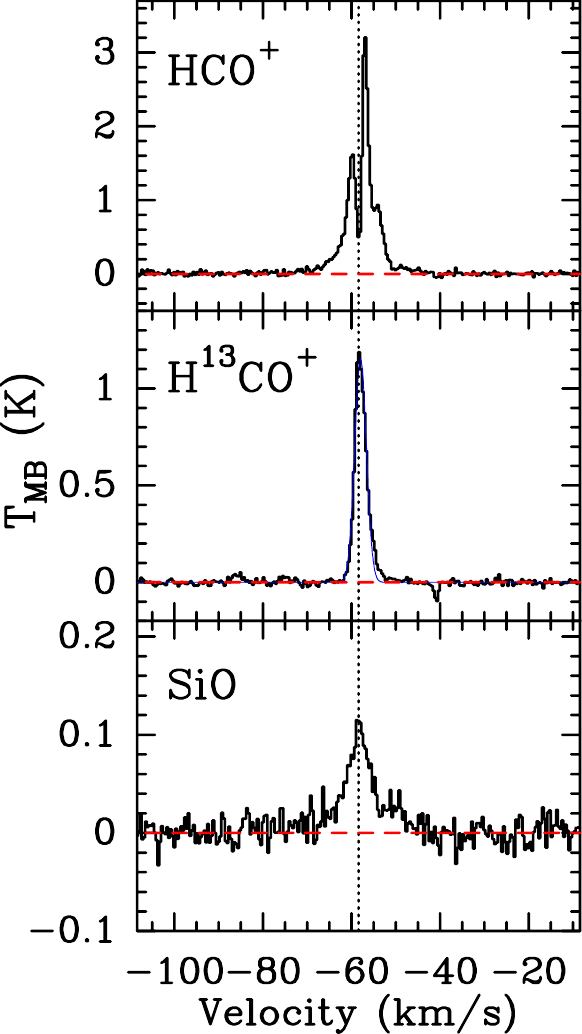}
			\caption{Spectral Profiles for the source HC09, Group B. The red dotted line is at FWZP, and the blue line in the H$^{13}$CO$^+$ panel shows the Gaussian fit. 
			}
			\label{fig: HC09 lines}
		\end{figure}
		
		\begin{figure}[h]
			\centering
			\includegraphics[width=0.3\textwidth]{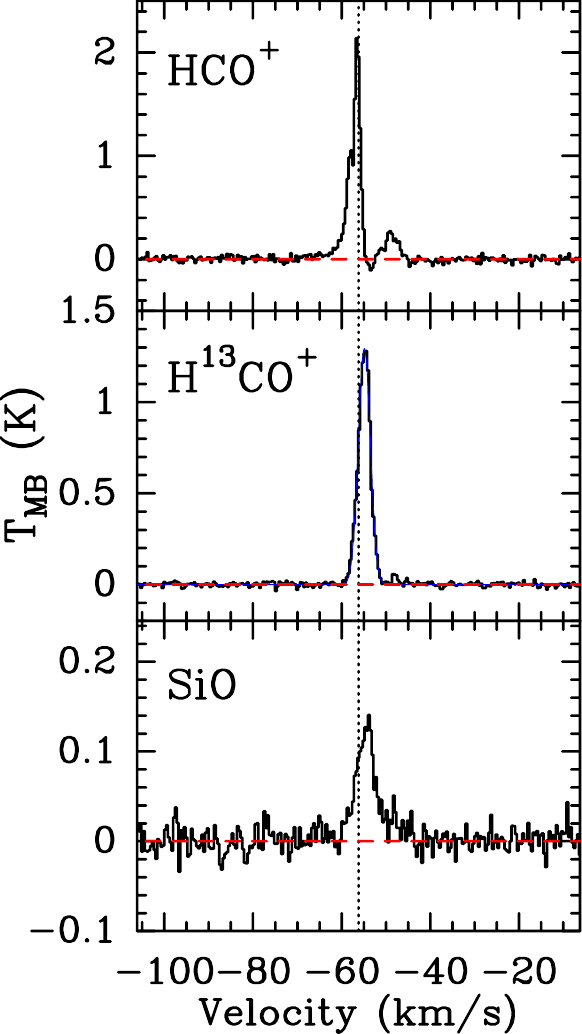}
			\caption{Spectral Profiles for the source HC10, Group B. The red dotted line is at FWZP, and the blue line in the H$^{13}$CO$^+$ panel shows the Gaussian fit. 
			}
			\label{fig: HC10 lines}
		\end{figure}
		
		\begin{figure}[h]
			\centering
			\includegraphics[width=0.3\textwidth]{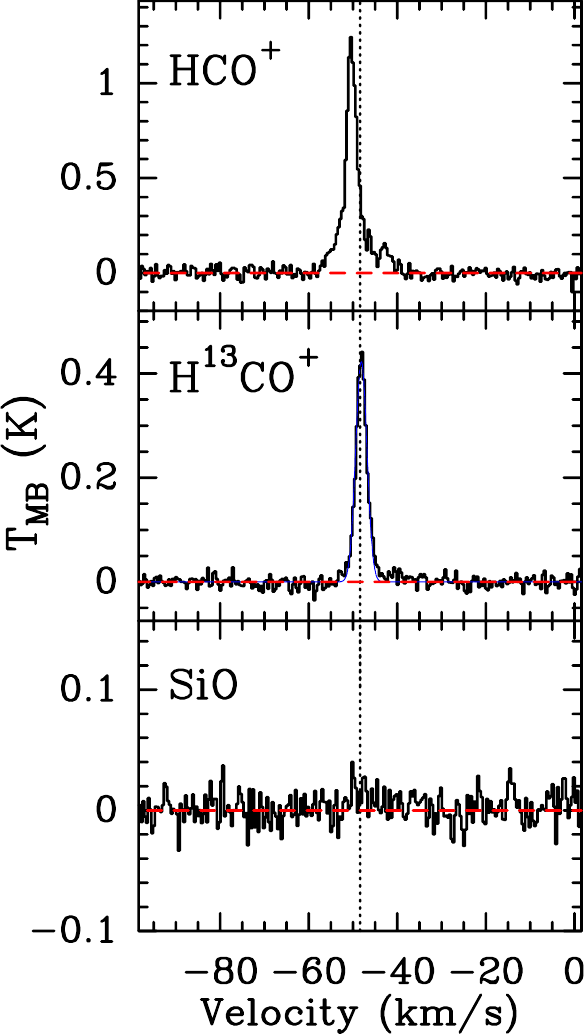}
			\caption{Spectral Profiles for the source HC11, Group C. The red dotted line is at FWZP, and the blue line in the H$^{13}$CO$^+$ panel shows the Gaussian fit. 
			}
			\label{fig: HC11 lines}
		\end{figure}
		
		\begin{figure}[h]
			\centering
			\includegraphics[width=0.3\textwidth]{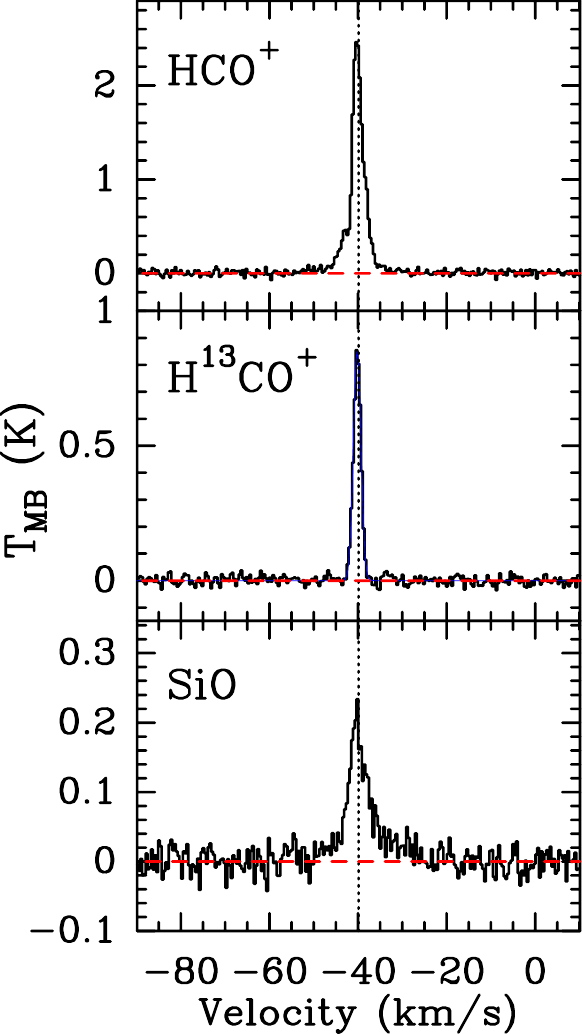}
			\caption{Spectral Profiles for the source HC12, Group B. The red dotted line is at FWZP, and the blue line in the H$^{13}$CO$^+$ panel shows the Gaussian fit. 
			}
			\label{fig: HC12 lines}
		\end{figure}
		
		\begin{figure}[h]
			\centering
			\includegraphics[width=0.3\textwidth]{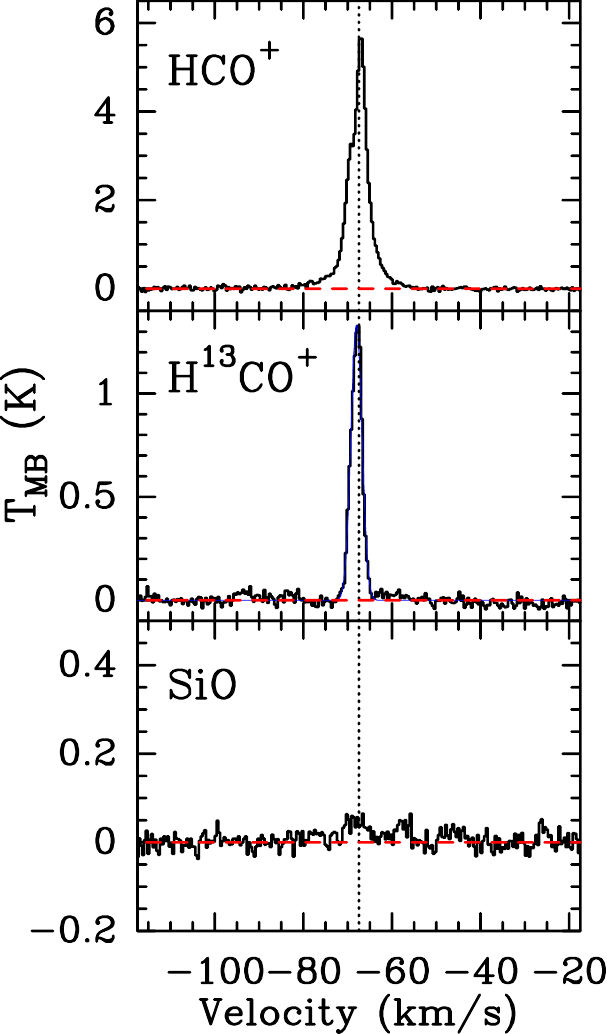}
			\caption{Spectral Profiles for the source HC13, Group C. The red dotted line is at FWZP, and the blue line in the H$^{13}$CO$^+$ panel shows the Gaussian fit. 
			}
			\label{fig: HC13 lines}
		\end{figure}
		
		\begin{figure}[h]
			\centering
			\includegraphics[width=0.3\textwidth]{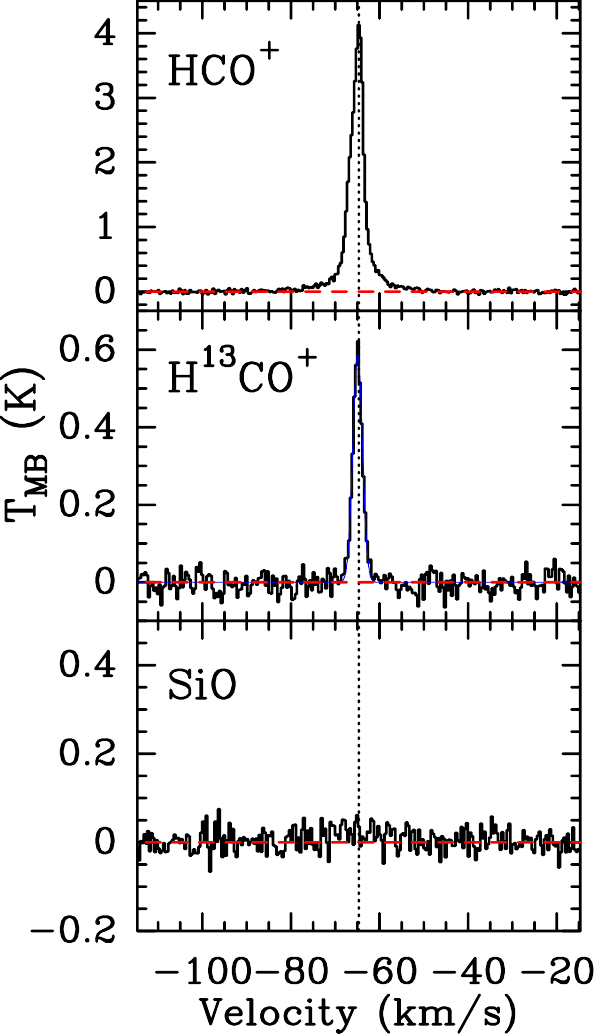}
			\caption{Spectral Profiles for the source HC14, Group C. The red dotted line is at FWZP, and the blue line in the H$^{13}$CO$^+$ panel shows the Gaussian fit. 
			}
			\label{fig: HC14 lines}
		\end{figure}
		
		\begin{figure}[h]
			\centering
			\includegraphics[width=0.3\textwidth]{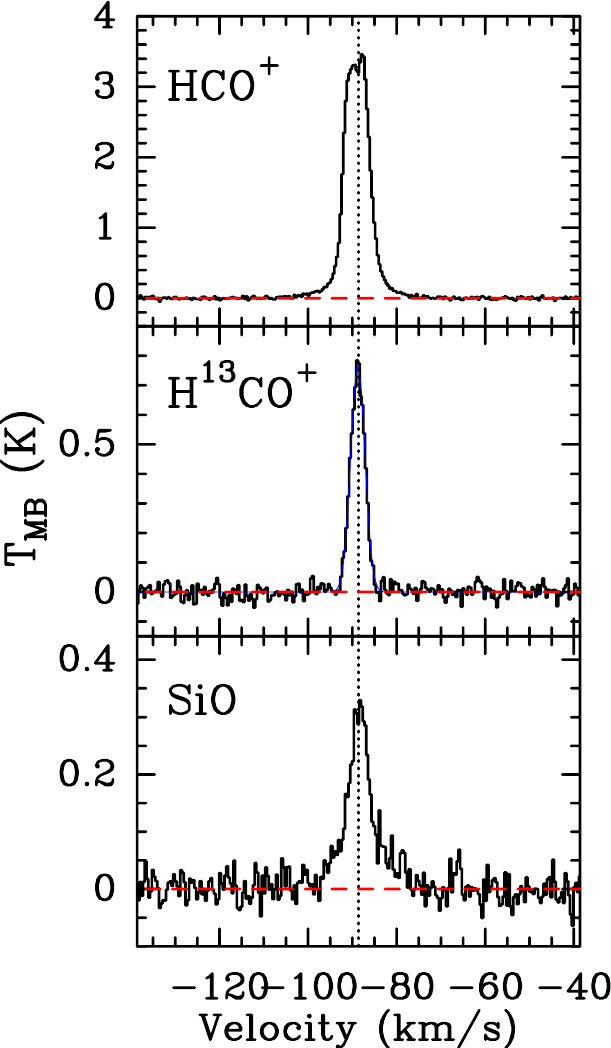}
			\caption{Spectral Profiles for the source HC15, Group B. The red dotted line is at FWZP, and the blue line in the H$^{13}$CO$^+$ panel shows the Gaussian fit. 
			}
			\label{fig: HC15 lines}
		\end{figure}
		
		\begin{figure}[h]
			\centering
			\includegraphics[width=0.3\textwidth]{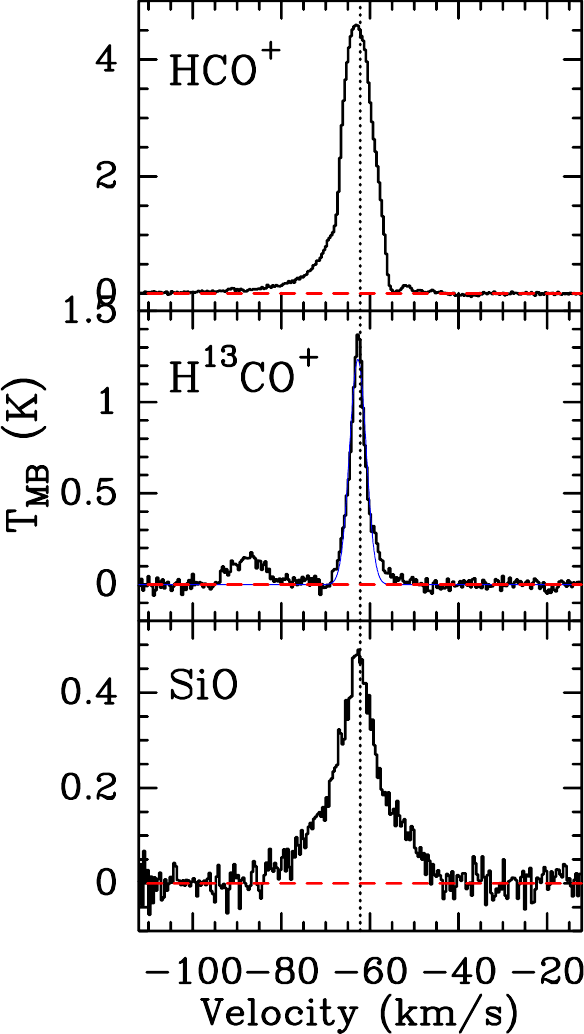}
			\caption{Spectral Profiles for the source HC16, Group A. The red dotted line is at FWZP, and the blue line in the H$^{13}$CO$^+$ panel shows the Gaussian fit. 
			}
			\label{fig: HC16 lines}
		\end{figure}
		
		\begin{figure}[h]
			\centering
			\includegraphics[width=0.3\textwidth]{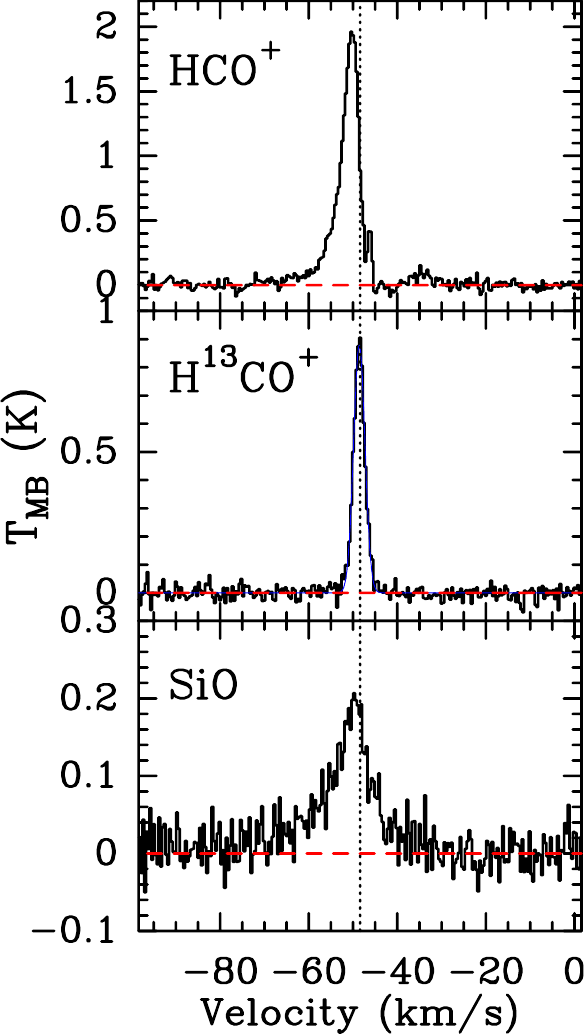}
			\caption{Spectral Profiles for the source HC18, Group A. The red dotted line is at FWZP, and the blue line in the H$^{13}$CO$^+$ panel shows the Gaussian fit. 
			}
			\label{fig: HC18 lines}
		\end{figure}
		
		\begin{figure}[h]
			\centering
			\includegraphics[width=0.3\textwidth]{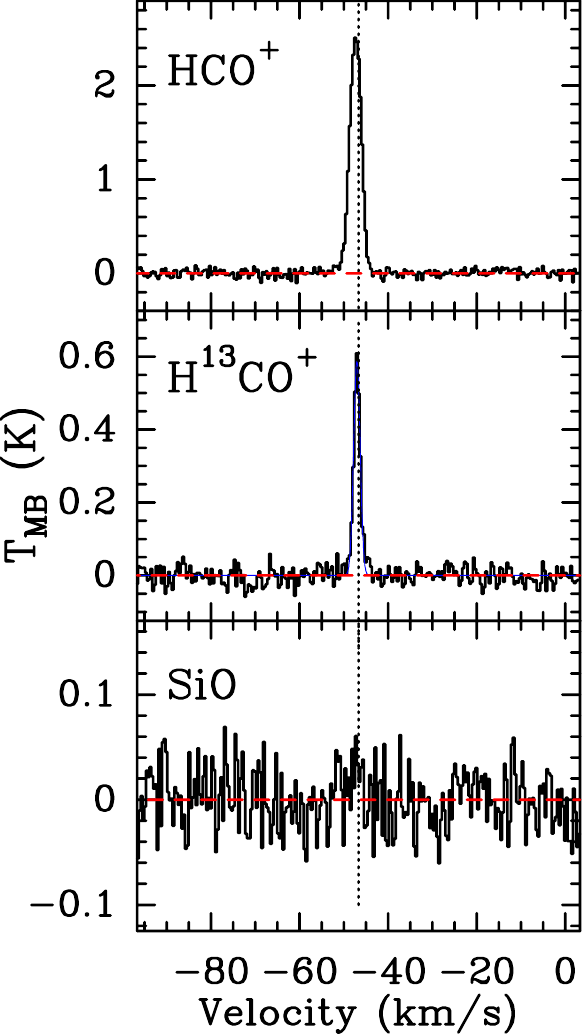}
			\caption{Spectral Profiles for the source HC19, Group C. The red dotted line is at FWZP, and the blue line in the H$^{13}$CO$^+$ panel shows the Gaussian fit. 
			}
			\label{fig: HC19 lines}
		\end{figure}
		
		\begin{figure}[h]
			\centering
			\includegraphics[width=0.3\textwidth]{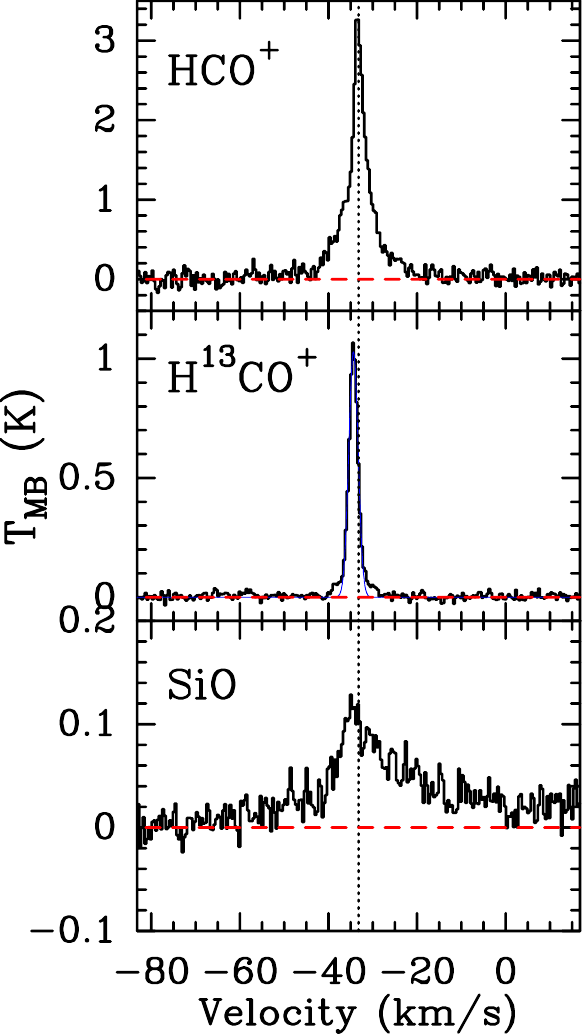}
			\caption{Spectral Profiles for the source HC22, Group A. The red dotted line is at FWZP, and the blue line in the H$^{13}$CO$^+$ panel shows the Gaussian fit. 
			}
			\label{fig: HC22 lines}
		\end{figure}
		
		\begin{figure}[h]
			\centering
			\includegraphics[width=0.3\textwidth]{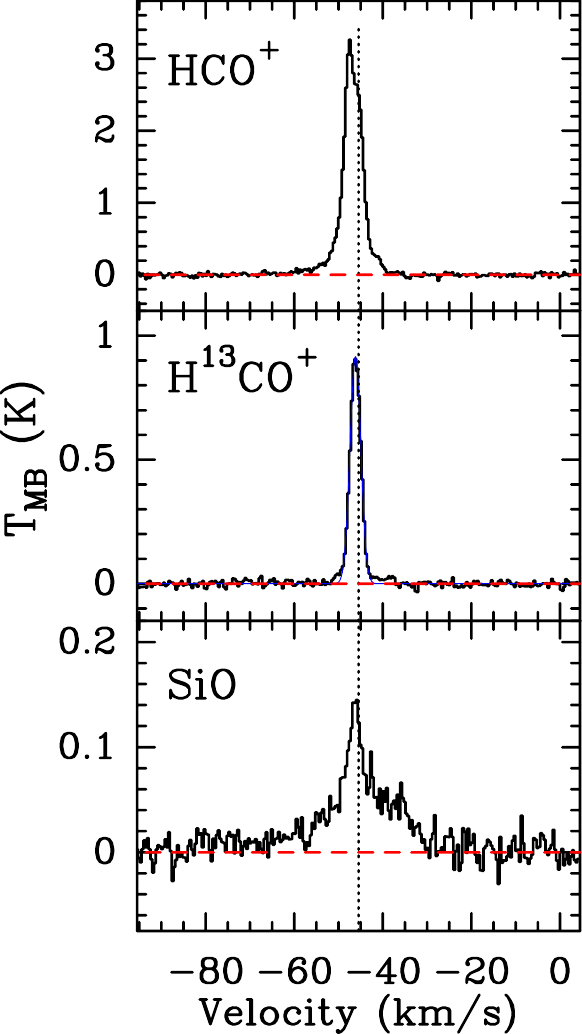}
			\caption{Spectral Profiles for the source HC23, Group B. The red dotted line is at FWZP, and the blue line in the H$^{13}$CO$^+$ panel shows the Gaussian fit. 
			}
			\label{fig: HC23 lines}
		\end{figure}
		
		\begin{figure}[h]
			\centering
			\includegraphics[width=0.3\textwidth]{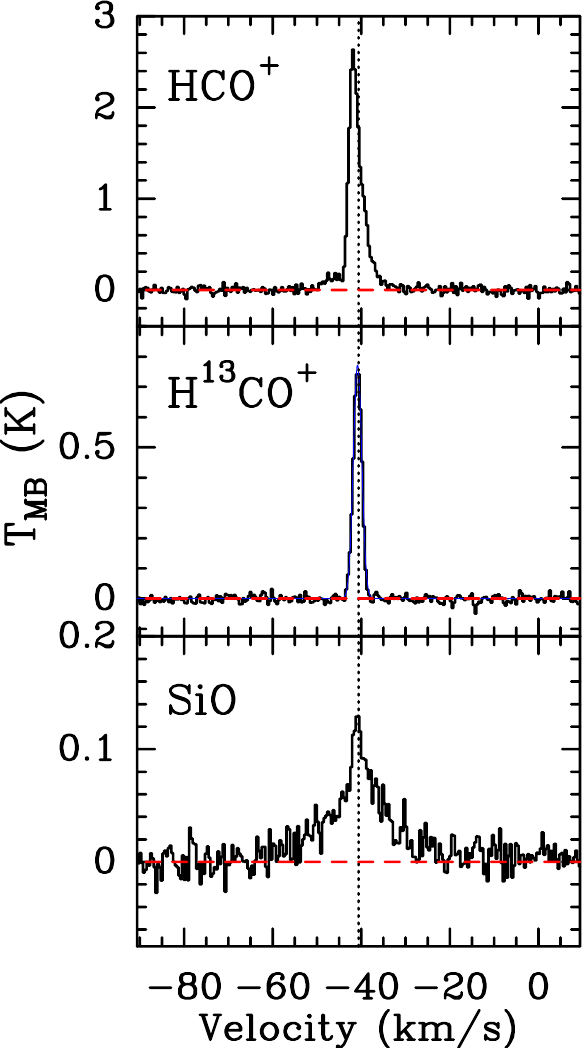}
			\caption{Spectral Profiles for the source HC24, Group B. The red dotted line is at FWZP, and the blue line in the H$^{13}$CO$^+$ panel shows the Gaussian fit. 
			}
			\label{fig: HC24 lines}
		\end{figure}
		
		\begin{figure}[h]
			\centering
			\includegraphics[width=0.3\textwidth]{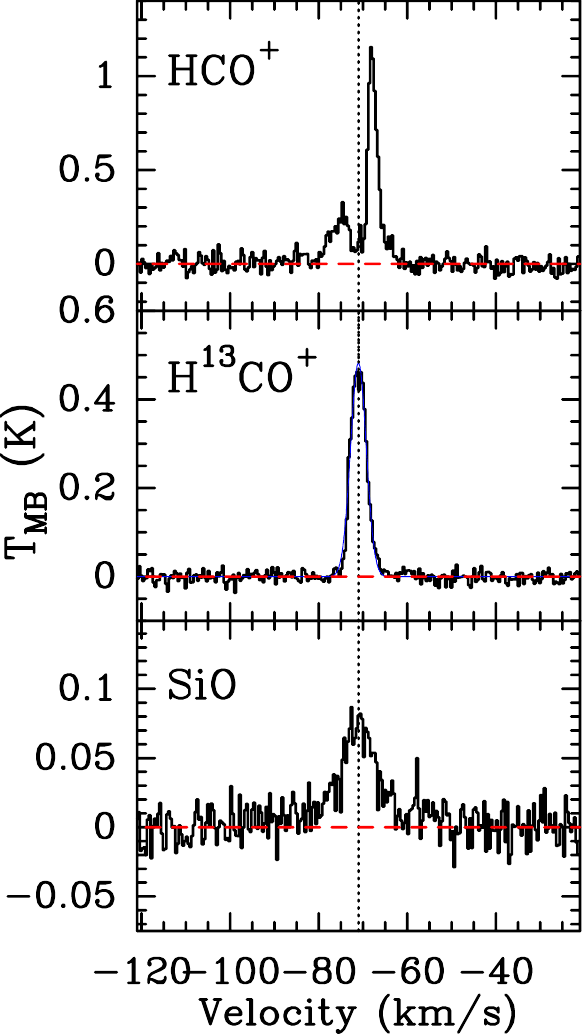}
			\caption{Spectral Profiles for the source HC25, Group B. The red dotted line is at FWZP, and the blue line in the H$^{13}$CO$^+$ panel shows the Gaussian fit. 
			}
			\label{fig: HC25 lines}
		\end{figure}
		
		\begin{figure}[h]
			\centering
			\includegraphics[width=0.3\textwidth]{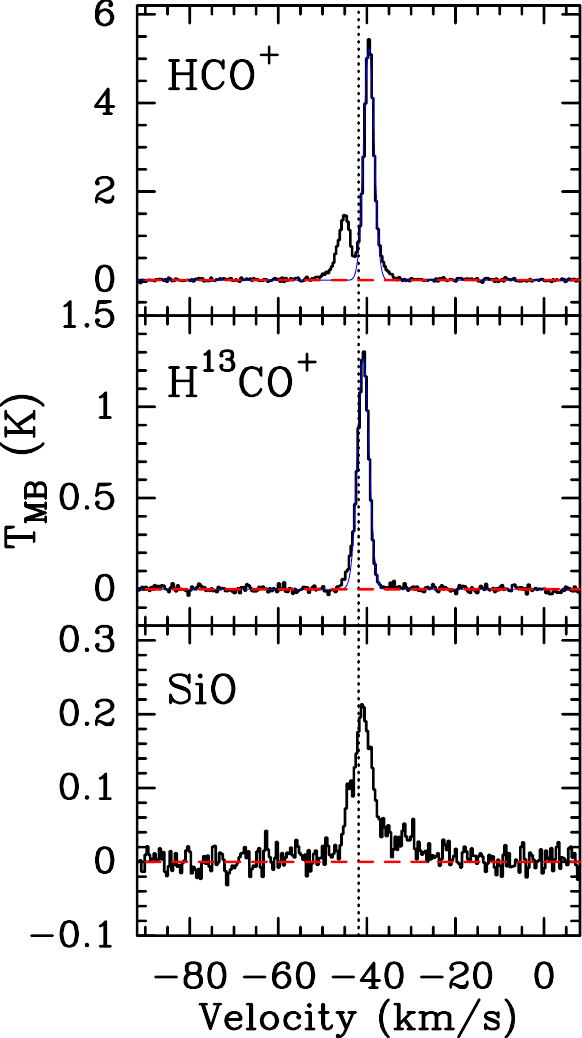}
			\caption{Spectral Profiles for the source HC26, Group B. The red dotted line is at FWZP, and the blue line in the H$^{13}$CO$^+$ panel shows the Gaussian fit. 
			}
			\label{fig: HC26 lines}
		\end{figure}
		
		\begin{figure}[h]
			\centering
			\includegraphics[width=0.3\textwidth]{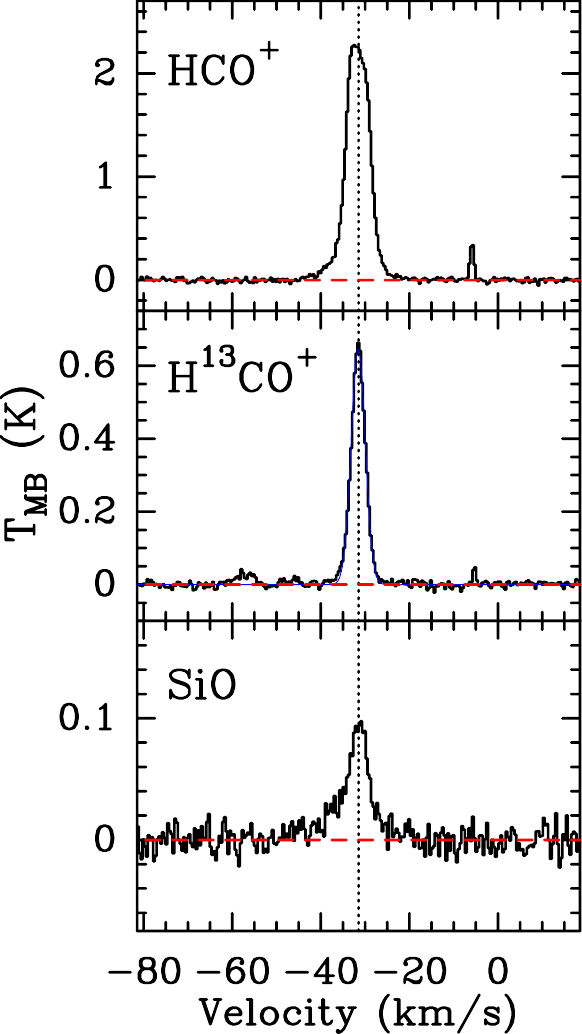}
			\caption{Spectral Profiles for the source HC28, Group B. The red dotted line is at FWZP, and the blue line in the H$^{13}$CO$^+$ panel shows the Gaussian fit. 
			}
			\label{fig: HC28 lines}
		\end{figure}
		
		\begin{figure}[h]
			\centering
			\includegraphics[width=0.3\textwidth]{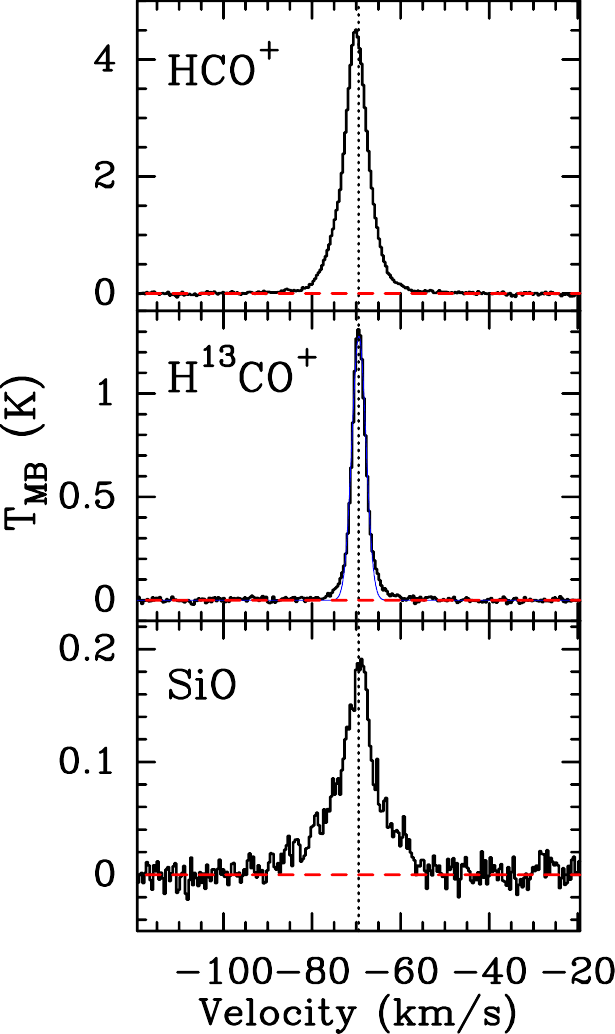}
			\caption{Spectral Profiles for the source HC29, Group A. The red dotted line is at FWZP, and the blue line in the H$^{13}$CO$^+$ panel shows the Gaussian fit. 
			}
			\label{fig: HC29 lines}
		\end{figure}
		
		\begin{figure}[h]
			\centering
			\includegraphics[width=0.3\textwidth]{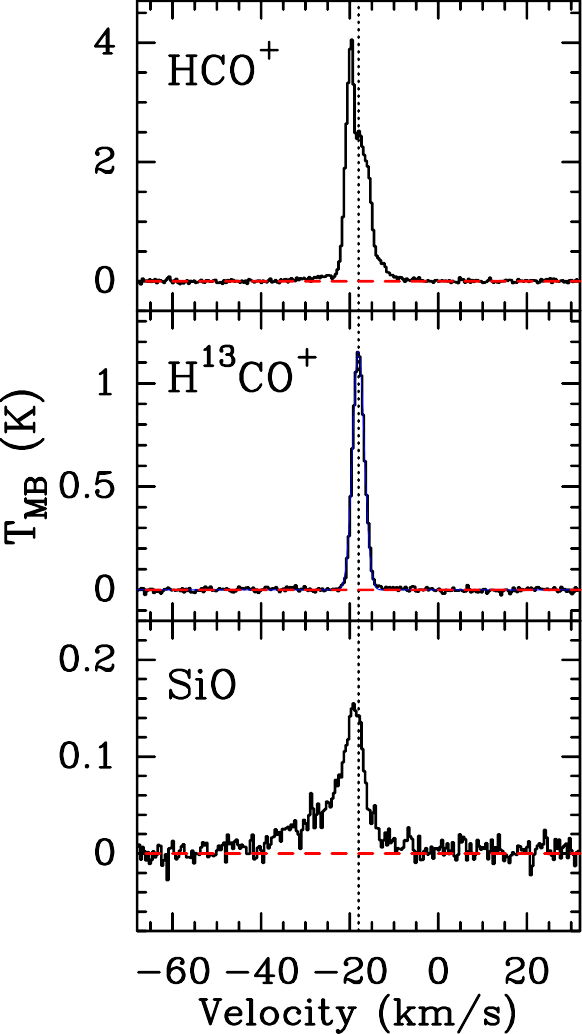}
			\caption{Spectral Profiles for the source HC30, Group B. The red dotted line is at FWZP, and the blue line in the H$^{13}$CO$^+$ panel shows the Gaussian fit. 
			}
			\label{fig: HC30 lines}
		\end{figure}
		
		\begin{figure}[h]
			\centering
			\includegraphics[width=0.3\textwidth]{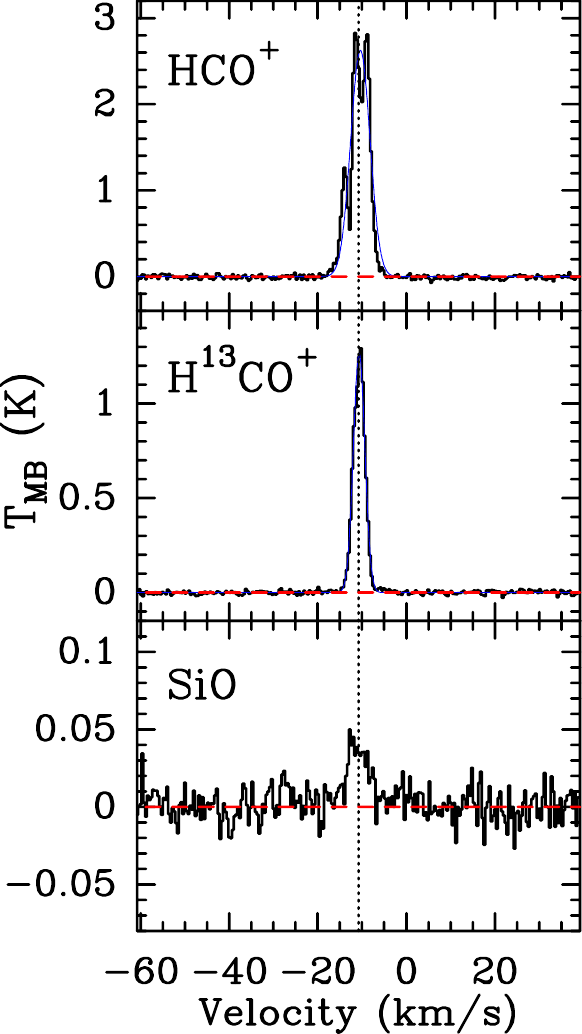}
			\caption{Spectral Profiles for the source HC31, Group C. The red dotted line is at FWZP, and the blue line in the H$^{13}$CO$^+$ panel shows the Gaussian fit. 
			}
			\label{fig: HC31 lines}
		\end{figure}
		
		\begin{figure}[h]
			\centering
			\includegraphics[width=0.3\textwidth]{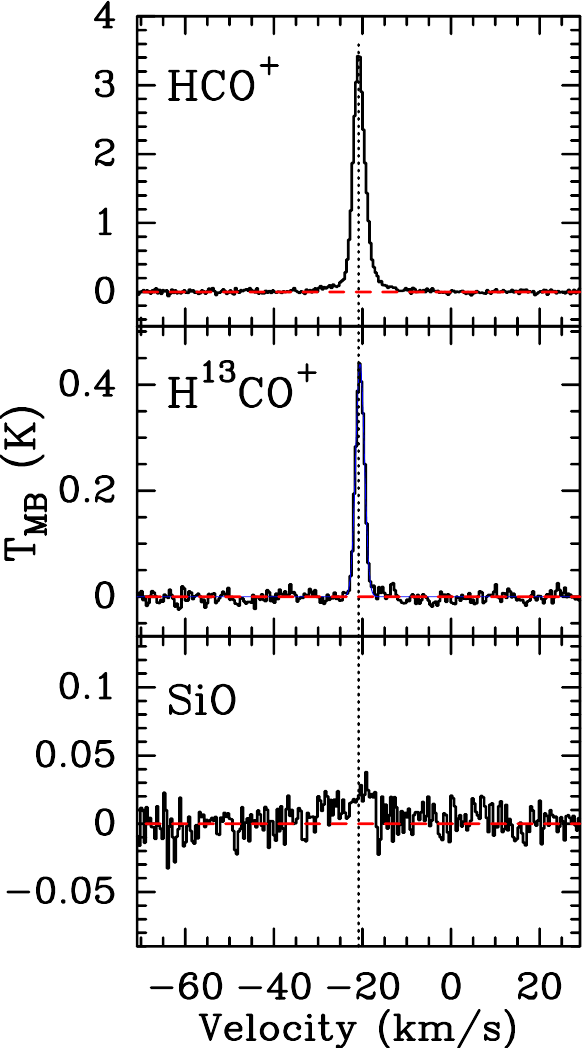}
			\caption{Spectral Profiles for the source HC32, Group C. The red dotted line is at FWZP, and the blue line in the H$^{13}$CO$^+$ panel shows the Gaussian fit. 
			}
			\label{fig: HC32 lines}
		\end{figure}
		
	\end{appendix}
	
\end{document}